\documentclass[journal,10pt]{IEEEtran}

\ifCLASSINFOpdf
 
\else
  
\fi

\usepackage{amsmath,amssymb,amsthm}
\usepackage{color}
\usepackage[noadjust]{cite}
\usepackage{graphicx}
\usepackage{epsfig}
\usepackage{breqn}
\usepackage[caption=false,font=footnotesize]{subfig}
\usepackage{array}

\newtheorem{theorem}{Theorem}[section]
\newtheorem{lemma}[theorem]{Lemma}

\newtheorem{prop}[theorem]{Proposition}

\newtheorem{fact}[theorem]{Fact}

\allowdisplaybreaks
\newcommand\given[1][]{\:#1\vert\:}

\begin{document}
\bstctlcite{IEEEexample:BSTcontrol}

\title{$k$-connectivity of inhomogeneous random key graphs with unreliable links}

\author{Rashad~Eletreby,~\IEEEmembership{Student Member,~IEEE,}
        and~Osman~Ya\u{g}an,~\IEEEmembership{Member,~IEEE,}
\thanks{R. Eletreby and O. Ya\u{g}an are with the Department
of Electrical and Computer Engineering and CyLab, Carnegie Mellon University, Pittsburgh,
PA, 15213 USA. E-mail: reletreby@cmu.edu, oyagan@ece.cmu.edu.}}


\maketitle

\begin{abstract}
We consider secure and reliable connectivity in wireless sensor networks that utilize a heterogeneous random key predistribution scheme. We model the unreliability of wireless links by an on-off channel model that induces an Erd\H{o}s-R\'enyi graph, while the 
 heterogeneous scheme induces an inhomogeneous random key graph. The overall network can thus be modeled by the intersection of both graphs.
We present conditions (in the form of zero-one laws) on how to scale the parameters of the intersection model so that with high probability i) all of its nodes are connected to at least $k$ other nodes; i.e., the minimum node degree of the graph is no less than $k$ and ii) the graph is $k$-connected, i.e., the graph remains connected even if {\em any} $k-1$ nodes leave the network. We also present numerical results to support these conditions in the finite-node regime. Our results are shown to complement and generalize several previous work in the literature. 
\end{abstract}


\IEEEpeerreviewmaketitle

\section{Introduction}
\subsection{Motivation and Background}
Wireless sensor networks (WSNs) enable a broad range of applications including military, health, and environmental monitoring, among others \cite{Akyildiz_2002}. A typical WSN consists of hundreds, thousands, or hundreds of thousands of nodes that are often deployed randomly in hostile environments. The ease of deployment, low cost, low power consumption, and small size have paved the way for the proliferation of WSNs, but also rendered them vulnerable to various types of attacks. In fact, security of WSNs is a key challenge given their unique features \cite{security_survey}; e.g., limited computational capabilities, limited transmission power, and vulnerability to node capture attacks. Random key predistribution schemes were proposed to tackle those limitations, and they are currently regarded as the most feasible solutions for securing WSNs; e.g., see \cite[Chapter~13]{Raghavendra_2004} and \cite{camtepe_2005}, and references therein.

Random key predistribution schemes were first introduced in the pioneering work of Eschenauer and Gligor \cite{Gligor_2002}. Their scheme, hereafter referred to as the EG scheme, operates as follows: prior to deployment, each sensor node is assigned a {\em random} set of $K$ cryptographic keys, selected from a key pool of size $P$ (without replacement). After deployment, two nodes can communicate {\em securely} over an existing channel {\em if} they share at least one key. 
The EG scheme led the way to several other variants, including the $q$-composite scheme  \cite{Haowen_2003}, and the random pairwise scheme \cite{Haowen_2003} among others.

Recently, a new variation of the EG scheme, referred to as the heterogeneous key predistribution scheme, was introduced \cite{Yagan/Inhomogeneous}. The heterogeneous scheme considers the case when the network includes sensor nodes with varying levels of resources, features, or connectivity requirements (e.g., regular nodes vs. cluster heads); it is in fact envisioned  \cite{Yarvis_2005} that many WSN applications will be heterogeneous. The scheme is described as follows. 
Given $r$ classes, each sensor is independently classified as a class-$i$ node with probability $\mu_i>0$ 
for each $i=1,\ldots, r$.
Then, sensors in class-$i$ are each assigned $K_i$ keys selected uniformly at random (without replacement) from a key pool of size $P$. Similar to the EG scheme, nodes that share key(s) can communicate securely over an available channel after the deployment; see Section \ref{sec:Model} for details. 

In \cite{Rashad/Inhomo}, the authors considered the reliability of secure WSNs under the heterogeneous key predistribution scheme; namely, when each wireless link fails with probability $1-\alpha$ independently from other links. From a wireless communication perspective, this is similar with investigating the secure connectivity of a WSN under an on/off channel model, wherein each wireless channel is on with probability $\alpha$ independently from other links. There, we established critical conditions on the probability distribution $\pmb{\mu}=\{\mu_1,\mu_2,\ldots,\mu_r \}$, and scaling of the key ring sizes $\pmb{K}=\{K_1,K_2,\ldots,K_r\}$, the key pool size $P$, and the channel parameter $\alpha$ as a function of network size $n$, so that the resulting WSN is securely connected with high probability. 
Although these results form a crucial starting point towards the analysis of the heterogeneous key predistribution scheme, there remains to establish several important properties of the scheme to obtain a full understanding of its performance in securing WSNs. In particular, the connectivity results given in \cite{Rashad/Inhomo} do not guarantee that the network would remain connected when sensors fail due to battery depletion or get captured by an adversary. Moreover, the results are not applicable for {\em mobile} WSNs; wherein, the mobility of sensor nodes may render the network disconnected. In essence, sharper results that guarantee network connectivity in the aforementioned scenarios are needed.

\subsection{Contributions}

The objective of our paper is to address the limitations of the results in \cite{Rashad/Inhomo}. We consider the heterogeneous key predistribtuion scheme under an on/off communication model consisting of independent wireless channels each of which is either on (with probability $\alpha$), or off (with probability $1-\alpha$). We focus on the $k$-connectivity property which implies that the network connectivity is preserved despite the failure of any $(k-1)$ nodes or links \cite{PenroseBook}. Accordingly, $k$-connectivity provides a guarantee of network reliability against the potential failures of sensors or links. Moreover, for a $k$-connected {\em mobile} WSN, any $(k-1)$ nodes are free to move anywhere while the rest of the network remains at least $1$-connected.

Our approach is based on modeling the WSN by an appropriate random graph and then establishing scaling conditions on the model parameters such that  certain desired properties hold with high probability (whp) as the number of nodes $n$ gets large.
The heterogeneous key predistribution scheme induces an inhomogeneous random key graphs \cite{Yagan/Inhomogeneous}, denoted hereafter by $\mathbb{K}(n,\pmb{\mu},\pmb{K},P)$, while the on-off communication model leads to a standard Erd\H{o}s-R\'enyi (ER) graph \cite{ER}, denoted by $\mathbb{G}(n,\alpha)$. Hence, the appropriate overall random graph model is the intersection of an inhomogeneous random key graph with an ER graph, denoted $\mathbb{K} (n;\pmb{\mu},\pmb{K},P) \cap \mathbb{G}(n;\alpha)$. 

We establish two main results for the intersection model $\mathbb{K} (n;\pmb{\mu},\pmb{K},P) \cap \mathbb{G}(n;\alpha)$; namely, i) a zero-one law for the minimum node degree of $\mathbb{K} (n;\pmb{\mu},\pmb{K},P) \cap \mathbb{G}(n;\alpha)$ to be no less than $k$ for any non-negative integer $k$ and ii) a zero-one law for the $k$-connectivity property of $\mathbb{K} (n;\pmb{\mu},\pmb{K},P) \cap \mathbb{G}(n;\alpha)$ for any non-negative integer $k$. More precisely, we present conditions on how to scale the parameters of $\mathbb{K} (n;\pmb{\mu},\pmb{K},P) \cap \mathbb{G}(n;\alpha)$ so that i) its minimum node degree is no less than $k$ and ii) it is $k$-connected, both with high probability when the number of nodes $n$ gets large. Furthermore, we show by simulations that minimum node degree being no less than $k$ and $k$-connectivity properties exhibit almost equal (empirical) probabilities. Not only do our results complement and generalize several previous work in the literature, but they also have broad range of applications to other interesting problems (See Section~\ref{sec:results} for details). 

\subsection{Notation and Conventions}
All limiting statements, including asymptotic equivalence are considered with the number of sensor nodes $n$ going to infinity. The random variables (rvs) under consideration are all defined on the same probability triple $(\Omega,\mathcal{F},\mathbb{P})$. Probabilistic statements are made with respect to this probability measure $\mathbb{P}$, and we denote the corresponding expectation by $\mathbb{E}$. The indicator function of an event $E$ is denoted by $\pmb{1}[E]$. We say that an event holds with high probability (whp) if it holds with probability $1$ as $n \rightarrow \infty$. For any event $E$, we let $\overline{E}$ denote the complement of $E$. For any discrete set $S$, we write $|S|$ for its cardinality. For sets $S_a$ and $S_b$, the relative compliment of $S_a$ in $S_b$ is given by $S_a \setminus S_b$. 
In comparing
the asymptotic behaviors of the sequences $\{a_n\},\{b_n\}$,
we use
$a_n = o(b_n)$,  $a_n=w(b_n)$, $a_n = O(b_n)$, $a_n = \Omega(b_n)$, and
$a_n = \Theta(b_n)$, with their meaning in
the standard Landau notation. 
Namely, we write 
$a_n = o(b_n)$ as a shorthand for the relation
 $\lim_{n \to \infty} \frac{a_n}{b_n}=0$, whereas $a_n = O(b_n)$
means that there exists $c>0$ such that $a_n \leq c b_n$ for all
$n$ sufficiently large. Also, we have $a_n = \Omega(b_n)$ if
$b_n=O(a_n)$, or equivalently, if there exists $c
> 0$ such that $a_n \geq c  b_n$ for all $n$ sufficiently
large. Finally, we write $a_n = \Theta(b_n)$ if we have $a_n =
O(b_n)$ and $a_n = \Omega(b_n)$ at the same time.
We also use $a_n \sim b_n$ to denote the
asymptotic equivalence $\lim_{n \to \infty} {a_n}/{b_n}=1$.

\section{The Model}
\label{sec:Model}
We consider a network consisting of $n$ sensor nodes labeled as $v_1, v_2, \ldots,v_n$. Each sensor is assigned to one of the  $r$ possible classes (e.g., priority levels) according to a probability distribution $\pmb{\mu}=\{\mu_1,\mu_2,\ldots,\mu_r\}$ with $\mu_i >0$
for each  $i=1,\ldots,r$; clearly it is also needed that $\sum_{i=1}^r \mu_i=1$. Prior to deployment, each class-$i$ node is given $K_i$ cryptographic keys selected uniformly at random from a pool of size $P$. 
Hence, the key ring $\Sigma_x$ of node $v_x$ is a $\mathcal{P}_{K_{t_x}}$-valued random variable (rv) where $\mathcal{P}_A$ denotes the collection of all subsets of $\{1,\ldots,P\}$ with exactly $A$ elements and $t_x$ denotes the class of node $v_x$. The rvs $\Sigma_1, \Sigma_2, \ldots, \Sigma_n$ are then i.i.d. with
\begin{equation}
\mathbb{P}[\Sigma_x=S \mid t_x=i]= \binom P{K_i}^{-1}, \quad S \in \mathcal{P}_{K_i}.
\nonumber
\end{equation}
After the deployment, two sensors can communicate securely over an existing communication channel if they have at least one key in common.

Throughout, we let $\pmb{K}=\{K_1,K_2,\ldots,K_r\}$, and assume without loss of generality that $K_1 \leq K_2 \leq \ldots \leq K_r$. 
Consider a random graph $\mathbb{K}$ induced on the vertex set $\mathcal{V}=\{v_1,\ldots,v_n\}$ such that  distinct nodes $v_x$ and $v_y$ are adjacent in $\mathbb{K}$, denoted by the event $K_{xy}$, if they have at least one cryptographic key in common, i.e.,
\begin{equation}
K_{xy} :=\left[\Sigma_x \cap \Sigma_y \neq \emptyset\right].
\label{adjacency_condition}
\end{equation}
The adjacency condition (\ref{adjacency_condition}) characterizes the inhomogeneous random key graph  $\mathbb{K}(n;\pmb{\mu},\pmb{K},P)$ that has been introduced recently in \cite{Yagan/Inhomogeneous}.
This model is also known in the literature  as
the {\em general random intersection graph}; e.g., see \cite{Zhao_2014,Rybarczyk,Godehardt_2003}.

The inhomogeneous random key graph models the {\em cryptographic} connectivity of the underlying WSN.
In particular, the probability $p_{ij}$ that a class-$i$ node and a class-$j$ have a common key, and thus are adjacent in $\mathbb{K}(n;\pmb{\mu},\pmb{K},P)$, is given by
\begin{equation}
p_{ij}= \mathbb{P}[K_{xy}] = 1-{\binom {P-K_i}{K_j}}\Bigg/{\binom {P}{K_j}}
\label{eq:osy_edge_prob_type_ij}
\end{equation}
as long as $K_i + K_j \leq P$; otherwise if $K_i +K_j > P$, we clearly have $p_{ij}=1$.
We also find it useful define the \textit{mean} probability $\lambda_i$ of edge occurrence for a class-$i$ node in $\mathbb{K}(n;\pmb{\mu},\pmb{K},P)$. With arbitrary nodes $v_x$ and $v_y$, we have
\begin{align}
\lambda_i&=\mathbb{P}[K_{xy} \given[\big] t_x=i ] 
=\sum_{j=1}^r p_{ij} \mu_j,  \quad i=1,\ldots, r,
 \label{eq:osy_mean_edge_prob_in_RKG}
\end{align}
as we condition on the class $t_y$ of node $v_y$.

In  this work, we consider the communication topology of the WSN as consisting of independent 
channels that are either {\em on} (with probability $\alpha$) or {\em off} (with probability $1-\alpha$). 
More precisely, let $\{B_{ij}(\alpha), 1 \leq i < j \leq n\}$ denote i.i.d Bernoulli rvs, each with success probability $\alpha$. The communication channel between two distinct nodes $v_x$ and $v_y$ is on (respectively, off) if $B_{xy}(\alpha)=1$ (respectively if $B_{xy}(\alpha)=0$). 
This simple on-off channel model captures the unreliability of wireless links and enables a comprehensive analysis of the properties of interest of the resulting WSN, e.g., its connectivity. It was also shown that on-off channel model provides a good approximation of the more realistic disk model \cite{Gupta99} in many similar settings and for similar properties of interest; e.g., see \cite{Yagan/EG_intersecting_ER,YaganPairwise}.
The on/off channel model induces  a standard Erd\H{o}s-R\'enyi (ER) graph $\mathbb{G}(n;\alpha)$  \cite{Bollobas}, defined  on the vertices $\mathcal{V}=\{v_1,\ldots,v_n\}$ such that 
  $v_x$ and $v_y$ are adjacent, denoted $C_{xy}$,  
if $B_{xy}(\alpha)=1$.


We model the overall topology of a WSN by the intersection of an inhomogeneous random key graph $\mathbb{K}(n;\pmb{\mu},\pmb{K},P)$ and an ER graph $\mathbb{G}(n;\alpha)$. Namely, nodes  $v_x$ and $v_y$ are adjacent in $\mathbb{K} (n;\pmb{\mu},\pmb{K},P) \cap \mathbb{G}(n;\alpha)$, denoted $E_{xy}$, if and only if they are adjacent in both $\mathbb{K}$ \textit{and} $\mathbb{G}$. In other words, the edges in the intersection graph
$\mathbb{K} (n;\pmb{\mu},\pmb{K},P) \cap \mathbb{G}(n;\alpha)$ represent pairs of sensors that
can securely communicate as they have i) a communication link in between
that is {\em on}, and ii) a shared cryptographic key.
Therefore, studying the connectivity properties of $\mathbb{K} (n;\pmb{\mu},\pmb{K},P) \cap \mathbb{G}(n;\alpha)$ amounts to studying the secure connectivity of heterogenous WSNs under the on/off channel model.


Hereafter, we denote the intersection graph $\mathbb{K} (n;\pmb{\mu},\pmb{K},P) \cap \mathbb{G}(n;\alpha)$ by the graph $\mathbb{H}(n;\pmb{\mu},\pmb{K},P,\alpha)$. To simplify the notation, we let $\pmb{\theta}=(\pmb{K},P)$, and $\pmb{\Theta}=(\pmb{\theta},\alpha)$. The probability of edge existence between a class-$i$ node $v_x$ and a class-$j$ node $v_y$ in $\mathbb{H}(n;\pmb{\Theta})$ is given by
\begin{equation} \nonumber
\mathbb{P}[E_{xy} \given[\Big] t_x=i,t_y=j]=\mathbb{P}[K_{xy} \cap C_{xy} \given[\big] t_x=i,t_y=j]=\alpha p_{ij}
\end{equation}
by independence. Similar to (\ref{eq:osy_mean_edge_prob_in_RKG}), the mean edge probability for a class-$i$ node in $\mathbb{H}(n;\pmb{\mu},\pmb{\Theta})$ as $\Lambda_i$ is given by
\begin{align} 
\Lambda_i = \sum_{j=1}^r \mu_j \alpha p_{ij} = \alpha \lambda_i, \quad i=1,\ldots, r.
\label{eq:osy_mean_edge_prob_in_system}
\end{align}

Throughout, we assume that the number of classes $r$ is fixed and does not scale with $n$, and so are the probabilities $\mu_1, \ldots,\mu_r$. All of the remaining parameters are assumed to be scaled with $n$.

We close this section with some additional notation that will be useful in the rest of the paper. For any three distinct nodes $v_x$ , $v_y$ and $v_j$, we define $E_{xj \cap yj}:=E_{xj} \cap E_{yj}$, $E_{xj \cap \overline{yj}}:=E_{xj} \cap \overline{E_{yj}}$, $E_{\overline{xj} \cap yj}:=\overline{E_{xj}} \cap E_{yj}$, and $E_{\overline{xj} \cap \overline{yj}}:=\overline{E_{xj}} \cap \overline{E_{yj}}$.

\section{Main Results and Discussion}
\label{sec:results}

\subsection{Results}
We refer to a mapping $K_1,\ldots,K_r,P: \mathbb{N}_0 \rightarrow \mathbb{N}_0^{r+1}$ as a \textit{scaling} (for the inhomogeneous random key graph) as long as the conditions
\begin{equation}
2 \leq K_{1,n} \leq K_{2,n} \leq \ldots \leq K_{r,n} \leq P_n/2
\label{scaling_condition_K}
\end{equation}
are satisfied  for all $n=2,3,\ldots$. Similarly any mapping $\alpha: \mathbb{N}_0 \rightarrow (0,1)$ defines a scaling for the ER graphs. As a result, a mapping $\pmb{\Theta} : \mathbb{N}_0 \rightarrow \mathbb{N}_0^{r+1} \times (0,1)$ defines a scaling for the intersection graph $\mathbb{H}(n;\pmb{\mu},\pmb{\Theta}_n)$ given that condition (\ref{scaling_condition_K}) holds. We remark that under (\ref{scaling_condition_K}), the edge probabilities $p_{ij}$ will be given by
(\ref{eq:osy_edge_prob_type_ij}).

We first present a zero-one law for the minimum node degree being no less than $k$ in the inhomogeneous random key graph intersecting ER graph.

\begin{theorem}
\label{theorem:min_node_degree}
{\sl
Consider a probability distribution $\pmb{\mu}=\{\mu_1,\ldots,\mu_r\}$ with $\mu_i >0$ for $i=1,\ldots,r$ and a scaling $\pmb{\Theta}: \mathbb{N}_0 \rightarrow \mathbb{N}_0^{r+1} \times (0,1)$. Let the sequence $\gamma: \mathbb{N}_0 \rightarrow \mathbb{R}$ be defined through 
\begin{equation}
\Lambda_1(n)=\alpha_n \lambda_1(n) = \frac{\log n + (k-1)\log \log n+\gamma_n}{n}, \label{scaling_condition_KG}
\end{equation}
for each $n=1,2, \ldots$. 

(a) If $\lambda_1(n)=o(1)$, we have
\begin{equation} \nonumber
\lim_{n \to \infty} \mathbb{P} \left[ \begin{split} &\text{ Minimum node degree}\\ &\text{ of } \mathbb{H}(n;\pmb{\mu},\pmb{\Theta}_n) \geq k \end{split} \right]= 0   \quad \text{ if } \lim_{n \to \infty} \gamma_n=-\infty
\end{equation}

(b) We have
\begin{equation} \nonumber
\lim_{n \to \infty} \mathbb{P} \left[ \begin{split} &\text{ Minimum node degree}\\ &\text{ of } \mathbb{H}(n;\pmb{\mu},\pmb{\Theta}_n) \geq k \end{split} \right]=1  \quad \text{ if } \lim_{n \to \infty} \gamma_n=\infty.
\end{equation}
}
\end{theorem}

Next, we present a zero-one law for the $k$-connectivity of $\mathbb{H}(n;\pmb{\mu},\pmb{\Theta})$.

\begin{theorem}
\label{theorem:kconnectivity}
{\sl
Consider a probability distribution $\pmb{\mu}=\{\mu_1,\ldots,\mu_r\}$ with $\mu_i >0$ for $i=1,\ldots,r$ and a scaling $\pmb{\Theta}: \mathbb{N}_0 \rightarrow \mathbb{N}_0^{r+1} \times (0,1)$. Let the sequence $\gamma: \mathbb{N}_0 \rightarrow \mathbb{R}$ be defined through (\ref{scaling_condition_KG}) for each $n=1,2, \ldots$. 

(a) If $\lambda_1(n)=o(1)$, we have
\begin{equation} \nonumber
\lim_{n \to \infty} \mathbb{P} \left[\mathbb{H}(n;\pmb{\mu},\pmb{\Theta}_n) \text{ is }k\text{-connected}  \right]= 0   \quad \text{ if } \lim_{n \to \infty} \gamma_n=-\infty
\end{equation}

(b) If 
\begin{align}
P_n &= \Omega(n), \label{eq:conn_Pn} \\
\frac{K_{r,n}}{P_n}&=o(1), \label{eq:conn_KrPn} \\
\frac{K_{r,n}}{K_{1,n}} &=o(\log n), \label{eq:conn_Kr_K1} 
\end{align}
we have
\begin{equation} 
\lim_{n \to \infty} \mathbb{P} \left[\mathbb{H}(n;\pmb{\mu},\pmb{\Theta}_n) \text{ is }k\text{-connected}  \right]=1  \quad \text{ if } \lim_{n \to \infty} \gamma_n=\infty.
\label{eq:kconn_OneLaw_Statement}
\end{equation}
}
\end{theorem}
In words, Theorem~\ref{theorem:min_node_degree} (respectively Theorem~\ref{theorem:kconnectivity}) states that the minimum node degree in $\mathbb{H}(n;\pmb{\mu},\pmb{\Theta}_n)$ is greater than or equal to $k$ (respectively $\mathbb{H}(n;\pmb{\mu},\pmb{\Theta}_n)$ is $k$-connected) whp if the mean degree of class-$1$ nodes, i.e., $n \Lambda_1(n)$, is scaled as $\left(\log n+(k-1) \log \log n+\gamma_n\right)$ for some sequence $\gamma_n$ satisfying $\lim_{n \to \infty} \gamma_n=\infty$. On the other hand, if the sequence $\gamma_n$ satisfies $\lim_{n \to \infty} \gamma_n=-\infty$, then whp $\mathbb{H}(n;\pmb{\mu},\pmb{\Theta}_n)$ has at least one node with degree strictly less than $k$, and hence is {\em not} $k$-connected. This shows that the critical scaling for the minimum node degree of 
$\mathbb{H}(n;\pmb{\mu},\pmb{\Theta}_n)$ being greater than or equal to $k$ (respectively for $\mathbb{H}(n;\pmb{\mu},\pmb{\Theta}_n)$ to be $k$-connected) is given by $\Lambda_1(n)=\frac{\log n+(k-1)\log \log n}{n}$, with the sequence $\gamma_n:\mathbb{N}_0 \rightarrow \mathbb{R}$ measuring the deviation of $\Lambda_1(n)$ from the critical scaling.

The scaling condition (\ref{scaling_condition_KG}) can be given a more explicit form under some additional constraints. In particular, it was shown in \cite[Lemma 4.2]{Yagan/Inhomogeneous} that if $\lambda_1(n) = o(1)$ then
\begin{equation}
\lambda_1(n) \sim \frac{K_{1,n} K_{\textrm{avg},n}}{P_n}
\label{eq:lambda_1_asymp}
\end{equation}
 where $K_{\textrm{avg},n} = \sum_{j=1}^{r}\mu_j K_{j,n}$  denotes the {\em mean} key ring size in the network. This shows that the minimum key ring size $K_{1,n}$ is of paramount importance in controlling the connectivity and reliability of the WSN; as explained previously, it then also controls the number of {\em mobile} sensors that can be accommodated in the network. For example, 
 with the mean number $K_{\textrm{avg},n}$ of keys per sensor is fixed, we see that  
 reducing $K_{1,n}$ by half means that the smallest $\alpha_n$ (that gives the largest link failure probability $1-\alpha_n$) for which the network remains $k$-connected whp  is increased by two-fold for any given $k$;
e.g., see Figure \ref{fig:3} for a numerical example demonstrating this.

\subsection{Comments on the additional technical conditions}
We first comment on the additional technical condition $\lambda_1(n)=o(1)$. This is enforced here mainly for technical reasons for the proof of the zero-law of Theorem~\ref{theorem:min_node_degree} (and thus of Theorem \ref{theorem:kconnectivity}) to work. A similar condition was also required in \cite[Thm 1]{Jun/K-Connectivity} for establishing the zero-law for the minimum node degree being no less than $k$ in the {\em homogeneous} random key graph intersecting ER graph. In view of (\ref{eq:lambda_1_asymp}), this condition is equivalent to
\begin{equation}
K_{1,n} K_{\textrm{avg},n} = o(P_n).
\label{eq:extra_cond_1_equiv}
\end{equation}
In real-world WSN applications the
key pool size $P_n$ is envisioned to be orders of magnitude larger than any key ring size in the network \cite{Gligor_2002,DiPietroTissec}.  As discussed below in more details, this is 
needed to ensure the resilience of the network against adversarial attacks. 
Concluding, 
(\ref{eq:extra_cond_1_equiv}) (and thus $\lambda_1(n)=o(1)$) is indeed likely to hold in most applications.

Conditions (\ref{eq:conn_Pn}) and (\ref{eq:conn_KrPn}) are also likely to be needed in practical WSN implementations in order to ensure the {\em resilience} of the network against node capture attacks; e.g., see \cite{Gligor_2002,DiPietroTissec}. To see this, assume that an adversary captures a number of sensors, compromising all the keys that belong to the captured nodes. If $P_n = O(K_{r,n})$ contrary to (\ref{eq:conn_KrPn}), then it would be possible for the adversary to compromise a positive fraction of the key pool (i.e., $\Omega(P_n)$ keys) by capturing only a constant number of sensors that are of type $r$. Similarly, if $P_n = o(n)$, contrary to  (\ref{eq:conn_Pn}), then again it would be possible for the adversary to compromise $\Omega(P_n)$ keys by capturing only  $o(n)$ sensors (whose type does not matter in this case). In both cases, the WSN would fail to exhibit the {\em unassailability} property \cite{MeiPanconesiRadhakrishnan2008,YM_ToN} and would be deemed as vulnerable against adversarial attacks.
We remark that both (\ref{eq:conn_Pn}) and (\ref{eq:conn_KrPn}) were required in \cite{Jun/K-Connectivity,Yagan/Inhomogeneous} for obtaining the one-law for connectivity and $k$-connectivity, respectively, in similar settings to ours.

Finally, the condition (\ref{eq:conn_Kr_K1}) is enforced mainly for technical reasons and takes away from the flexibility of assigning very small key rings to a certain fraction of sensors when $k$-connectivity is considered; we remark that (\ref{eq:conn_Kr_K1}) is not needed for the minimum node degree result given at Theorem \ref{theorem:min_node_degree}. An equivalent condition was also needed in \cite{Yagan/Inhomogeneous} for establishing the one-law for connectivity in inhomogeneous random key graphs. We refer the reader to \cite[Section 3.2]{Yagan/Inhomogeneous} for an extended discussion on the feasibility of (\ref{eq:conn_Kr_K1}) for real-world WSN implementations, as well as possible ways to replace it with milder conditions. 

We close by providing a concrete example that demonstrates how all the conditions required by Theorem \ref{theorem:kconnectivity}
can be met in a real-world implementation. Consider any number $r$ of sensor types, and pick any probability distribution $\pmb{\mu}=\{\mu_1, \ldots, \mu_r\}$ with $\mu_i > 0$ for all $i=1,\ldots, r$. For any channel probability $\alpha_n = \Omega(\frac{\log n}{n})$, set $P_n = n \log n$ and use 
\[
K_{1,n} = \frac{(\log n)^{1/2+\varepsilon}}{\sqrt{\alpha_n}} \quad \textrm{and} \quad K_{r,n} = \frac{(1 + \varepsilon)(\log n)^{3/2-\varepsilon}}{\mu_r \sqrt{\alpha_n}}
\]
 with any $\varepsilon > 0$. Other key ring sizes $K_{1,n} \leq K_{2,n}, \ldots, K_{r-1,n} \leq K_{r,n}$ can be picked arbitrarily. In view of Theorem \ref{theorem:kconnectivity} and the fact \cite[Lemma 4.2]{Yagan/Inhomogeneous} that $\lambda_1(n) \sim \frac{K_{1,n} K_{\textrm{avg},n}}{P_n}$, the resulting network will be $k$-connected whp for any $k=1, 2, \ldots$. Of course, there are many other parameter scalings that one can choose.



\subsection{Comparison with related work}

In comparison with the existing literature on similar models, our result can be seen to extend the work by Zhao et al. \cite{Jun/K-Connectivity} on the homogeneous random key graph intersecting ER graph to the heterogeneous setting. There, zero-one laws for the property that the minimum node degree is no less than $k$ and the property that the graph is $k$-connected were established for $\mathbb{H}(n,K,P,\alpha_n)$. With $r=1$, i.e., when all nodes belong to the same class and thus receive the same number $K$ of keys, Theorem~\ref{theorem:min_node_degree} and Theorem~\ref{theorem:kconnectivity} recover the result of Zhao et al. (See \cite[Theorems~1-2]{Jun/K-Connectivity}).

Our paper also extends the work by Ya\u{g}an \cite{Yagan/Inhomogeneous} who  considered the inhomogeneous random key graph $\mathbb{K}(n,\pmb{\mu},\pmb{K},P)$ under {\em full} visibility; i.e., when all pairs of nodes have a communication channel in between. There,  Ya\u{g}an established zero-one laws for the absence of isolated nodes (i.e., absence of nodes with degree zero) and $1$-connectivity. Our work generalizes Ya\u{g}an's results on two fronts. Firstly, we consider more practical WSN scenarios where the unreliability of wireless communication channels are taken into account through the on/off channel model. Secondly, in addition to the properties that the graph has no isolated nodes (i.e., the minimum node degree is no less than $1$) and is $1$-connected, we consider general minimum node degree and connectivity values, $k=0,1,\ldots$. 

Finally,  our work (with $\alpha_n=1$ for each $n=2, 3, \ldots$) improves upon the results by Zhao et al. \cite{Zhao_2014}; therein, this model was referred to as the general random intersection graph. Our main argument is that the additional conditions required by their main result renders them inapplicable in practical WSN implementations.
This issue is discussed at length in \cite[Section 3.3]{Yagan/Inhomogeneous}, but we give a summary here for completeness.
With $X_n$ denoting the random variable representing the number of keys assigned to an arbitrary node in the network, the main result in  \cite{Zhao_2014} requires
\begin{equation}
\textrm{var} [X_n]=o \left( \frac{\left( \mathbb{E}[X_n] \right)^2}{n \left( \log n \right)^2} \right)
\label{eq:jun_condition}
\end{equation}
that puts a  prohibitively stringent limit on the variance of the key ring sizes. For instance, it precludes 
using $K_{2,n} = c K_{1,n}$ for some $c>1$, and forces key ring sizes to be asymptotically equivalent; i.e., $K_{r,n} \sim K_{1,n}$.
In fact, under (\ref{eq:jun_condition}), even the simplest case where key ring sizes vary by a constant is possible only when
$\mathbb{E}[X_n]=\omega \left(\sqrt{n} \log n \right)$. 
Put differently, the results in \cite{Zhao_2014} are useful only if the mean number of keys assigned to a sensor node is much larger than $\sqrt{n} \log n$; and even then only small variations among key ring sizes would be possible.
However, in most WSN applications, sensor nodes will have very limited memory and computational capabilities \cite{Akyildiz_2002} and such large key ring sizes are not likely to be feasible; typically key rings on the order of $\log n$ are envisioned in applications  \cite{Gligor_2002,DiPietroTissec}. These arguments show that conditions enforced in   \cite{Zhao_2014}  are not likely to hold in practice. In contrast, our results allow for much larger variation in key ring sizes and require parameter conditions that are likely to hold in practice; e.g., we only need $\mathbb{E}[X_n] = o({P_n})$.


%

\section{Numerical Results}
\label{sec:numerical}
We now present numerical results to support Theorems \ref{theorem:min_node_degree} and \ref{theorem:kconnectivity} in the finite node regime. In all experiments, we fix the number of nodes at $n = 500$ and the size of the key pool at $P = 10^4$. To help better visualize the results,  we use the curve fitting tool of MATLAB.

\begin{figure}[t]
\centerline{\includegraphics[scale=0.45]{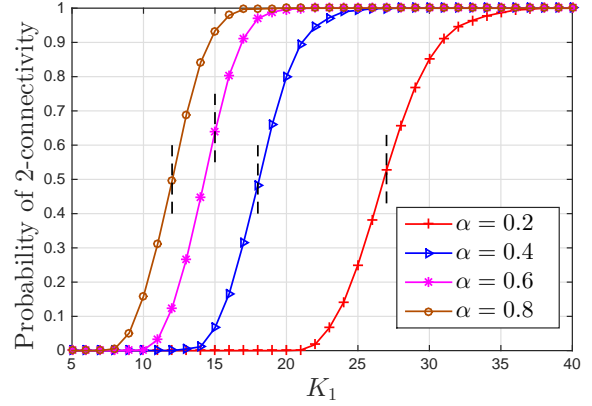}}
\caption{Empirical probability that $\mathbb{H}(n;\pmb{\mu},\pmb{\theta},\alpha)$ is $2$-connected as a function of $\pmb{K}$ for $\alpha = 0.2$, $\alpha = 0.4$, $\alpha = 0.6$, $\alpha = 0.8$ with $n = 500$ and $P = 10^4$; in each case, the empirical probability value is obtained by averaging over $200$ experiments. Vertical dashed lines stand for the critical threshold of connectivity asserted by Theorem~\ref{theorem:kconnectivity}.}
\label{fig:1}
\end{figure}

In Figure~\ref{fig:1}, we consider the channel parameters $\alpha = 0.2$, $\alpha = 0.4$, $\alpha = 0.6$, and $\alpha = 0.8$, while varying the parameter $K_1$, i.e., the smallest key ring size, from $5$ to $40$. The number of classes is fixed to $2$, with $\pmb{\mu}=\{0.5,0.5\}$. For each value of $K_1$, we set $K_2=K_1+10$. For each parameter pair $(\pmb{K}, \alpha)$, we generate $200$ independent samples of the graph $\mathbb{H}(n;\pmb{\mu},\pmb{\Theta})$ and count the number of times (out of a possible 200) that the obtained graphs i) have minimum node degree no less than $2$ and ii) are $2$-connected. Dividing the counts by $200$, we obtain the (empirical) probabilities for the events of interest.  In all cases considered here, we observe that $\mathbb{H}(n;\pmb{\mu},\pmb{\Theta})$ is $2$-connected whenever it has minimum node degree no less than $2$ yielding the same empirical probability for both events. This supports the fact that the 
properties of $k$-connectivity and minimum node degree being larger than $k$
are asymptotically equivalent in $\mathbb{H}(n;\pmb{\mu},\pmb{\Theta}_n)$.

In Figure~\ref{fig:1} as well as the ones that follow we show the critical threshold of connectivity \lq\lq predicted" by Theorem~\ref{theorem:kconnectivity} by a vertical dashed line. More specifically, the vertical dashed lines stand for the minimum integer value of $K_1$ that satisfies
\begin{equation}
\hspace{-.6mm}\lambda_1(n)\hspace{-1mm}=\hspace{-1mm}\sum_{j=1}^2 \mu_j \hspace{-1mm}\left( \hspace{-.2mm} 1- \frac{\binom{P-K_j}{K_1}}{\binom{P}{K_1}}  \hspace{-.2mm}\right) \hspace{-.1mm}>  \hspace{-.1mm}\frac{1}{\alpha} \frac{\log n+(k-1)\log \log n}{n}
\label{eq:numerical_critical}
\end{equation}
with any given $k$ and $\alpha$.
We see from Figure~\ref{fig:1} that the probability of $k$-connectivity transitions from zero to one within relatively small variations in $K_1$. Moreover, the critical values of $K_1$ obtained by (\ref{eq:numerical_critical}) lie within the transition interval.  

\begin{figure}[t]
\centerline{\includegraphics[scale=0.45]{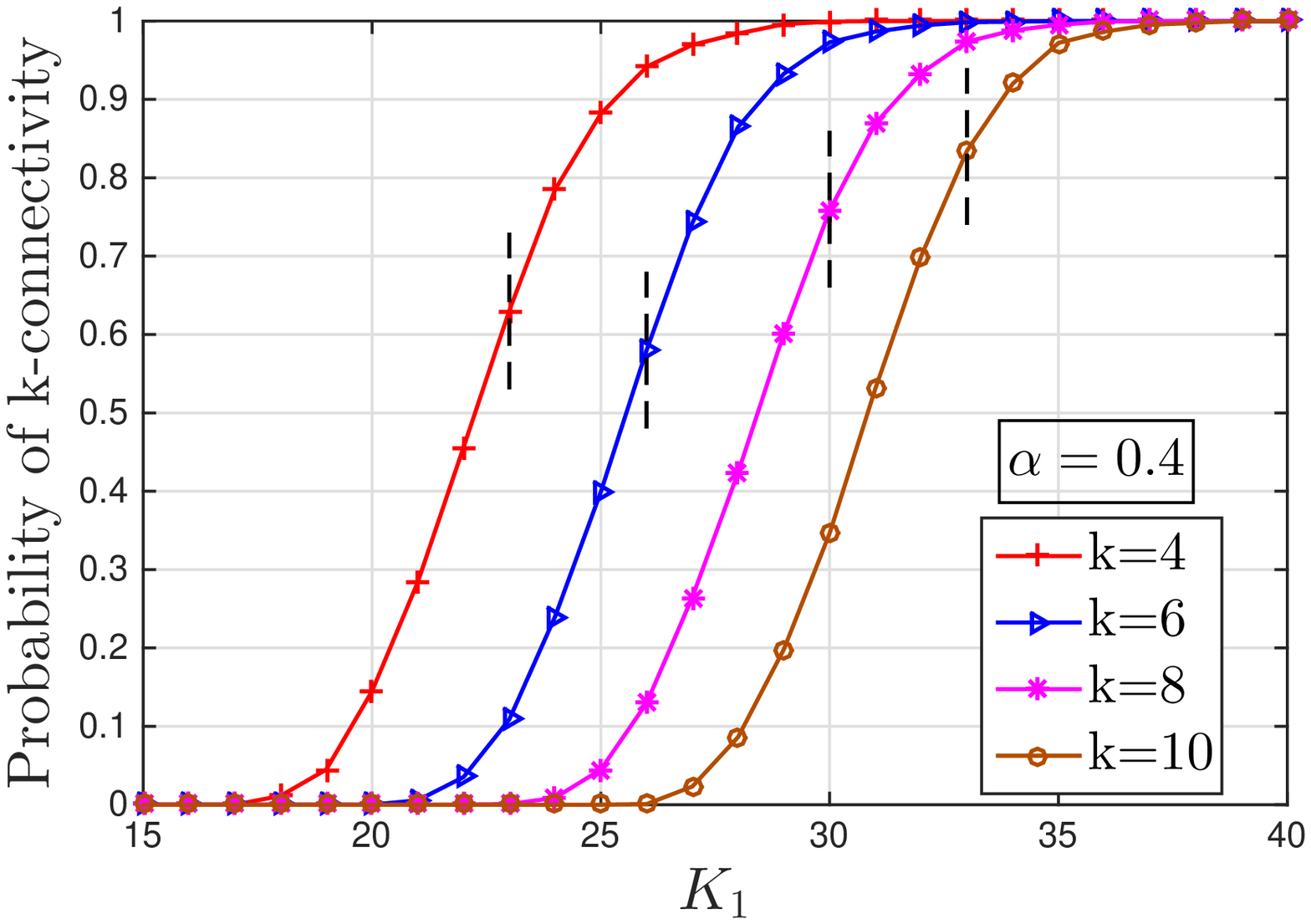}}
\caption{Empirical probability that $\mathbb{H}(n;\pmb{\mu},\pmb{\theta},\alpha)$ is $k$-connected as a function of $K_1$ for $k=4$, $k=6$, $k=8$, and $k=10$, with $n = 500$ and $P = 10^4$; in each case, the empirical probability value is obtained by averaging over $200$ experiments. Vertical dashed lines stand for the critical threshold of connectivity asserted by Theorem~\ref{theorem:kconnectivity}.}
\label{fig:2}
\end{figure}

In Figure~\ref{fig:2}, we consider four different values for $k$, namely we set $k = 4$, $k = 6$, $k = 8$, and $k=10$ while varying $K_1$ from $15$ to $40$ and fixing $\alpha$ to $0.4$. The number of classes is fixed to $2$ with $\pmb{\mu}=\{0.5,0.5\}$ and we set $K_2=K_1+10$ for each value of $K_1$. Using the same procedure that produced Figure~\ref{fig:1}, we obtain the empirical probability that $\mathbb{H}(n;\pmb{\mu},\pmb{\theta},\alpha)$ is $k$-connected versus $K_1$. The critical threshold of connectivity asserted by Theorem~\ref{theorem:kconnectivity} is shown by a vertical dashed line in each curve.
Again, we see that numerical results are in parallel with Theorem~\ref{theorem:kconnectivity}.

\begin{figure}[t]
\centerline{\includegraphics[scale=0.45]{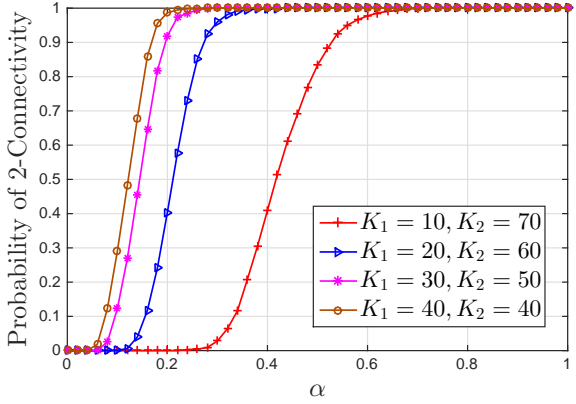}}
\caption{Empirical probability that $\mathbb{H}(n;\pmb{\mu},\pmb{\theta},\alpha)$ is $2$-connected with $n = 500, \pmb{\mu} = (1/2,1/2)$, and $P = 10^4$; we consider four choices of $\pmb{K} = (K_1,K_2)$ each with the same mean.}
\label{fig:3}
\end{figure}

Figure~\ref{fig:3} is generated in a similar manner with Figure~\ref{fig:1}, this time with an eye towards understanding the impact of the minimum key ring size $K_1$ on network connectivity. To that end, we fix the number of classes at $2$ with $\pmb{\mu}=\{0.5,0.5\}$ and consider 
four different key ring sizes $\pmb{K}$
each with mean $40$; we consider
$\pmb{K} = \{10,70\}$, $\pmb{K} = \{20,60\}$, $\pmb{K} = \{30,50\}$, and $\pmb{K} = \{40,40\}$.
We compare the probability of $2$-connectivity in the resulting networks while varying $\alpha$ from zero to one. We see that although the average number of keys per sensor is kept constant in all four cases, network connectivity improves dramatically as the minimum key ring size $K_1$ increases; e.g., with $\alpha=0.2$, the probability of connectivity is one when $K_1=K_2=40$ while it drops to zero if we set $K_1=10$ while increasing $K_2$ to $70$ so that the mean key ring size is still 40. 

\begin{figure}[t]
\centerline{\includegraphics[scale=0.45]{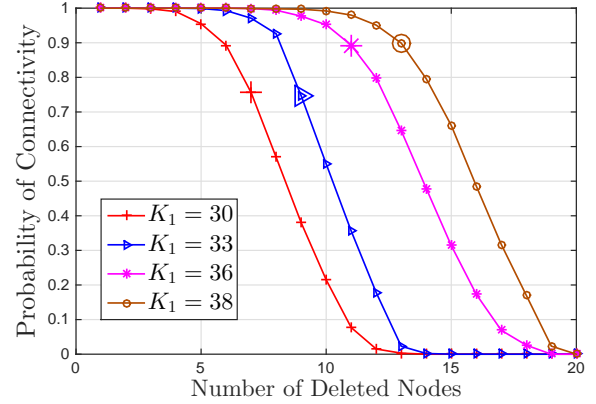}}
\caption{Empirical probability that $\mathbb{H}(n;\pmb{\mu},\pmb{\theta},\alpha)$ remains connected after deleting nodes from the {\em minimum vertex cut} set. We fix $n = 500, \pmb{\mu}=(1/2,1/2), \alpha=0.4,P = 10^4$, and choose $K_1$ and $K_2=K_1+10$
from (\ref{eq:numerical_critical}) for each  $k=8$, $k=10$, $k=12$, and $k=14$;
i.e., we use $K_1=30,33,36,38$, respectively.
}\label{fig:4}
\end{figure}

Finally, we examine the reliability of $\mathbb{H}(n;\pmb{\mu},\pmb{\theta},\alpha)$ 
by looking at the probability of 1-connectivity as the number of  deleted (i.e., failed) nodes increases. 
From a mobility perspective, this is equivalent to investigating the probability of a WSN remaining connected 
as the number of {\em mobile} sensors leaving the network increases. In Figure~\ref{fig:4}, we set $n = 500, \pmb{\mu}=\{1/2,1/2\}, \alpha=0.4,P = 10^4$, and select $K_1$ and $K_2=K_1+10$ from (\ref{eq:numerical_critical}) for
$k=8$, $k=10$, $k=12$, and $k=14$. With these settings, we would expect (for very large $n$) the network to remain connected whp after the deletion of up to 7, 9, 11, and 13 nodes, respectively.
Using the same procedure that produced Figure~\ref{fig:1}, we obtain the empirical probability that $\mathbb{H}(n;\pmb{\mu},\pmb{\theta},\alpha)$ is connected as a function of  the number of deleted nodes\footnote{We choose the nodes to be deleted from the {\em minimum vertex cut} of $\mathbb{H}$, defined as the minimum cardinality set whose removal renders it disconnected. This captures the worst-case nature of the $k$-connectivity property in a computationally efficient manner (as compared to searching over all $k$-sized subsets and deleting the one that gives maximum damage).} in each case.  We see that even with $n=500$ nodes, the resulting reliability is close to the levels expected to be attained asymptotically as $n$ goes to infinity. In particular, we see that the probability of remaining connected when $(k-1)$ nodes leave the network is around $0.75$ for the first two cases and around  $0.90$ for the other two cases.

\section{Preliminaries}
A number of technical results are collected here for easy referencing.
\begin{prop} [{\cite[Proposition~4.1]{Yagan/Inhomogeneous}}]
For any scaling $K_1,K_2,\ldots,K_r,P:\mathbb{N}_0 \rightarrow \mathbb{N}_0^{r+1}$, we have
\begin{equation}
\lambda_1(n) \leq \lambda_2(n) \leq \ldots \leq \lambda_r(n),  \qquad n=2,3,\ldots.
\label{eq:ordering_of_lambda}
\end{equation}
\label{prop:ordering_of_lambda}
\end{prop}
In view of (\ref{eq:osy_mean_edge_prob_in_system}), Proposition~\ref{prop:ordering_of_lambda} implies that
\begin{equation}
\Lambda_1(n) \leq \Lambda_2(n) \leq \ldots \leq \Lambda_r(n), \qquad n=2,3,\ldots.
\label{eq:ordering_of_capital_lambda}
\end{equation}

%



{\prop
Consider a scaling $K_1, K_2, \ldots,K_r,P:\mathbb{N}_0 \rightarrow \mathbb{N}_0^{r+1}$ and a scaling $\alpha:\mathbb{N}_0 \rightarrow (0,1)$. Let the sequence $\gamma: \mathbb{N}_0 \rightarrow \mathbb{R}$ be defined through (\ref{scaling_condition_KG}) for each $n=1,2, \ldots$. Under (\ref{eq:conn_Pn}) and (\ref{eq:conn_Kr_K1}), we have
\begin{equation}
K_{1,n}=\omega(1)
\label{eq:conn_K1_omega_1}
\end{equation}
when $\lim_{n \to \infty} \gamma_n=+\infty$. 
\label{prop:conn_K1_omega_1}
}

\begin{proof}
From (\ref{scaling_condition_KG}), we clearly have
\begin{equation} 
\lambda_1(n) > \frac{\log n}{n \alpha_n}
\label{eq:lower_bound_on_lambda}
\end{equation}
for all $n$ sufficiently large when $\lim_{n \to \infty} \gamma_n=+\infty$. We also know from \cite[Lemmas~7.1-7.2]{yagan2012zero} that
\begin{equation} \nonumber
p_{1j}(n) \leq \frac{K_{1,n}K_{j,n}}{P_n-K_{j,n}}\leq 2 \frac{K_{1,n}K_{j,n}}{P_n}, ~~ j=1,\ldots, r
\end{equation}
where the last bound follows from (\ref{scaling_condition_K}). This leads to
\begin{equation} 
\hspace{-.2mm}\lambda_1(n) =\hspace{-.1mm} \sum_{j=1}^r \hspace{-.1mm} \mu_j p_{1j} \leq 2\sum_{j=1}^r \hspace{-.1mm} \mu_j \frac{K_{1,n}K_{j,n}}{P_n} \leq  2\frac{K_{1,n}K_{r,n}}{P_n} 
\label{eq:upper_bound_on_lambda}
\end{equation}
Combining (\ref{eq:lower_bound_on_lambda}) and (\ref{eq:upper_bound_on_lambda}) we get
\begin{equation} \nonumber
K_{1,n}^2 \frac{K_{r,n}}{K_{1,n}} > \frac{P_n}{2} \frac{\log n}{n \alpha_n}
\end{equation}
for all $n$ sufficiently large. Under (\ref{eq:conn_Pn}) and (\ref{eq:conn_Kr_K1}),  this immediately establishes (\ref{eq:conn_K1_omega_1}) since $\alpha_n \leq 1$.
\end{proof}

{
\fact
\label{fact:monotone_decreasing}
For any positive constants $\ell_1,\ell_2$, the function
\begin{equation}
f(x)=x^{\ell_1} (1-x)^{n-\ell_2}, \quad x \in (0,1)
\label{eq:monotone_decreasing}
\end{equation}
is monotone decreasing in $x$ for all $n$ sufficiently large.
}

\begin{proof}
Differentiating $f(x)$ with respect to $x \in (0,1)$, we get
\begin{align}
\frac{d}{dx} f(x)&=\ell_1  x^{\ell_1-1} (1-x)^{n-\ell_2} - (n-\ell_2) x^{\ell_1} (1-x)^{n-\ell_2-1}\nonumber \\
&=x^{\ell_1 -1} (1-x)^{n-\ell_2-1} (\ell_1(1-x) - (n-\ell_2) x). \nonumber
\end{align}
The conclusion follows since $(\ell_1(1- x) - (n-\ell_2) x)<0$ for 
all $n$ sufficiently large, for any positive $\ell_1,\ell_2$ and $x \in (0,1)$.
\end{proof}

{
\fact [{\cite[Lemma~8]{Jun/K-Connectivity}}]
\label{fact:asym_eq}
Given (\ref{scaling_condition_K}), If $p_{1i}(n)=o(1)$ or $\frac{K_{1,n} K_{i,n}}{P_n}=o(1)$, then $\frac{K_{1,n} K_{i,n}}{P_n}=p_{1i}(n) \pm O\left( \left( p_{1i}(n)\right)^2 \right)$
}

{
\fact [{\cite[Fact~3]{Jun/K-Connectivity}}]
\label{fact:3}
Let $x$ and $y$ be both positive functions of $n$. If $x=o(1)$, and $x^2y=o(1)$ hold, then
\begin{equation} \nonumber
\left(1-x\right)^y \sim e^{-xy}
\end{equation}
}



{
\lemma [{\cite[Lemma~7.1]{yagan2012zero}}]
\label{lemma:bounding_fac}
For positive integers $K$, $L$, and $P$ such that $K+L \leq P$, we have
\begin{equation} \nonumber
\left( 1- \frac{L}{P-K}\right)^K \leq \frac{\binom{P-L}{K}}{\binom{P}{K}} \leq \left( 1- \frac{L}{P}\right)^K
\end{equation}
}

We will use several bounds given below throughout the paper:
\begin{align}
&(1 \pm x) \leq e^{ \pm x}, \quad x \in (0,1)
\label{eq:exp_bound}
\\
& \left(x+y \right)^p \leq 2^{p-1} \left(x^p + y^p \right)
\label{eq:conn_bounds_3}
\\
& \binom{n}{\ell} \leq \left( \frac{en}{\ell} \right)^\ell, \quad \ell=1,\ldots,n, \quad n=1,2,\ldots
\label{eq:conn_bounds_4}
\\
& \sum_{\ell=2}^{\left \lfloor{\frac{n}{2}}\right \rfloor} \binom{n}{\ell} \leq 2^n
\label{eq:conn_bounds_5}
\\
& \binom{n}{\ell} \leq n^\ell, \quad \ell=1,\ldots,n, \quad n=1,2,\ldots
\label{eq:conn_bounds_6}
\end{align}

\section{Proof of Theorem~\ref{theorem:min_node_degree}}
\subsection{Establishing the one-law}
\label{subsec:one_law_min_node_degree}
The proof of Theorem~\ref{theorem:min_node_degree} relies on the method of first and second moments applied to the number of nodes with degree $\ell$ in $\mathbb{H}(n;\pmb{\mu},\pmb{\Theta}_n)$. Let $X_{\ell}(n;\pmb{\mu},\pmb{\Theta}_n)$ denote the total number of nodes with degree $\ell$ in $\mathbb{H}(n;\pmb{\mu},\pmb{\Theta}_n)$, namely,
\begin{equation}
X_{\ell}(n;\pmb{\mu},\pmb{\Theta}_n)=\sum_{i=1}^n \pmb{1}\left[v_i \text{ is of degree } \ell \text{ in } \mathbb{H}(n;\pmb{\mu},\pmb{\Theta}_n)\right]
\label{eq:isolated_Xl}
\nonumber
\end{equation}
The method of first moment \cite[Eqn. (3.10), p.
55]{JansonLuczakRucinski} gives
\begin{equation}
\mathbb{P}\left[X_{\ell}(n;\pmb{\mu},\pmb{\Theta}_n)=0\right] \geq 1-\mathbb{E}\left[X_{\ell}(n;\pmb{\mu},\pmb{\Theta}_n)\right] 
\label{eq:degree_3_pt1}
\end{equation}
The one-law states that the minimum node degree in $\mathbb{H}(n;\pmb{\mu},\pmb{\Theta}_n)$ is no less than $k$ asymptotically almost surely (a.a.s.); i.e.,
$ \lim_{n \to \infty} \mathbb{P}\left[X_{\ell}(n;\pmb{\mu},\pmb{\Theta}_n)=0\right]=1$,  for all $\ell =0,1, \ldots, k-1$.
 Thus, the one-law will follow if we show that
\begin{equation}
\lim_{n \to \infty} \mathbb{E}\left[X_{\ell}(n;\pmb{\mu},\pmb{\Theta}_n)\right]=0, \quad  \ell =0,1, \ldots, k-1.
\label{eq:node_isol_one_law_to_show}
\end{equation}

We let $D_{i,\ell}(n;\pmb{\mu},\pmb{\Theta}_n)$ denote the event that node $v_i$ in $\mathbb{H}(n;\pmb{\mu},\pmb{\Theta}_n)$ has degree $\ell$ for each $i=1,2,\ldots,n$. Throughout, we simplify the notation by writing $D_{i,\ell}$ instead of $D_{i,\ell}(n;\pmb{\mu},\pmb{\Theta}_n)$. By definition, we have $X_{\ell}(n;\pmb{\mu},\pmb{\Theta}_n)=\sum_{i=1}^n \pmb{1}\left[D_{i,\ell}\right]
$
and it follows that
\begin{align} 
\mathbb{E}\left[X_{\ell}(n;\pmb{\mu},\pmb{\Theta}_n)\right]=\sum_{i=1}^n \mathbb{P}\left[D_{i,\ell} \right] =n \mathbb{P}\left[D_{x,\ell}\right]
\label{eq:degree_4}
\end{align}
by the exchangeability of the indicator rvs $\left\{ \pmb{1}\left[ D_{i,\ell} \right];i=1,\ldots,n \right\}$. 


In view of (\ref{eq:degree_3_pt1}) and (\ref{eq:degree_4}), we see that (\ref{eq:node_isol_one_law_to_show}) and hence the one-law
would follow upon showing
\begin{equation}
\lim_{n \to \infty} n \mathbb{P}\left[D_{x,\ell}\right]=0, \quad  \ell = 0, 1, \ldots, k-1.
\label{eq:to_show1}
\end{equation}

%
%
%
We start by deriving the probability of $D_{x,\ell}$. For any node $v_x$, the events $E_{1x}, E_{2x},\ldots,E_{(x-1)x},E_{(x+1)x},\ldots,E_{nx}$ are mutually independent
{\em conditionally} on the type $t_x$.
It follows from (\ref{eq:osy_mean_edge_prob_in_system}) that the degree of a node $v_x$, i.e., $D_x$, is conditionally binomial leading to
\begin{align} \nonumber
D_x  =_{st} \text{Bin}(n-1,\Lambda_i), ~~ \text{ with probability }\mu_i, \quad i=1,\ldots,r \nonumber
\end{align}
Thus, we get
\begin{align}
\mathbb{P}\left[ D_{x,\ell} \right]&=\sum_{i=1}^r \mu_i \mathbb{P}\left[ D_{x,\ell} \given[\big] t_x=i\right] \nonumber \\
&=\sum_{i=1}^r \mu_i \binom{n-1}{\ell} \left( \Lambda_i(n) \right)^\ell \left(1-\Lambda_i(n) \right)^{n-\ell-1} \nonumber \\
&\leq \left( \sum_{i=1}^r \mu_i \left( n\Lambda_i(n) \right)^\ell \left(1-\Lambda_i(n) \right)^{n-\ell-1}\right) \nonumber \\
&\leq \left( \ell! \right)^{-1} \left( n \Lambda_1(n) \right)^\ell  \left( 1-\Lambda_1(n) \right)^{n-\ell-1}\nonumber \\
& \leq \left( \ell! \right)^{-1} \left( n \Lambda_1(n) \right)^\ell  e^{- (n-\ell-1) \Lambda_1(n)}
\nonumber
\end{align}
for all $n$ sufficiently large, as we invoke Fact~\ref{fact:monotone_decreasing} together with (\ref{eq:ordering_of_capital_lambda}), and note that $\ell$ is a non-negative integer constant.
Combining (\ref{scaling_condition_KG}) and (\ref{eq:conn_bounds_3}), and using the fact that $\Lambda_1(n) \leq 1$, we see that
\begin{align}
&n\mathbb{P}\left[ D_{x,\ell} \right] \nonumber \\
&\leq n\left( \ell! \right)^{-1} \left( \log n+(k-1) \log \log n+\gamma_n \right)^\ell  \cdot\nonumber \\
& \cdot e^{-\log n - (k-1) \log \log n - \gamma_n} e^{(\ell+1) \Lambda_1(n)}\nonumber \\
&\leq 2^{\ell-1} \left(\left(\log n\right)^\ell \left(1+ o(1)\right)^\ell + \gamma_n^\ell \right) e^{-(k-1)\log \log n - \gamma_n} e^{O(1)}\nonumber \\
&=  O(1)e^{-(k-1-\ell)\log \log n - \gamma_n}  + O(1) \gamma_n^\ell  e^{-(k-1)\log \log n - \gamma_n}. \nonumber  
\end{align}
When $\lim_{n \to \infty} \gamma_n=\infty$, we readily get the desired conclusion
(\ref{eq:to_show1}).
This establishes the one-law.

\subsection{Establishing the zero-law}
Our approach in establishing the zero-law relies on the method of second moment applied to a variable that counts the number of nodes in $\mathbb{H}(n;\pmb{\mu},\pmb{\Theta}_n)$ that are \textit{class-$1$} and with degree $\ell$. Similar to the discussion given before, we let $Y_{\ell}(n;\pmb{\mu},\pmb{\Theta}_n)$ denote the total number of nodes that are class-$1$ and with degree $\ell$ in $\mathbb{H}(n;\pmb{\mu},\pmb{\Theta}_n)$, namely,
\begin{align}
&Y_{\ell}(n;\pmb{\mu},\pmb{\Theta}_n)\label{eq:isolated_Yl}
 \\ \nonumber
&=\sum_{i=1}^n \pmb{1}\left[v_i \text{ is class } 1 \text{ and has degree } \ell \text{ in } \mathbb{H}(n;\pmb{\mu},\pmb{\Theta}_n)\right]
\end{align}
Clearly, if we can show that whp there exists at least one class-$1$ node with a degree strictly less than $k$ under the enforced assumptions (with $\lim_{n \to \infty} \gamma_n=-\infty$) then the zero-law immediately follows. 

With a slight abuse of notations, we let ${D}_{i,\ell}(n;\pmb{\mu},\pmb{\Theta}_n)$ denote the event that node $v_i$ in $\mathbb{H}(n;\pmb{\mu},\pmb{\Theta}_n)$ is class-$1$ and has degree $\ell$ for each $i=1,2,\ldots,n$. Throughout, we simplify the notation by writing ${D}_{i,\ell}$ instead of ${D}_{i,\ell}(n;\pmb{\mu},\pmb{\Theta}_n)$. Thus, we have
$Y_{\ell}(n;\pmb{\mu},\pmb{\Theta}_n)=\sum_{i=1}^n \pmb{1}\left[{D}_{i,\ell}\right]$.
The method of second moments
\cite[Remark 3.1, p. 55]{JansonLuczakRucinski} gives
\begin{equation}
\mathbb{P}\left[Y_{\ell}(n;\pmb{\mu},\pmb{\Theta}_n)=0\right] \leq 1-\frac{\mathbb{E}\left[Y_{\ell}(n;\pmb{\mu},\pmb{\Theta}_n)\right]^2}{\mathbb{E}\left[Y_{\ell}(n;\pmb{\mu},\pmb{\Theta}_n)^2\right]}.
\label{eq:degree_7}
\end{equation}
We have
$
\mathbb{E}\left[Y_{\ell}(n;\pmb{\mu},\pmb{\Theta}_n)\right]=n \mathbb{P}\left[ {D}_{x,\ell} \right]
$
and
\begin{equation}
\mathbb{E}\left[Y_{\ell}(n;\pmb{\mu},\pmb{\Theta}_n)^2 \right]=n \mathbb{P}\left[ {D}_{x,\ell} \right]+n(n-1) \mathbb{P}\left[ {D}_{x,\ell} \cap {D}_{y,\ell} \right],
\nonumber
\end{equation}
whence
\begin{equation}
\frac{\mathbb{E}\left[Y_{\ell}(n;\pmb{\mu},\pmb{\Theta}_n)^2\right]}{\mathbb{E}\left[Y_{\ell}(n;\pmb{\mu},\pmb{\Theta}_n)\right]^2}=\frac{1}{n \mathbb{P}\left[ {D}_{x,\ell} \right]}+\frac{n-1}{n} \frac{\mathbb{P}\left[ {D}_{x,\ell} \cap {D}_{y,\ell} \right]}{\left( \mathbb{P}\left[ {D}_{x,\ell} \right]\right)^2}.
\label{eq:degree_10}
\end{equation}

From (\ref{eq:degree_7}) and (\ref{eq:degree_10}), we see that the zero-law will follow if we show that
\begin{equation}
\lim_{n \to \infty} n \mathbb{P}\left[ {D}_{x,\ell} \right]=\infty,
\label{eq:degree_11}
\end{equation}
and
\begin{equation}
\mathbb{P}\left[ {D}_{x,\ell} \cap {D}_{y,\ell} \right] \sim \left(\mathbb{P}\left[ {D}_{x,\ell} \right]\right)^2
\label{eq:degree_12}
\end{equation}
for some $\ell=0,1,\ldots,k-1$ under the enforced assumptions. The next two results will help establish (\ref{eq:degree_11}) and (\ref{eq:degree_12}). 


{
\lemma 
\label{lemma:2}
If $\Lambda_1(n)=o\left( \frac{1}{\sqrt n} \right)$, then for any non-negative integer constant $\ell$ and any node $v_x$, we have
\begin{equation}
\mathbb{P}\left[ {D}_{x,\ell} \right] \sim \mu_1 \left( \ell ! \right)^{-1} \left( n \Lambda_1(n) \right)^\ell e^{-n \Lambda_1(n)}
\label{eq:degree_13}
\end{equation}
}
\begin{proof}
Considering any class-$1$ node $v_i$, and recalling (\ref{eq:osy_mean_edge_prob_in_system}), we know that the events $E_{1i}, E_{2i},\ldots,E_{(i-1)i},E_{(i+1)i},\ldots,E_{ni}$ are mutually independent. Thus, it follows that the degree of a given node $v_i$, conditioned on being class-$1$, follows a Binomial distribution Bin$\left(n-1,\Lambda_1(n) \right)$. Thus,
\begin{align} 
\mathbb{P}\left[ {D}_{i,\ell} \right]&=\mu_1 \mathbb{P}\left[ D_{i,\ell} \given[\big] t_i=1 \right] \nonumber \\
&=\mu_1 \binom{n-1}{\ell} \Lambda_1(n)^\ell \left(1-\Lambda_1(n) \right)^{n-\ell-1} \nonumber
\end{align}

Next, given that $\Lambda_1(n)=o\left( \frac{1}{\sqrt n} \right)$ and $\ell$ is constant, it follows that $\Lambda_1(n)=o(1)$ and $\Lambda_1(n)^2 (n-\ell-1)=o(1)$. Invoking Fact~\ref{fact:3}, and the fact that $\binom{n-1}{\ell} \sim \left( \ell ! \right)^{-1} n^\ell$, the conclusion (\ref{eq:degree_13}) follows.
\end{proof}

{
\lemma
\label{lemma:3}
Consider scalings $K_1,\ldots,K_r,P: \mathbb{N}_0 \rightarrow \mathbb{N}_0^{r+1}$ and $\alpha: \mathbb{N}_0 \rightarrow (0,1)$, such that  $\lambda_1(n)=o(1)$ and
(\ref{scaling_condition_KG}) holds with $\lim_{n \to \infty} \gamma_n=-\infty$. The following two properties hold

(a) If $n \Lambda_1(n) = \Omega(1)$,
then for any non-negative integer constant $\ell$ and any two distinct nodes $v_x$ and $v_y$, we have
\begin{equation}
\mathbb{P}\left[ {D}_{x,\ell} \cap {D}_{y,\ell} \right] \sim \mu_1^2 \left( \ell ! \right)^{-2} \left( n \Lambda_1(n) \right)^{2\ell} e^{-2n \Lambda_1(n)}
\label{eq:lemma3_conc1}
\end{equation}

(b) For any two distinct nodes $v_x$ and $v_y$, we have
\begin{equation}
\mathbb{P}\left[ {D}_{x,0} \cap {D}_{y,0} \right] \sim  \mu_1^2 e^{-2n \Lambda_1(n)}
\label{eq:lemma3_conc2}
\end{equation}
}

The proof of Lemma~\ref{lemma:3} is given in Appendix~\ref{app:proof_of_lemma_3}. We now show why the zero-law follows from Lemma~\ref{lemma:2} and Lemma~\ref{lemma:3}
by means of establishing (\ref{eq:degree_11}) and (\ref{eq:degree_12}) for some $\ell=0, 1, \ldots, k-1$. 
 First, we see from (\ref{scaling_condition_KG}) that $\Lambda_1(n) \leq \frac{\log n+(k-1)\log \log n}{n}=o\left( \frac{1}{\sqrt n}\right)$
 when $\lim_{n \to \infty} \gamma_n=-\infty$.
  Invoking Lemma~\ref{lemma:2}, this gives
\begin{equation}
n \mathbb{P}\left[ {D}_{x,\ell} \right] \sim n \mu_1 \left( \ell ! \right)^{-1} \left( n \Lambda_1(n) \right)^\ell e^{-n \Lambda_1(n)}
\label{eq:finalizing_1}
\end{equation}
for each $\ell=0,1,\ldots$. 
We will obtain (\ref{eq:degree_11}) and (\ref{eq:degree_12})
using subsubsequence principle \cite[p. 12]{JansonLuczakRucinski} and considering the cases where $n \Lambda_1(n) = \Omega(1)$ and $n \Lambda_1(n)=o(1)$ separately.

\subsubsection{The case where there exists an $\epsilon>0$ such that $n \Lambda_1(n)>\epsilon$ for all $n$ sufficiently large}
\label{subsection:1}
In this case we will establish (\ref{eq:degree_11}) and (\ref{eq:degree_12}) for $\ell=k-1$. Setting $\ell=k-1$ and substituting (\ref{scaling_condition_KG}) into (\ref{eq:finalizing_1}), we get
\begin{align}
&n \mathbb{P}\left[ {D}_{x,\ell} \right] \nonumber\\
 &\sim n \mu_1 \left[ \left(k-1\right)! \right]^{-1} \left( n \Lambda_1(n) \right)^{k-1} e^{-\log n-(k-1) \log \log n - \gamma_n} \nonumber \\
&=\mu_1 \left[ \left(k-1\right)! \right]^{-1} \left( \log n+(k-1) \log \log n +\gamma_n \right)^{k-1} \cdot \nonumber \\
&\quad \cdot  e^{-(k-1) \log \log n - \gamma_n}
\label{eq:finalizing_2}
\end{align}
Let
\begin{align} 
&f_n(k;\gamma_n) \nonumber \\
&:=\left( \log n+(k-1) \log \log n +\gamma_n \right)^{k-1} e^{-(k-1) \log \log n - \gamma_n},\nonumber
\end{align}
and note that $\left( \log n+(k-1) \log \log n +\gamma_n \right) \geq \epsilon$ for all $n$ sufficiently large by virtue of the fact that $n \Lambda_1(n)>\epsilon$. Fix $n$ sufficiently large, pick $\zeta \in (0,1)$ and consider the cases when $\gamma_n \leq -(1-\zeta) \log n$ and $\gamma_n > -(1-\zeta) \log n$, separately. In the former case, we get
\begin{equation} \nonumber
f_n(k;\gamma_n) \geq \epsilon e^{-(k-1)\log \log n+(1-\zeta)\log n},
\end{equation}
and in the latter case, we get
\begin{equation} \nonumber
f_n(k;\gamma_n) \geq \left( \zeta \log n \right)^{k-1} e^{-(k-1)\log \log n - \gamma_n}=\zeta^{k-1} e^{-\gamma_n}.
\end{equation}
Thus, for all $n$ sufficiently large, we have
\begin{equation} \nonumber
f_n(k;\gamma_n) \geq \min \left\{ \epsilon e^{-(k-1)\log \log n+(1-\zeta)\log n}, \zeta^{k-1} e^{-\gamma_n}\right\}.
\end{equation}
It is now clear that 
\begin{equation}
\lim_{n \to \infty} f_n(k;\gamma_n) = \infty,
\label{eq:finalizing_3}
\end{equation}
since $\zeta \in (0,1)$ and $\lim_{n \to \infty} \gamma_n=-\infty$. Reporting (\ref{eq:finalizing_3}) into (\ref{eq:finalizing_2}), we establish (\ref{eq:degree_11}). Furthermore, from Lemma~\ref{lemma:2} and Lemma~\ref{lemma:3}, it is clear that (\ref{eq:degree_12}) follows for $\ell=k-1$.

\subsubsection{The case where $\lim_{n \to \infty} n \Lambda_1(n)=0$}
\label{subsection:2}
In this case, we will establish (\ref{eq:degree_11}) and (\ref{eq:degree_12}) for $\ell=0$. Setting $\ell=0$ in (\ref{eq:finalizing_1}), we obtain
\begin{equation} \nonumber
n \mathbb{P}\left[ {D}_{x,0} \right] \sim n \mu_1 e^{n\Lambda_1(n)} \sim n \mu_1
\end{equation}
by virtue of the fact that $n \Lambda_1(n)= o(1)$. This readily gives (\ref{eq:degree_11}). Furthermore, from Lemma~\ref{lemma:2} (with $\ell=0$) and Lemma~\ref{lemma:3}, (\ref{eq:degree_12}) immediately follows. 

The two cases considered cover all the possibilities for the limit of $n \Lambda_1(n)$. By virtue of the subsubsequence principle \cite[p. 12]{JansonLuczakRucinski}, we get
(\ref{eq:degree_11}) and (\ref{eq:degree_12}) without any condition on the sequence $n \Lambda_1(n)$; i.e., we obtain the zero-law even when the sequence $n \Lambda_1(n)$ does not have a limit!

\section{Proof of Theorem \ref{theorem:kconnectivity}}
\subsection{Establishing the zero-law}
Let $\kappa$ denote the the vertex connectivity of $\mathbb{H}(n,\pmb{\mu},\pmb{\Theta}_n)$, i.e., the minimum number of nodes to be deleted to make the graph disconnected. 
Also, let $\delta$ denote the minimum node degree in $\mathbb{H}(n,\pmb{\mu},\pmb{\Theta}_n)$.
It is clear that if a random graph is $k$-connected, meaning that $\kappa \geq k$, then it does not have any node with degree less than $k$. Thus
$\left[ \kappa \geq k \right] \subseteq \left[ \delta \geq k  \right]
$
and the conclusion
\begin{equation}
\mathbb{P}[\kappa \geq k] \leq \mathbb{P}[\delta \geq k ]
\label{eq:conn_ZeroLaw}
\end{equation}
immediately follows. In view of (\ref{eq:conn_ZeroLaw}), we obtain the zero-law for $k$-connectivity, i.e., that
\begin{equation} \nonumber
\lim_{n\to\infty} \mathbb{P}[\mathbb{H}(n;\pmb{\mu},\pmb{\Theta}_n) \text{ is }k \text{-connected}]= 0,
\end{equation}
when $\lim_{n \to \infty} \gamma_n=-\infty$ from the zero-law part of Theorem~\ref{theorem:min_node_degree}. Put differently, the conditions that lead to the zero-law part of Theorem~\ref{theorem:min_node_degree}, i.e., $\lambda_1(n)=o(1)$ and $\lim_{n \to \infty} \gamma_n=- \infty$, automatically lead to the zero-law part of Theorem~\ref{theorem:kconnectivity}.

\subsection{Establishing the one-law}
An important step towards establishing the one-law of Theorem~\ref{theorem:kconnectivity}
is presented in Appendix~\ref{subsection:confining}. There, we show that it suffices to establish the one law in Theorem \ref{theorem:kconnectivity} under the additional condition that $\gamma_n=o \left(\log n \right)$, which leads to a number of useful consequences. 
Let a sequence $\beta_{\ell,n}: \mathbb{N} \times \mathbb{N}_0 \rightarrow \mathbb{R}$ be defined through the relation
\begin{equation}
\Lambda_1(n)=\frac{\log n+ \ell \log \log n+\beta_{\ell,n}}{n}
\label{eq:kconn_new_scaling}
\end{equation}
for each $n \in \mathbb{N}_0$ and $\ell \in \mathbb{N}$.
In view of the arguments in 
Appendix~\ref{subsection:confining},
the one-law (\ref{eq:kconn_OneLaw_Statement}) follows from the next result.
{
\theorem
\label{lemma_key_lemma}
Let $\ell$ be a non-negative constant integer. Under (\ref{eq:conn_Pn}), (\ref{eq:conn_KrPn}), (\ref{eq:conn_Kr_K1}), and (\ref{eq:kconn_new_scaling}) with $\beta_{\ell,n}=o \left( \log n \right)$ and $\lim_{n \to \infty} \beta_{\ell,n}= +\infty$, we have

\begin{equation} \nonumber
\lim_{n \to \infty} \mathbb{P}\left[ \kappa=\ell \right]=0.
\end{equation}
}

Before we give a formal proof, we first explain why the one-law (\ref{eq:kconn_OneLaw_Statement}) follows from Theorem~\ref{lemma_key_lemma}. Comparing (\ref{eq:kconn_new_scaling}) with (\ref{scaling_condition_KG}) and noting that $\gamma_n=o \left(\log n \right)$, we get
\begin{align}
\beta_{\ell,n}&=(k-1-\ell) \log \log n + \gamma_n =o \left( \log n \right)
\label{eq:KeyLemma_1}
\end{align}
Moreover, for $\ell=0,1,\ldots,k-1$, we have
\begin{equation}
\lim_{n \to \infty} \beta_{\ell,n}=+ \infty
\label{eq:KeyLemma_2}
\end{equation}
by recalling the fact that $\lim_{n \to \infty} \gamma_n=+\infty$. Recalling (\ref{eq:KeyLemma_1}) and (\ref{eq:KeyLemma_2}), we notice that the conditions needed for Theorem~\ref{lemma_key_lemma} are met when $\ell=0,1,\ldots,k-1$; thus, we have $\mathbb{P}\left[ \kappa=\ell \right]=o(1)$ for $\ell=0,1,\ldots,k-1$, which in turn implies that $\lim_{n \to \infty} \mathbb{P}\left[ \kappa \geq k \right]=1$, i.e., the one-law.

We now give a road map to the proof of Theorem~\ref{lemma_key_lemma}. By a simple union bound, we get
\begin{equation} \nonumber
\mathbb{P}\left[ \kappa=\ell \right] \leq \mathbb{P}\left[ \delta \leq \ell \right] + \mathbb{P}\left[ \left(\kappa=\ell \right) \cap (\delta > \ell) \right].
\end{equation}
It is now immediate that Theorem~\ref{lemma_key_lemma} is established once we show that
\begin{equation}
\lim_{n \to \infty} \mathbb{P}\left[ \delta \leq \ell \right]=0
\label{eq:kconn_union_bound_part1}
\end{equation}
and
\begin{equation} 
\lim_{n \to \infty} \mathbb{P}\left[ \left(\kappa=\ell \right) \cap (\delta > \ell) \right]=0
\label{eq:kconn_union_bound_part2}
\end{equation}
under the enforced assumptions of Theorem~\ref{lemma_key_lemma}. We start by establishing (\ref{eq:kconn_union_bound_part1}). Following the analysis of Section~\ref{subsec:one_law_min_node_degree}, it is easy to see that
\begin{align}
n\mathbb{P}\left[ D_{x,\ell} \right] &\leq  2^{\ell-1} \left(\left(\log n\right)^\ell \left(1+ o(1)\right)^\ell +\beta_{\ell,n}^\ell \right) \cdot \nonumber \\
& \quad \quad \cdot e^{- \ell \log \log n - \beta_{\ell,n}} e^{O(1)}\nonumber \\
& = O(1)  e^{- \beta_{\ell,n}} + O(1) \beta_{\ell,n}^\ell  e^{-\ell \log \log n - \beta_{\ell,n}}, \nonumber
\end{align}
and it follows that
$\lim_{n \to \infty} n \mathbb{P}\left[ D_{x,\ell} \right]=0$ 
as long as $\lim_{n \to \infty} \beta_{\ell,n}=+ \infty$. From (\ref{eq:degree_3_pt1}) and (\ref{eq:degree_4}), this yields 
\begin{equation}
\lim_{n \to \infty} \mathbb{P} \left[ \delta=\ell \right]=0 \quad \textrm{when $\lim_{n \to \infty} \beta_{\ell,n}=+ \infty$}
\label{eq:MinDegree_OneLaw_1}
\end{equation}
However, from (\ref{eq:kconn_new_scaling}) it is easy to see that  $\beta_{\ell,n}$ is monotonically decreasing in $\ell$. Thus, the fact that $\lim_{n \to \infty} \beta_{\ell,n}=+ \infty$ for some $\ell$ implies
\[
\lim_{n \to \infty} \beta_{\hat{\ell},n}=+ \infty, \quad \hat{\ell} = 0, 1, \ldots, \ell
\]  
From (\ref{eq:MinDegree_OneLaw_1}) this in turn implies
that $\mathbb{P} [ \delta=\hat{\ell} ]=o(1)$ for $\hat{\ell}=0,1,\ldots,\ell$, or equivalently 
(\ref{eq:kconn_union_bound_part1}).

We now focus on establishing (\ref{eq:kconn_union_bound_part2}) under the enforced assumptions of Theorem~\ref{lemma_key_lemma}. The proof is based on finding a tight upper bound on the probability $\mathbb{P}\left[ \left(\kappa=\ell \right) \cap \delta > \ell \right]$ and showing that this bound goes to zero as $n$ goes to infinity. Let $\mathcal{N}$ denote the collection of all non-empty subsets of $\{v_1, v_2, \ldots,v_n \}$. Define $\mathcal{N}_*=\{ T : T \in \mathcal{N}, ~|T| \geq 2 \}$ and 
\begin{equation} \nonumber
\mathcal{E}(\pmb{J})=\cup_{T \in \mathcal{N}_*} \left[ |\cup_{v_i \in T} \Sigma_i| \leq J_{|T|} \right]
\end{equation}
where $\pmb{J}=\left[ J_2,J_3,\ldots,J_n \right]$ is an $(n-1)$-dimensional integer-valued array. $\mathcal{E}(\pmb{J})$ encodes the event that for at least one $|T| = 2,\ldots,n$, the total number of distinct keys held by at least one set of $|T|$ sensors is less than or equal to $J_{|T|}$. Now, define 
\begin{align}
m_n &:=\min \left( \left \lfloor{\frac{P_n}{K_{1,n}}}\right \rfloor, \left \lfloor{\frac{n}{2}}\right \rfloor \right)
\label{eq:defn_m_n}
\end{align}
 and let
\begin{equation}
J_i=
\begin{cases} 
       \max \left( \left \lfloor{(1+\epsilon) K_{1,n}}\right \rfloor, \left \lfloor{i \zeta K_{1,n}}\right \rfloor \right)    \hfill & i=2,\ldots,m_n \\
       \left \lfloor{\psi P_n}\right \rfloor \hfill & i=m_n+1,\ldots,n \\
\end{cases}
\label{eq:conn_X}
\end{equation}
for some $\zeta, \psi$ in $(0,1)$ to be specified later at (\ref{eq:conn_zeta}) and (\ref{eq:conn_psi}), respectively.
A crude bounding argument gives
\begin{equation} \nonumber
\mathbb{P}\left[ \left(\kappa=\ell \right) \cap \delta > \ell \right] \leq \mathbb{P}\left[ \mathcal{E}(\pmb{J})\right] + \mathbb{P}\left[ \left(\kappa=\ell \right) \cap \delta > \ell \cap \overline{\mathcal{E}(\pmb{J})}\right]
\end{equation}

Hence, establishing (\ref{eq:kconn_union_bound_part2}) consists of establishing the following two results.
{
\prop
\label{prop_kconn2}
Let $\ell$ be a non-negative constant integer. Assume that (\ref{eq:kconn_new_scaling}) holds with $\beta_{\ell,n}>0$, and that we have (\ref{eq:conn_KrPn}) and (\ref{eq:conn_Kr_K1}). Also, assume that  (\ref{eq:conn_Pn}) holds such that
\[
P_n \geq \sigma n
\]
for some $\sigma>0$ for all $n$ sufficiently large.
Then
\begin{equation} \nonumber
\lim_{n \to \infty} \mathbb{P}\left[ \mathcal{E}(\pmb{J})\right]=0,
\end{equation}
where $\pmb{J}$ is as defined in (\ref{eq:conn_X}) with arbitrary $\epsilon \in (0,1)$, constant $\zeta \in (0,\frac{1}{2})$ selected small enough such that
\begin{equation}
\max  \left( 2 \zeta \sigma, \zeta \left( \frac{e^2}{\sigma} \right)^{\frac{\zeta}{1-2 \zeta}} \right) < 1
\label{eq:conn_zeta}
\end{equation}
and $\psi \in (0,\frac{1}{2})$ selected small enough such that
\begin{equation}
\max  \left( 2 \left( \sqrt{\psi} \left( \frac{e}{\psi} \right)^\psi \right)^\sigma, \sqrt{\psi} \left( \frac{e}{\psi} \right)^\psi \right) < 1
\label{eq:conn_psi}
\end{equation}
}
\begin{proof}
The proof follows the same steps with 
\cite[Proposition 7.2]{Yagan/Inhomogeneous}
to show that it suffices to establish Proposition \ref{prop_kconn2} for the homogenous case where all key rings are of the same size $K_{1,n}$. This is evident upon realizing that
 with $U_\ell (\pmb{\mu},\pmb{\theta})=| \cup_{i=1}^\ell \Sigma_i|$ and $U_\ell (K_{1,n},P_n)=_{st} U_\ell (\pmb{\mu}=\{1,0,\ldots,0\},\pmb{\theta})$, we have 
 \[
 U_\ell (K_{1,n},P_n) \preceq U_\ell (\pmb{\mu},\pmb{\theta}),
 \] 
 where $\preceq$ denotes the usual stochastic ordering. 
After this reduction, the proof reduces to \cite[Proposition~3]{Jun/K-Connectivity}. Results only require conditions (\ref{eq:conn_Pn}), (\ref{eq:conn_K1_omega_1}), and $K_{1,n}=o(P_n)$ to hold. We note that $K_{1,n}=o(P_n)$ follows from (\ref{eq:conn_KrPn}) and the fact that $K_{1,n} \leq K_{r,n}$. Also, (\ref{eq:conn_K1_omega_1}) follows under the enforced assumptions as shown in Proposition \ref{prop:conn_K1_omega_1}. \end{proof}

{
\prop
\label{prop_kconn3}
Let $\ell$ be a non-negative constant integer. Under (\ref{eq:conn_Pn}), (\ref{eq:conn_KrPn}), (\ref{eq:conn_Kr_K1}),  and (\ref{eq:kconn_new_scaling}) with $\beta_{\ell,n}=o \left( \log n \right)$ and $\lim_{n \to \infty} \beta_{\ell,n}= +\infty$, we have

\begin{equation} \nonumber
\lim_{n \to \infty} \mathbb{P}\left[ \left(\kappa=\ell \right) \cap \left( \delta > \ell \right) \cap \overline{\mathcal{E}(\pmb{J})}\right]=0
\end{equation}
}
The proof of Proposition~\ref{prop_kconn3} is given in Section~\ref{sec:prop_kconn3}.
Proposition~\ref{prop_kconn2} and Proposition~\ref{prop_kconn3} establish (\ref{eq:kconn_union_bound_part2}) which, combined with (\ref{eq:kconn_union_bound_part1}), establish Theorem~\ref{lemma_key_lemma}. We remark that Theorem~\ref{lemma_key_lemma} establishes the one-law.

\section{Proof of Proposition~\ref{prop_kconn3}}
\label{sec:prop_kconn3}
For notation simplicity, we denote $\mathbb{H}(n;\pmb{\mu},\pmb{K},P,\alpha)$ by $\mathbb{H}$. Let $\mathbb{H}(U)$ be a subgraph of $\mathbb{H}$ restricted to the vertex set $U$. For any subset of nodes $U$, define $U^c:=\{v_1,\ldots,v_n\} \setminus U$. We also let $\mathcal{N}_{U^c}$ denote the collection of all non-empty subsets of $\{v_1,v_2,\ldots,v_n\} \setminus U$. We note that a subset $T$ of $\mathcal{N}_{U^c}$ is isolated in $\mathbb{H}(U^c)$ if there are no edges in $\mathbb{H}$ between nodes in $T$ and nodes in $U^c \setminus T$, i.e., 
\begin{equation*}
\overline{E_{ij}}, \quad v_i \in T, \quad v_j \in U^c \setminus T.
\end{equation*}

Next, we present key observations that pave the way to establishing Proposition~\ref{prop_kconn3}. If $\kappa=\ell$ but $\delta>\ell$, then there exists subsets $U$ and $T$ of nodes with $U \in \mathcal{N}$, $|U|=\ell$, $T \in \mathcal{N}_{U^c}$, $|T| \geq 2$ such that $\mathbb{H}(T)$ is connected while $T$ is isolated in $\mathbb{H}(U^c)$. This ensures that $\mathbb{H}$ can be disconnected by deleting a properly selected set of $\ell$ nodes, i.e., the set $U$. This would not be possible for sets $T \in \mathcal{N}_{U^c}$ with $|T|=1$ since we have $\delta \geq \ell+1$ which implies that the single node in $T$ is connected to at least one node in $U^c \setminus T$. Finally, having $\kappa=\ell$ ensures that $\mathbb{H}$ remains connected after removing $(\ell-1)$ nodes. Then, if there exists a subset $U$ with $|U|=\ell$ such that some $T \in \mathcal{N}_{U^c}$ is isolated in  $\mathbb{H}(U^c)$, each node in $U$ must be connected to at least one node in $T$ and at least one node in $U^c \setminus T$. This can be proved by contradiction. Consider subsets $U \in \mathcal{N}$ with $|U|=\ell$, and $T \in \mathcal{N}_{U^c}$ with $|T| \geq 2$, such that $T$ is isolated from $U^c \setminus T$. Suppose there exists a node $v_i \in U$ such that $v_i$ is connected to at least one node in $T$ but not connected to any node in $U^c \setminus T$. In this case, it is easy to see that there are no edges between nodes in $U^c \setminus T$ and nodes in $\{v_i\} \cup T$. Thus, the graph could have been made disconnected by removing nodes in $U \setminus \{v_i\}$. But $|U \setminus \{v_i\}|=\ell-1$, and this contradicts the fact that $\kappa=\ell$.

We now present several events that characterize the aforementioned observations. For each non-empty subset $T \subseteq U^c$, we define $\mathcal{C}_T$ as the event that $\mathbb{H}(T)$ is itself connected, and $\mathcal{D}_{U,T}$ as the event that $T$ is isolated in $\mathbb{H}(U^c)$, i.e.,
\begin{equation*}
\mathcal{D}_{U,T}:=\bigcap_{\substack{v_i \in T \\ v_j \in U^c \setminus T}} \overline{E_{ij}},
\end{equation*}
Moreover, we define $\mathcal{B}_{U,T}$ as the event that each node in $U$ has an edge with at least one node in $T$, i.e.,
\begin{equation*}
\mathcal{B}_{U,T}:=\bigcap_{v_i \in U} \bigcup_{v_j \in T} E_{ij},
\end{equation*}
and finally, we let $\mathcal{A}_{U,T}:=\mathcal{B}_{U,T} \cap \mathcal{D}_{U,T} \cap \mathcal{C}_T$. It is clear that $\mathcal{A}_{U,T}$ encodes the event that $\mathbb{H}(T)$ is itself connected, each node in $U$ has an edge with at least one node in $T$, but $T$ is isolated in $\mathbb{H}(U^c)$. 
 The aforementioned observations enable us to express the event $\left[ \left( \kappa=\ell \right) \cap \left( \delta>\ell \right) \right]$ in terms of the event sequence $\mathcal{A}_{U,T}$. In particular, we have
\begin{equation*}
\left[ \left( \kappa=\ell \right) \cap \left( \delta>\ell \right) \right] \subseteq \bigcup_{{U \in \mathcal{N}_{n,\ell}, T \in \mathcal{N}_{U^c}, |T| \geq 2}} \mathcal{A}_{U,T}
\end{equation*}
with $\mathcal{N}_{n,\ell}$ denoting the collection of all subsets of $\{v_1,\ldots,v_n\}$ with exactly $\ell$ elements. We also note that the union need only to be taken over all subsets $T$ with $2 \leq |T| \leq \left \lfloor{\frac{n-\ell}{2}}\right \rfloor$. This is because if the vertices in $T$ form a component then so do the vertices in $\mathcal{N}_{U^c} \setminus T$. Now, using a standard union bound, we obtain
\begin{align*}
&\mathbb{P}\left[ \left(\kappa=\ell \right) \cap \left( \delta > \ell \right) \cap \overline{\mathcal{E}(\pmb{J})}\right] \nonumber \\
&\leq \sum_{U \in \mathcal{N}_{n,\ell}, T \in \mathcal{N}_{U^c},2 \leq |T| \leq \left \lfloor{\frac{n-\ell}{2}}\right \rfloor} \mathbb{P}\left[ \mathcal{A}_{U,T} \cap \overline{\mathcal{E}(\pmb{J})} \right]\\
&= \sum_{m=2}^{\left \lfloor{\frac{n-\ell}{2}}\right \rfloor} \sum_{U \in \mathcal{N}_{n,\ell}, T \in \mathcal{N}_{U^c,m}} \mathbb{P}\left[ \mathcal{A}_{U,T} \cap \overline{\mathcal{E}(\pmb{J})} \right]
\end{align*}
where $\mathcal{N}_{U^c,m}$ denotes the collection of all subsets of $U^c$ with exactly $m$ elements. Now, for each $m=1,\ldots,n-\ell-1$, we simplify the notation by writing $\mathcal{A}_{\ell,m}:=\mathcal{A}_{\{ v_1,\ldots,v_\ell\},\{ v_{\ell+1},\ldots,v_{\ell+m} \}}$, $\mathcal{D}_{\ell,m}:=\mathcal{D}_{\{v_1,\ldots,v_\ell \},\{ v_{\ell+1},\ldots,v_{\ell+m} \}}$, $\mathcal{B}_{\ell,m}:=\mathcal{B}_{\{v_1,\ldots,v_\ell \},\{ v_{\ell+1},\ldots,v_{\ell+m} \}}$, and $\mathcal{C}_{m}:=\mathcal{C}_{\{ v_{\ell+1},\ldots,v_{\ell+m} \}}$. 
From exchangeability, we get
\begin{equation*}
\mathbb{P}\left[ \mathcal{A}_{U,T} \right] = \mathbb{P}\left[ \mathcal{A}_{\ell,m} \right], \quad U \in \mathcal{N}_{n,\ell}, \ \ T \in \mathcal{N}_{U^c,m}
\end{equation*}
and the key bound
\begin{align} 
&\mathbb{P}\left[ \left(\kappa=\ell \right) \cap \left( \delta > \ell \right) \cap \overline{\mathcal{E}(\pmb{J})}\right] \nonumber \\
&\leq \sum_{m=2}^{\left \lfloor{\frac{n-\ell}{2}}\right \rfloor} \binom{n}{\ell} \binom{n-\ell}{m} \mathbb{P}\left[ \mathcal{A}_{\ell,m} \cap \overline{\mathcal{E}(\pmb{J})} \right] 
\end{align}
is obtained readily upon noting that
 $|\mathcal{N}_{n,\ell}|=\binom{n}{\ell}$ and $|\mathcal{N}_{U^c,m}|=\binom{n-\ell}{m}$. 
 Thus, Proposition~\ref{prop_kconn3} will be established if we show that
\begin{equation}
\lim_{n \to \infty} \sum_{m=2}^{\left \lfloor{\frac{n-\ell}{2}}\right \rfloor} \binom{n}{\ell} \binom{n-\ell}{m} \mathbb{P}\left[ \mathcal{A}_{\ell,m} \cap \overline{\mathcal{E}(\pmb{J})} \right]=0.
\label{eq:kconn_3}
\end{equation}

We now derive bounds for the probabilities $\mathbb{P}\left[ \mathcal{A}_{\ell,m} \cap \overline{\mathcal{E}(\pmb{J})} \right]$. First, for $m=2,\ldots,n-\ell-1$, we have
\begin{equation} \label{eq:defn_D_l_m}
\mathcal{D}_{\ell,m}:= \bigcap_{j=m+\ell+1}^n \left[ \left( \cup_{i \in \nu_{m,j}} \Sigma_i \right) \cap \Sigma_j = \emptyset \right]
\end{equation}
where $\nu_{m,j}$ is defined as
\begin{equation} \nonumber
\nu_{m,j}:=\{i=\ell+1,\ldots,\ell+m : C_{ij}\}
\end{equation}
for each $j=1,\ldots, \ell$ and $j=m+\ell+1,\ldots,n$. Put differently, $\nu_{m,j}$ is the set of indices in $i=\ell+1,\ldots,\ell+m $ for which nodes $v_j$ and $v_i$ are adjacent in the ER graph $\mathbb{G}(n;\alpha_n)$. Then, (\ref{eq:defn_D_l_m})
follows from the fact that
for $v_j$ to be isolated from $\{v_{\ell+1}, \ldots, v_{\ell+m}\}$ 
in $\mathbb{H}$, $\Sigma_j$ needs to be disjoint from each of the key rings $\{\Sigma_i: i \in \nu_{m,j}\}$.

 Now, using the law of iterated expectation, we get
\begin{align}
&\mathbb{P}\left[ \mathcal{D}_{\ell,m} \given[\Big]  \Sigma_{\ell+1}, \ldots, \Sigma_{\ell+m} \right] \nonumber \\
&=\mathbb{E}\left[ \pmb{1}\left[\mathcal{D}_{\ell,m} \right]\given[\Big]  \Sigma_{\ell+1}, \ldots, \Sigma_{\ell+m} \right]  \nonumber \\
&=\mathbb{E}\left[ \mathbb{E} \left[ \pmb{1}\left[\mathcal{D}_{\ell,m} \right]  \given[\Big] \substack{ \Sigma_{\ell+1}, \ldots, \Sigma_{n} \\ C_{ij}, i=\ell+1, \ldots, \ell+m \\ \quad \ j=\ell+m+1,\ldots,n} \right]\given[\Bigg]  \Sigma_{\ell+1}, \ldots, \Sigma_{\ell+m} \right]\nonumber \\
&=\mathbb{E}\left[ \prod_{j=\ell+m+1}^n \left( \frac{\binom{P-|\cup_{i \in \nu_{m,j}} \Sigma_i|}{|\Sigma_j|}}{\binom{P}{|\Sigma_j|}} \right)  \given[\Bigg]  \Sigma_{\ell+1}, \ldots, \Sigma_{\ell+m} \right]\nonumber \\
&=\mathbb{E}\left[ \frac{\binom{P-|\cup_{i \in \nu_m} \Sigma_i|}{|\Sigma|}}{\binom{P}{|\Sigma|}} \given[\Bigg]  \Sigma_{\ell+1}, \ldots, \Sigma_{\ell+m} \right]^{n-\ell-m} \label{eq:osy_new_to_use1}
\end{align}
by independence of the random variables $\nu_{m,j}$ and $|\Sigma_j|$ for $j=\ell+m+1,\ldots,n$. Here we define $\nu_m$ and $\left|\Sigma\right|$ as generic random variables following the same distribution with  any of $\{\nu_{m,j}, j= \ell + m+ 1, \ldots, n\}$
and $\{|\Sigma_j|, j= \ell + m+ 1, \ldots, n\}$, respectively.
Put differently, $\nu_m$ is a Binomial rv with parameters $m$ and $\alpha$, while $\left|\Sigma\right|$ is a rv that takes the value $K_j$ with probability $\mu_j$. 

Next, we bound the probabilities $\mathbb{P}\left[ \mathcal{B}_{\ell,m} \right]$. We know that
\begin{equation} \nonumber
\mathcal{B}_{\ell,m}:=\cap_{i=1}^{\ell} \cup_{j=\ell+1}^{m} E_{ij}.
\end{equation}
Thus,
\begin{align}
&\mathbb{P}\left[ \mathcal{B}_{\ell,m} \given[\Big]  \Sigma_{\ell+1}, \ldots, \Sigma_{\ell+m} \right] \nonumber \\
&=\mathbb{E}\left[ \pmb{1}\left[\mathcal{B}_{\ell,m} \right]  \given[\Big]  \Sigma_{\ell+1}, \ldots, \Sigma_{\ell+m}\right]\nonumber \\
&=\mathbb{E}\left[ \mathbb{E} \left[ \pmb{1}\left[\mathcal{B}_{\ell,m} \right] \given[\Big] \substack{\Sigma_{1}, \ldots, \Sigma_{\ell+m} \\ C_{ij}, i=\ell+1, \ldots, \ell+m \\  j=1,\ldots,\ell} \right] \given[\Bigg]  \Sigma_{\ell+1}, \ldots, \Sigma_{\ell+m} \right]\nonumber \\
&=\mathbb{E}\left[ \prod_{j=1}^\ell \left( 1-\frac{\binom{P-|\cup_{i \in \nu_{m,j}} \Sigma_i|}{|\Sigma_j|}}{\binom{P}{|\Sigma_j|}} \right) \given[\Bigg]  \Sigma_{\ell+1}, \ldots, \Sigma_{\ell+m}  \right]\nonumber \\
&=\mathbb{E}\left[ 1-\frac{\binom{P-|\cup_{i \in \nu_m} \Sigma_i|}{|\Sigma|}}{\binom{P}{|\Sigma|}} \given[\Bigg]  \Sigma_{\ell+1}, \ldots, \Sigma_{\ell+m} \right]^{\ell} 
\label{eq:osy_new_to_use2}
\end{align}
by independence of the random variables $\nu_{m,j}$ and $|\Sigma_j|$ for $j=1,\ldots,\ell$. 

We note that, on the event $\overline{\mathcal{E}(\pmb{J})}$, we have
\begin{equation} \nonumber
|\cup_{i \in \nu_m} \Sigma_i| \geq \left( J_{|\nu_m|}+1 \right) 1\left[ |\nu_m|>1 \right]
\end{equation}
and it is always the case that
$|\cup_{i \in \nu_m} \Sigma_i| \geq K_1 1\left[ |\nu_m|>0 \right]
$
and
\begin{equation} \label{eq:osy_new_to_use3}
|\cup_{i \in \nu_m} \Sigma_i| \leq |\nu_m| K_r.
\end{equation}
Next, we define
\begin{equation} \nonumber
L(\nu_m)=\max \left( K_1 1\left[ |\nu_m|>0 \right], \left( J_{|\nu_m|}+1 \right) 1\left[ |\nu_m|>1 \right]\right)
\end{equation}
so that on $\overline{\mathcal{E}(\pmb{J})}$, we have
\begin{equation}
|\cup_{i \in \nu_m} \Sigma_i| \geq L(\nu_m).
\label{eq:osy_new_to_use4}
\end{equation}
%
Using (\ref{eq:osy_new_to_use4}) in (\ref{eq:osy_new_to_use1})
and (\ref{eq:osy_new_to_use3}) in (\ref{eq:osy_new_to_use2}), we get
\begin{align}
&\mathbb{P}\left[ \mathcal{A}_{\ell,m} \cap \overline{\mathcal{E}(\pmb{J})} \right] \label{eq:kconn_1}
 \\
&=\mathbb{E}\left[ \pmb{1}\left[\mathcal{C}_m\right]  \pmb{1} \left[ \mathcal{B}_{\ell,m} \right]  \pmb{1}\left[ \mathcal{D}_{\ell,m} \cap \overline{\mathcal{E}(\pmb{J})} \right] \right] \nonumber \\
&=\mathbb{E}\left[ \mathbb{E}\left[ \pmb{1}\left[\mathcal{C}_m\right]  \pmb{1} \left[ \mathcal{B}_{\ell,m} \right]  \pmb{1}[ \mathcal{D}_{\ell,m} \cap \overline{\mathcal{E}(\pmb{J})} ]  \given[\Big]  \substack{\Sigma_{\ell+1}, \ldots, \Sigma_{\ell+m} \\ C_{ij}, i,j=\ell+1, \ldots, \ell+m} \right]  \right] \nonumber \\
&\leq \mathbb{P}\left[ \mathcal{C}_m \right]  \mathbb{E}\left[ 1-\frac{\binom{P-|\nu_m|K_r}{|\Sigma|}}{\binom{P}{|\Sigma|}} \right]^{\ell}  \mathbb{E}\left[ \frac{\binom{P-L(\nu_m)}{|\Sigma|}}{\binom{P}{|\Sigma|}} \right]^{n-\ell-m} 
\nonumber
\end{align}
since $\mathcal{C}_m$ is fully determined by the rvs $\Sigma_{\ell+1}, \ldots, \Sigma_{\ell+m}$ and $\{C_{ij}, i,j=\ell+1, \ldots, \ell+m\}$ while $\mathcal{B}_{\ell,m}$, $\mathcal{D}_{\ell,m}$, and $\mathcal{E}(\pmb{J})$ are independent from $\{C_{ij}, i,j=\ell+1, \ldots, \ell+m\}$. Here, we also used the fact that given $\{\Sigma_{\ell+1}, \ldots, \Sigma_{\ell+m}\}$,
$\mathcal{D}_{\ell,m}$ is independent from $\mathcal{B}_{\ell,m}$.

The following lemma provides upper bounds for (\ref{eq:kconn_1}).
{
\lemma
\label{lemma_kconn_KeyLemma}
Let $\pmb{J}$ be defined as in (\ref{eq:conn_X}) for some $\epsilon \in (0,1)$, $\zeta \in \left(0,\frac{1}{2} \right)$ such that (\ref{eq:conn_zeta}) holds, $\psi \in \left(0,\frac{1}{2}\right)$ such that (\ref{eq:conn_psi}) holds. Assume that $\Lambda_1(n)=o(1)$ and  (\ref{eq:conn_Pn}), (\ref{eq:conn_KrPn}), and (\ref{eq:conn_Kr_K1}) hold. Then
 for all $n$ sufficiently large, and for each $m=2,3,\ldots, n$, we have
\begin{align}
&\mathbb{P}\left[ \mathcal{A}_{\ell,m} \cap \overline{\mathcal{E}(\pmb{J})} \right]  \label{eq:key_kconn_bound1} \\
&\leq \min\left \{1, m^{m-2} \left(\alpha_n p_{rr}(n) \right)^{m-1}\right \} \Bigg (\hspace{-.6mm}\pmb{1} \left[m \hspace{-.3mm}>\hspace{-.2mm} \left \lfloor{\frac{P_n-K_{r,n}}{2K_{r,n}}}\right \rfloor \right] 
\nonumber \\
& ~~
+\pmb{1} \left[m \leq \left \lfloor{\frac{P_n-K_{r,n}}{2K_{r,n}}}\right \rfloor \right]
 \left(1- e^{- 3 m \alpha_n p_{rr}(n) } \right)^\ell  \Bigg) \cdot
 \nonumber \\
& ~~~ \cdot \hspace{-.5mm}\Bigg(\hspace{-1mm}\min \Bigg\{ 1-\Lambda_1(n),e^{- \Big( 1+\frac{\epsilon}{2}\Big) \Lambda_1(n)}, e^{- \psi K_{1,n}} \pmb{1}\left[ m>m_n \right]  + \nonumber \\
& ~~~\quad  \min \Big\{ 1\hspace{-.3mm}-\hspace{-.3mm}\mu_r\hspace{-.5mm}+\hspace{-.5mm}\mu_r e^{-\alpha_n p_{1r}(n) \zeta m},e^{-\alpha_n p_{11}(n) \zeta m} \Big\}\hspace{-.5mm}  \Bigg\} \hspace{-.5mm}\Bigg)^{\hspace{-1mm} n-m-\ell} 
\nonumber
\end{align}

%
}

The proof of Lemma~\ref{lemma_kconn_KeyLemma} is given in Appendix~\ref{app:proof_of_KeyLemma}. Now, the proof of Proposition~\ref{prop_kconn3} will be completed upon establishing (\ref{eq:kconn_3}) by means of Lemma~\ref{lemma_kconn_KeyLemma}. We devote Section~\ref{sec:establishing} to establishing (\ref{eq:kconn_3}).

\section{Establishing (\ref{eq:kconn_3})}
\label{sec:establishing} 
In this section, we make several use of the following lemma.
{\lemma
\label{lemma:kconn_1}
Consider a scaling $K_1,K_2,\ldots,K_r,P:\mathbb{N}_0 \rightarrow \mathbb{N}_0^{r+1}$ and a scaling $\alpha: \mathbb{N}_0 \rightarrow (0,1)$ such that (\ref{eq:kconn_new_scaling}) holds with $\beta_{\ell,n}=o(\log n)$. We have
\begin{equation}
\frac{1}{2} \frac{\log n}{n} \leq \alpha_n p_{1r}(n) \leq \frac{2}{\mu_r} \frac{\log n}{n},
\label{eq:kconn_range2}
\end{equation}
for all $n$ sufficiently large, i.e., $\alpha_n p_{1r}(n)=\Theta\left(\frac{\log n}{n}\right)$. If in addition (\ref{eq:conn_Kr_K1}) holds, we have
\begin{equation}
\alpha_n p_{rr}(n) =o \left( \log n \right) \alpha_n p_{1r}(n)=o \left(\frac{(\log n)^2}{n} \right)
\label{eq:conn_prr_asym}
\end{equation}
and
\begin{equation}
\alpha_n p_{1r}(n) =o \left( \log n \right) \alpha_n p_{11}(n)
\label{eq:conn_p11_asym}
\end{equation}
}
The proof of Lemma~\ref{lemma:kconn_1} is given in Appendix~\ref{app:proof_of_lemma_A4}.

We now proceed with establishing (\ref{eq:kconn_3}). We start by defining $f_{n,\ell,m}$ as 
\begin{equation*}
f_{n,\ell,m}=\binom{n}{\ell} \binom{n-\ell}{m} \mathbb{P}\left[ \mathcal{A}_{\ell,m} \cap \overline{\mathcal{E}(\pmb{J})} \right]
\end{equation*}
Thus, establishing (\ref{eq:kconn_3}) becomes equivalent to showing
\begin{equation}
\lim_{n \to \infty} \sum_{m=2}^{\left \lfloor{\frac{n-\ell}{2}}\right \rfloor} f_{n,\ell,m}=0. 
\label{eq:kconn_4}
\end{equation}
We will establish (\ref{eq:kconn_4}) in several steps with each step focusing on a specific range of the summation over $m$. Throughout, we consider scalings $K_1,\ldots,K_r,P: \mathbb{N}_0 \rightarrow \mathbb{N}_0^{r+1}$ and $\alpha: \mathbb{N}_0 \rightarrow (0,1)$ such that (\ref{eq:kconn_new_scaling}) 
holds with $\lim_{n \to \infty} \beta_{\ell,n} = +\infty$ and $\beta_{\ell,n}=o(\log n)$,  and (\ref{eq:conn_Pn}), (\ref{eq:conn_KrPn}), (\ref{eq:conn_Kr_K1}) hold. We will make repeated use of the bounds (\ref{eq:conn_bounds_4}), (\ref{eq:conn_bounds_5}),  (\ref{eq:conn_bounds_6}), and  (\ref{eq:conn_prr_asym}).
\subsubsection{The case where $2 \leq m \leq M$}
This range considers fixed values of $m$. Pick an integer $M$ to be specified later at (\ref{eq:choosing_R}). We note that on this range we have
 $m \leq \lfloor{\frac{P_n-K_{r,n}}{2 K_{r,n}}}\rfloor$ for all $n$ sufficiently large by virtue of (\ref{eq:conn_KrPn}).
On the same range we also 
have
\begin{equation}
1- e^{-3m \alpha_n p_{rr}(n)} \leq 3m \alpha_n p_{rr}(n)
\label{eq:to_be_used_first_range}
\end{equation}
by virtue of (\ref{eq:conn_prr_asym}), (\ref{eq:exp_bound}), and the fact that $m$ is bounded. 

Using (\ref{eq:conn_bounds_6}), (\ref{eq:key_kconn_bound1}), (\ref{eq:conn_prr_asym}), and (\ref{eq:to_be_used_first_range}), and noting that
$\Lambda_1(n)=o(1)$ under (\ref{eq:kconn_new_scaling}) with  $\beta_{\ell,n}=o(\log n)$, we get
\begin{align*}
&f_{n,\ell,m} \nonumber \\
&\leq n^\ell n^m m^{m-2} \left(\alpha_n p_{rr}(n) \right)^{m-1} \left(3 m \right)^\ell \left(\alpha_n p_{rr}(n) \right)^\ell \cdot \nonumber \\
&\quad \cdot e^{- \left( 1+\frac{\epsilon}{2}\right) (n-m-\ell) \Lambda_1(n) } \nonumber \\
&=O(1) n^{\ell+m} \left(\alpha_n p_{rr}(n) \right)^{\ell+m-1} \cdot  e^{- \left( 1+\frac{\epsilon}{2}\right) (n-m-\ell) \Lambda_1(n) } \nonumber \\
& = o(1) n^{\ell+m} \left(\frac{(\log n)^2}{n} \right)^{\ell+m-1} \hspace{-2mm}e^{-\left( 1+\frac{\epsilon}{2} \right) \left(\log n + \ell \log \log n+ \beta_{\ell,n} \right)}
\nonumber \\ 
&={o(1)} n^{-\frac{\epsilon}{2}} \left( \log n \right)^{\ell \left( 1-\frac{\epsilon}{2} \right) + 2(m-1)} e^{-\left( 1+\frac{\epsilon}{2} \right) \beta_{\ell,n}} \nonumber \\
&=o(1)
\end{align*}
since $\ell$ is non-negative integer constant, $m$ is bounded, and $\lim_{n \to \infty} \beta_{\ell,n}=+\infty$.
This establishes
\begin{equation} \nonumber
\lim_{n \to \infty} \sum_{m=2}^{M} f_{n,\ell,m}=0.
\end{equation}

\subsubsection{The case where $M+1 \leq m \leq \min\{m_n,  \lfloor{\frac{\mu_r n}{2 \zeta \log n}} \rfloor \}$}
\label{subsection:geometric_series}
Our goal in this and the next subsubsection is to cover the range $M+1 \leq m \leq  \lfloor{\frac{\mu_r n}{2 \zeta \log n}} \rfloor $.
Since the bound given at (\ref{eq:key_kconn_bound1}) takes a different form when $m > m_n$ (with $m_n$ defined at (\ref{eq:defn_m_n})), we first consider the 
range $M+1 \leq m \leq \min\{m_n,  \lfloor{\frac{\mu_r n}{2 \zeta \log n}} \rfloor \}$;
we note from (\ref{eq:conn_KrPn}) and (\ref{scaling_condition_K}) that 
$\lim_{n \to \infty} m_n = \infty$.

On the range considered here, we have from (\ref{eq:conn_bounds_4}), (\ref{eq:conn_bounds_6}), and (\ref{eq:key_kconn_bound1}) that
\begin{align}
&\sum_{m=M+1}^{\min\{m_n, \lfloor{\frac{\mu_r n}{2 \zeta \log n}} \rfloor\}} f_{n,\ell,m}  \nonumber \\
&\leq \sum_{m=M+1}^{\min\{m_n, \lfloor{\frac{\mu_r n}{2 \zeta \log n}} \rfloor \}}  n^\ell \left( \frac{en}{m} \right)^m m^{m-2} \left( \alpha_n p_{rr}(n) \right)^{m-1} \cdot \nonumber \\
&\quad \cdot \left( 1-\mu_r \left( 1-e^{- \alpha_n p_{1r}(n) \zeta m} \right)\right)^{n-m-\ell}
\label{eq:kconn_11}
\end{align}
From the upper bound in (\ref{eq:kconn_range2})  and the fact that $m \leq \frac{\mu_r n}{ 2 \zeta \log n}$ for all $n$ sufficiently large, we have 
\begin{align*}
 \alpha_n p_{1r}(n) \zeta m \leq \frac{2  \log n }{ \mu_r n} \zeta \frac{\mu_r n}{2 \zeta \log n}  =  1.
\end{align*}
Using the fact that
$
1-e^{-x} \geq \frac{x}{2}$ for all $0 \leq x \leq 1$,
we  get
\begin{align}
1-\mu_r \left( 1-e^{- \alpha_n p_{1r}(n) \zeta m} \right) &\leq 1  -  \frac{\mu_r \alpha_n p_{1r}(n)\zeta m }{2} \nonumber \\
&\leq e^{- \zeta m \mu_r \frac{\log n}{4 n} }
\label{eq:conn_12}
\end{align}
as we invoke the lower bound in (\ref{eq:kconn_range2}).
Reporting this last bound and (\ref{eq:conn_prr_asym}) into (\ref{eq:kconn_11}), and noting that 
\begin{equation}
n-m-\ell \geq \frac{n-\ell}{2} \geq \frac{n}{3}, \qquad m = 2, 3, \ldots, \left \lfloor \frac{n-\ell}{2} \right \rfloor,
\label{eq:bounding_n_minus_l}
\end{equation}
 we get 
\begin{align}
&\sum_{m=M+1}^{\min\{m_n, \lfloor{\frac{\mu_r n}{2 \zeta \log n}} \rfloor\}} f_{n,\ell,m} \nonumber \\
&\leq \hspace{-1mm}\sum_{m=M+1}^{\min\{m_n, \lfloor{\frac{\mu_r n}{2 \zeta \log n}} \rfloor \}} \hspace{-1mm} n^{\ell+m} e^m  \left( \frac{(\log n)^2}{n} \right)^{\hspace{-1mm}m-1}   \hspace{-2mm} e^{-\zeta m \mu_r \log n \frac{n-m-\ell}{4n}} \nonumber \\
&\leq n^{\ell+1} \sum_{m=M+1}^\infty \left( e \left(\log n\right)^2 e^{-\zeta \frac{\mu_r}{12} \log n}  \right)^m
\label{eq:kconn_13}
\end{align}
for all $n$ sufficiently large. 
Given that 
 $\zeta,\mu_r>0$ we  have
\begin{equation}
 e \left(\log n\right)^2 e^{-\zeta \frac{\mu_r}{12} \log n}  = o(1).
 \label{eq:summable_sequence}
 \end{equation}
Thus, the geometric series in (\ref{eq:kconn_13}) is summable, and we have
\begin{align}
\sum_{m=M+1}^{\min\{m_n, \lfloor{\frac{\mu_r n}{2 \zeta \log n}} \rfloor\}} \hspace{-2mm}f_{n,\ell,m}  \leq O(1) n^{\ell+1-(M+1) \zeta \frac{\mu_r}{12}} \left(e \log n \right)^{2(M+1)} \nonumber
\end{align}
and it follows that
\begin{equation} \nonumber
\lim_{n \to \infty} \sum_{m=M+1}^{\min\{m_n, \lfloor{\frac{\mu_r n}{2 \zeta \log n}} \rfloor\}} f_{n,\ell,m}=0 
\end{equation}
for any positive integer $M$ with
\begin{equation} 
M>\frac{12 (\ell+1)}{\zeta \mu_r}. 
\label{eq:choosing_R}
\end{equation}
This choice is permissible given that $\zeta, \mu_r > 0$. 


\subsubsection{The case where $ \min \{ \lfloor{\frac{\mu_r n}{ 2 \zeta \log n}} \rfloor,m_n \} < m \leq  \lfloor{\frac{\mu_r n}{ 2 \zeta \log n}} \rfloor$}

Clearly, this range becomes obsolete if $m_n \geq  \lfloor{\frac{\mu_r n}{ 2 \zeta \log n}} \rfloor$. 
Thus, it suffices to consider the subsequences for which the range $m_n+1 \leq m \leq  \lfloor{\frac{\mu_r n}{2 \zeta \log n}} \rfloor$ is non-empty. On this range, following the same arguments
that lead to (\ref{eq:kconn_11}) and (\ref{eq:kconn_13}) gives
\begin{align}
&\sum_{m=m_n+1}^{ \lfloor{\frac{\mu_r n}{2 \zeta \log n}} \rfloor} f_{\ell,n,m} \label{eq:conn_newrange_2}
 \\
&\leq \sum_{m=m_n+1}^{ \lfloor{\frac{\mu_r n}{2 \zeta \log n}} \rfloor}  n^{\ell+1} \left( e (\log n)^2\right)^{m} \cdot \nonumber \\
& \quad \cdot \left( 1-\mu_r \left(1-e^{-\zeta m \alpha_n p_{1r}(n)} \right)+e^{-\psi K_{1,n}} \right)^{\frac{n}{3}}
\nonumber \\
& \leq n^{\ell+1} \sum_{m=m_n+1}^{ \lfloor{\frac{\mu_r n}{2 \zeta \log n}} \rfloor} \hspace{-3mm} \left(e \left(\log n \right)^2 \right)^m \hspace{-1mm} \left( e^{- \zeta m \mu_r \frac{\log n}{4n} } + e^{-\psi K_{1,n}}\right)^{\frac{n}{3}}
\nonumber
\end{align}
where in the last step we used
 (\ref{eq:conn_12}) in view of $m \leq \frac{\mu_r n}{2 \zeta \log n}$. Next, we write
\begin{align}
&e^{- \zeta m \mu_r \frac{\log n}{4n} } + e^{-\psi K_{1,n}} \nonumber \\
&=  e^{- \zeta m \mu_r \frac{\log n}{4n} } \left(1+e^{-\psi K_{1,n} + \zeta m \mu_r \frac{\log n}{4n}} \right) \nonumber \\
& \leq \exp \left\{- \zeta m \mu_r \frac{\log n}{4n} + e^{-\psi K_{1,n} + \zeta m \mu_r \frac{\log n}{4n}} \right\} \nonumber \\
& \leq \exp \left\{- \zeta m \mu_r \frac{\log n}{4n} \left(1 - \frac{e^{-\psi K_{1,n} + \frac{\mu_r^2 }{8}}}{\zeta m \mu_r \frac{\log n}{4n}} \right) \right\} \label{eq:intermediary_range_new}
\end{align}
where the last inequality is obtained from $m \leq {\frac{\mu_r n}{ 2 \zeta \log n}}$. 
Using the fact that $m > m_n=\min\{  \lfloor{\frac{P_n}{K_{1,n}}} \rfloor,  \lfloor{\frac{n}{2}} \rfloor \}$
 and that $P_n \geq \sigma n$ for some $\sigma>0$ under (\ref{eq:conn_Pn}), we have 
\begin{align}
\frac{e^{-\psi K_{1,n} + \frac{\mu_r^2}{8}}}{\zeta m \mu_r \frac{\log n}{4n}}  & \leq 
\max\left\{  \frac{K_{1,n}}{P_n},  \frac{2}{n} \right \}{4n}
\frac{e^{-\psi K_{1,n}}}{\zeta \mu_r \log n} \cdot e^{\frac{\mu_r^2}{8}}\nonumber \\
&\leq \max \left \{ \frac{4 K_{1,n} e^{-\psi K_{1,n}}}{\zeta \mu_r \sigma \log n},\frac{8 e^{-\psi K_{1,n}}}{\zeta \mu_r \log n} \right\}\cdot e^{\frac{\mu_r^2}{8}} \nonumber \\
&=o(1) \nonumber
\end{align}
by virtue of (\ref{eq:conn_K1_omega_1}) and  the facts that $\zeta, \mu_r, \sigma>0$. Reporting this
into (\ref{eq:intermediary_range_new}), we see that for
for any $\varepsilon>0$, there exists a finite integer $n^*(\varepsilon)$ such that 
\begin{equation}
\left( e^{- \zeta m \mu_r \frac{\log n}{4n} } + e^{-\psi K_{1,n}}\right) \leq e^{- \zeta m \mu_r \frac{\log n}{4n} (1-\varepsilon)} 
\label{eq:conn_newrange_3}
\end{equation}
for all $n \geq n^*(\varepsilon)$.
Using (\ref{eq:conn_newrange_3}) in (\ref{eq:conn_newrange_2}), we get
\begin{align}
&\sum_{m=m_n+1}^{ \lfloor{\frac{\mu_r n}{2 \zeta \log n}} \rfloor} f_{\ell,n,m} \nonumber \\
&\leq n^{\ell+1} \sum_{m=m_n+1}^{ \lfloor{\frac{\mu_r n}{2 \zeta \log n}} \rfloor} \hspace{-3mm}  \left(e \left(\log n \right)^2 \right)^m \left( e^{- \zeta m \mu_r \frac{\log n}{4n} (1-\varepsilon)} \right)^{\frac{n}{3}}  \nonumber \\
&\leq n^{\ell+1} \sum_{m=m_n+1}^{\infty} \left( e \left(\log n\right)^2 e^{- \zeta \mu_r \frac{\log n}{12} (1-\varepsilon)}  \right)^m
\label{eq:conn_newrange_4}
\end{align}
Similar to (\ref{eq:summable_sequence}), we have
$e \left(\log n\right)^2 e^{- \zeta \mu_r \frac{\log n}{12} (1-\varepsilon)} = o(1)$ so that the sum 
in (\ref{eq:conn_newrange_4}) converges. 
Following a similar approach to that in Section~\ref{subsection:geometric_series}, we then see that 
\begin{equation*}
 \sum_{m=m_n+1}^{ \lfloor{\frac{\mu_r n}{2 \zeta \log n}} \rfloor} \hspace{-1.4mm}f_{n,\ell,m} \hspace{-.4mm}= 
 O(1) n^{\ell+1-m_n  \hspace{-.4mm}\frac{\zeta\mu_r (1-\varepsilon)}{12}}\hspace{-.4mm} (e\log n)^{2(m_n+1)} \hspace{-.6mm}=\hspace{-.4mm}o(1)
\end{equation*}
since $\lim_{n \to \infty} m_n = \infty$ under the enforced assumptions. 

\subsubsection{The case where $\lfloor{\frac{\mu_r n}{2 \zeta \log n}} \rfloor+1 \leq m \leq \left \lfloor{\nu n}\right \rfloor $} 
\label{subsec:range_4}
 We  consider $\lfloor{\frac{\mu_r n}{2 \zeta \log n}} \rfloor+1 \leq m \leq \left \lfloor{\nu n}\right \rfloor $ for some $\nu \in (0,\frac{1}{2})$ to be specified later at (\ref{eq:conn_16}). Recalling (\ref{eq:conn_bounds_4}), (\ref{eq:conn_bounds_6}), (\ref{eq:key_kconn_bound1}), (\ref{eq:kconn_range2}), and (\ref{eq:bounding_n_minus_l}), and noting that $\binom{n}{m}$ is monotone increasing in $m$ when $0 \leq m \leq \left \lfloor{\frac{n}{2}}\right \rfloor$, we get 
\begin{align}
&\sum_{m= \lfloor{\frac{\mu_r n}{2 \zeta \log n}} \rfloor +1}^{ \left \lfloor{\nu n}\right \rfloor} f_{n,\ell,m} \nonumber \\
&\leq  \sum_{m= \lfloor{\frac{\mu_r n}{ 2 \zeta \log n}} \rfloor+1}^{\left \lfloor{\nu n} \right \rfloor} n^\ell \binom{n}{\left \lfloor{\nu n}\right \rfloor} \cdot \nonumber \\
&\quad \cdot \left( 1-\mu_r+\mu_re^{- \zeta m \alpha_n p_{1r}(n)} + e^{- \psi K_{1,n}}\right)^{\frac{n}{3}}\nonumber \\
& \leq n^\ell \sum_{m= \lfloor{\frac{\mu_r n}{2 \zeta \log n}} \rfloor +1}^{ \left \lfloor{\nu n}\right \rfloor} \left( \frac{e}{\nu} \right)^{\nu n} \cdot \nonumber \\
& \quad \cdot \bigg( 1-\mu_r +\mu_re^{-   \zeta  \frac{\mu_r n}{ 2 \zeta \log n}  \frac{ \log n}{ 2 n  }} + e^{- \psi K_{1,n}}\bigg)^{\frac{n}{3}} \nonumber \\
&\leq n^\ell \left( \frac{e}{\nu} \right)^{\nu n}  \left( 1-\mu_r+\mu_re^{- \frac{\mu_r}{4}} + e^{- \psi K_{1,n}} \right)^{\frac{n}{3}}
\nonumber \\
&= n^\ell \left( \left( \frac{e}{\nu} \right)^{3\nu} \left( 1-\mu_r+\mu_re^{- \frac{\mu_r}{4}} + e^{-\psi K_{1,n} }\right)\right)^{\frac{n}{3}}
\label{eq:conn_15}
\end{align}
 for all $n$ sufficiently large.

We have $1-\mu_r+\mu_re^{- \frac{\mu_r}{4}}<1$ from $\mu_r>0$ and 
$e^{- \psi K_{1,n}}=o(1)$ from
 (\ref{eq:conn_K1_omega_1}). Also,  it holds that $\lim_{\nu \to 0} \left( \frac{e}{\nu}\right)^{3\nu}=1$.
Thus, if we pick $\nu$ small enough to ensure that
\begin{equation}
\left( \frac{e}{\nu} \right)^{3\nu} \left( 1-\mu_r+\mu_re^{- \frac{\mu_r}{4}} \right) < 1,
\label{eq:conn_16}
\end{equation}
then for any $0< \varepsilon < 1- \left( {e}/{\nu} \right)^{3\nu} \left( 1-\mu_r+\mu_re^{- \frac{\mu_r}{4}} \right)$ there exists a finite integer $n^{\star}(\varepsilon)$  such that
\[
 \left( \frac{e}{\nu} \right)^{3\nu} \left( 1-\mu_r+\mu_re^{- \frac{\mu_r}{4}} + e^{-\psi K_{1,n} }\right) \leq 1- \varepsilon, \quad  \forall n \geq n^{\star}(\varepsilon).
\]
Reporting this into (\ref{eq:conn_15}), we get
\begin{equation} \nonumber
\lim_{n \to \infty} \sum_{m= \lfloor{\frac{\mu_r n}{2 \zeta \log n}} \rfloor +1}^{ \left \lfloor{\nu n}\right \rfloor} f_{n,\ell,m}=0
\end{equation}
since $\lim_{n \to \infty} n^\ell (1-\varepsilon)^{n/2} = 0$ for any positive integer $\ell$.

\subsubsection{The case where $\left \lfloor{\nu n}\right \rfloor+1 \leq m \leq \lfloor{\frac{n-\ell}{2}} \rfloor $}
\label{subsec:range_5}
In this range, we use (\ref{eq:conn_bounds_5}), (\ref{eq:conn_bounds_6}), (\ref{eq:key_kconn_bound1}), and (\ref{eq:bounding_n_minus_l}) to get
 \begin{align*}
&\sum_{m=\left \lfloor{\nu n}\right \rfloor+1}^{\left \lfloor{\frac{n-\ell}{2}}\right \rfloor} f_{n,\ell,m} \nonumber \\
&\leq n^\ell  \sum_{m=\left \lfloor{\nu n}\right \rfloor+1}^{\left \lfloor{\frac{n-\ell}{2}}\right \rfloor} \binom{n}{m}  \left( e^{- \zeta m \alpha_n p_{11}(n)} + e^{- \psi K_{1,n}}\right)^{\frac{n}{3}}\nonumber \\
&\leq n^\ell \left(\sum_{m=\left \lfloor{\nu n}\right \rfloor+1}^{\left \lfloor{\frac{n-\ell}{2}}\right \rfloor} \binom{n}{m} \right) \left( e^{- \zeta \nu n \alpha_n p_{11}(n)} + e^{- \psi K_{1,n}}\right)^{\frac{n}{3}} \nonumber
\\
& \leq  n^\ell \left( 8 e^{- \zeta \nu n \alpha_n p_{11}(n)} + 8 e^{- \psi K_{1,n}}\right)^{\frac{n}{3}} 
\end{align*}
 
Noting that $\zeta,\nu,\psi>0$ and recalling (\ref{eq:conn_p11_asym}) and the lower bound of (\ref{eq:kconn_range2}), we get 
\begin{align}
e^{- \zeta \nu n \alpha_n p_{11}(n)} &=e^{- \zeta \nu n \frac{w_n}{\log n} \alpha_n p_{1r}(n)}  \leq e^{- \frac{\zeta \nu w_n}{2}} \nonumber
\end{align}
for some sequence $w_n$ satisfying  $\lim_{n \to \infty} w_n=+\infty$. It is now obvious that $e^{- \zeta \nu n \alpha_n p_{11}(n)}=o(1)$. Moreover, we have $e^{- \psi K_{1,n}}=o(1)$ from
 (\ref{eq:conn_K1_omega_1}).
The conclusion
\begin{equation} \nonumber
\lim_{n \to \infty} \sum_{m=\left \lfloor{\nu n}\right \rfloor+1}^{\left \lfloor{\frac{n-\ell}{2}}\right \rfloor} f_{n,\ell,m}=0
\end{equation}
immediately follows and
 the proof of  one-law is completed.

\section*{Acknowledgment}
This work has been supported in part by National Science Foundation through grants CCF-1617934 and CCF-1422165
and in part by the start-up funds from the Department of Electrical and Computer Engineering at Carnegie Mellon University (CMU). 

\bibliographystyle{IEEEtran}
\bibliography{IEEEabrv,TMC}

\newpage
\appendices
\numberwithin{equation}{section}

\section{Additional Preliminaries}

\label{appendix_lemmas}
{\prop[{\cite[Proposition~4.4]{Yagan/Inhomogeneous}}]
For any set of positive integers $K_1,\ldots,K_r,P$ and any scalar $a \geq 1$, we have
\begin{equation}
\frac{\binom {P-\left \lceil{aK_i}\right \rceil }{K_j}}{\binom P{K_j}} \leq
\left(\frac{\binom {P-K_i}{K_j}}{\binom P{K_j}}\right)^a, \quad i,j=1,\ldots,r
\label{eq:combinatorial_bound}
\end{equation}
}

{
\prop
\label{prop:popo}
Consider a random variable $Z$ defined as
\begin{equation} \nonumber
Z=1-p_{1i}=\frac{\binom{P-K_1}{K_i}}{\binom{P}{K_i}}, \quad \textrm{with probability } \mu_i, \quad i=1,\ldots,r.
\end{equation}
We have
$
\textrm{var} \left[ Z \right] \leq \frac{1}{4} \left( p_{1r}\right)^2.
$}

\begin{proof}
Recalling (\ref{eq:ordering_of_lambda}), we see that  $p_{ij}$ increases with both $i$ and $j$, and it follows that
\begin{equation} \nonumber
1-p_{1r} \leq Z \leq 1-p_{11},
\end{equation}
From Popoviciu's inequality \cite[pp.~9]{jensen_99}, we see that
\begin{align} \nonumber
\textrm{var} \left[ Z \right] &\leq \frac{1}{4} \left( Z_{\max}-Z_{\min} \right)^2 
= \frac{1}{4} \left( p_{1r}-p_{11} \right)^2 
 \leq\frac{1}{4} \left( p_{1r}\right)^2 
\end{align}
since $p_{1r} \geq p_{11} \geq 0$.
\end{proof}

{
\fact
\label{fact:lambda_o(1)}
If $\lambda_1(n)=o(1)$, then
\begin{equation} \nonumber
p_{1i}(n)=o(1), \quad i=1,\ldots,r
\end{equation}
}

\begin{proof}
Recalling (\ref{eq:osy_mean_edge_prob_in_RKG}), we obtain
\begin{equation} \nonumber
p_{1i}(n) \leq \left( \frac{1}{\mu_i} \right) \lambda_1(n)=O\left(\lambda_1(n)\right)=o(1)
\end{equation}
under the given assumption that $\lambda_1(n)=o(1)$.
\end{proof}

{
\fact
\label{fact:2}
For $0 \leq x \leq 1$, the following properties hold.

(a) {\cite[Fact~2]{Jun/K-Connectivity}} If $0 < y <1$, then $\left( 1- x \right)^y \leq 1-xy$.

(b) Let $a>1$. Then,
$1-x^a \leq a(1-x)$.
}
\begin{proof}
By a crude bounding, we have
\begin{equation} \nonumber
1-x^a=\int_x^1 at^{a-1} \ dt \leq \int_x^1 a \ dt=a(1-x).
\end{equation} 
\end{proof}

\setcounter{page}{1}

{
\fact [{\cite[Fact~5]{Jun/K-Connectivity}}]
\label{fact:fact5}
Let $a$, $x$, and $y$ be positive integers satisfying $y\geq (2a+1)x$. Then,
\begin{equation} \nonumber
\frac{\binom{y-ax}{x}}{\binom{y}{x}} \geq \left[ \frac{\binom{y-x}{x}}{\binom{y}{x}}\right]^{2a}
\end{equation}
}

{
\fact
\label{fact:myFact1}
Let $x \in (0,1)$ and $a>1$. Then,
\begin{equation} \nonumber
1-x^a \leq a(1-x)
\end{equation}
}

{
\lemma 
\label{lemma:lemma9}
Consider a scaling $K_1,\ldots,K_r,P: \mathbb{N}_0 \rightarrow \mathbb{N}_0^{r+1}$ such that (\ref{scaling_condition_K}) holds, a scaling $\alpha: \mathbb{N}_0 \rightarrow (0,1)$, and $\Lambda_1(n)=\frac{\log n+(k-1)\log \log n+\gamma_n}{n}$. The following properties hold for any three distinct nodes $v_x$, $v_y$, and $v_j$.

(a) We have
\begin{equation}
\mathbb{P} \left[ \left( K_{xj} \cap K_{yj}\right) \given[\big] \overline{K_{xy}},t_x=1,t_y=1 \right] \leq \left( 1+\frac{1}{4 \mu_r^2} \right) \lambda_1(n)^2 
\label{eq:lemma9_6}
\end{equation}

(b) If $\lambda_1(n)=o(1)$, then for any $u=0,1,\ldots,K_{1,n}$, we have
\begin{align} 
&\mathbb{P} \left[ \left( K_{xj} \cap K_{yj}\right) \given[\big] \left( |S_{xy}|=u \right) ,t_x=1,t_y=1\right] \nonumber \\
&= \frac{u}{K_{1,n}} \lambda_1(n) \pm O\left( \left( \lambda_1(n) \right)^2 \right), \nonumber
\end{align}
and
\begin{align}
&\mathbb{P} \left[ E_{xj \cup yj} \given[\big] \left( |S_{xy}|=u \right) ,t_x=1,t_y=1\right] \nonumber \\
&= 2 \Lambda_1(n) - \frac{\alpha_n u}{K_{1,n}} \Lambda_1(n) \pm O\left( \left( \Lambda_1(n) \right)^2 \right)
\label{eq:lemma9_13}
\end{align}
}

\begin{proof}
We know that
\begin{align}
&\mathbb{P} \left[ \left( K_{xj} \cap K_{yj}\right) \given[\bigg] \left( |S_{xy}|=u \right) ,t_x=1,t_y=1\right] \nonumber \\
&=1-\mathbb{P} \left[ \left( \overline{K_{xj}} \cup \overline{K_{yj}}\right) \given[\bigg] \left( |S_{xy}|=u \right) ,t_x=1,t_y=1\right] \nonumber \\
&=1-\mathbb{P} \left[  \overline{K_{xj}} \given[\bigg] \left( |S_{xy}|=u \right) ,t_x=1,t_y=1\right] \label{eq:from_lemma_1} \\
& ~~ - \mathbb{P} \left[  \overline{K_{yj}} \given[\bigg] \left( |S_{xy}|=u \right) ,t_x=1,t_y=1\right] \nonumber \\
& ~~ + \mathbb{P} \left[ \left( \overline{K_{xj}} \cap \overline{K_{yj}}\right) \given[\bigg] \left( |S_{xy}|=u \right),t_x=1,t_y=1 \right] \nonumber
\end{align}

It is easy to see that
\begin{align}
& \hspace{-5mm} \mathbb{P} \left[  \overline{K_{xj}} \given[\bigg] \left( |S_{xy}|=u \right),t_x=1,t_y=1 \right] \nonumber \\
&=\mathbb{P} \left[  \overline{K_{xj}} \given[\Big] t_x=1\right] \nonumber \\
&=\sum_{i=1}^r \mu_i \left(1-p_{1i}(n)\right)\nonumber \\
&=1-\lambda_1(n)
\label{eq:lemma9_3}
\end{align}

Similarly, it is easy to see that
\begin{align}
\mathbb{P} \left[  \overline{K_{yj}} \given[\bigg] \left( |S_{xy}|=u \right) ,t_x=1,t_y=1 \right]=1-\lambda_1(n)
\label{eq:lemma9_4}
\end{align}

Next, by recalling (\ref{eq:combinatorial_bound}), we observe that
\begin{align}
&\mathbb{P} \left[ \left( \overline{K_{xj}} \cap \overline{K_{yj}}\right) \given[\bigg] \overline{K_{xy}},t_x=1,t_y=1 \right] \nonumber \\
&=\mathbb{P} \left[ \Sigma_j \in \mathcal{P} \setminus \left\{ \Sigma_x \cup \Sigma_y \right\} \given[\Big] t_x=1,t_y=1 \right] \nonumber \\
&= \sum_{i=1}^r \mu_i \frac{\binom{P_n-2K_{1,n}}{K_{i,n}}}{\binom{P_n}{K_{i},n}} \nonumber \\
&\leq \sum_{i=1}^r \mu_i \left( \frac{\binom{P_n-K_{1,n}}{K_{i,n}}}{\binom{P_n}{K_{i,n}}} \right)^2 \nonumber \\
&=\mathbb{E}\left[  Z_{n} \left(\pmb{\mu},\pmb{\theta}_n \right)^2 \right] \nonumber \\
&=\left( \mathbb{E}\left[  Z_{n} \left(\pmb{\mu},\pmb{\theta}_n \right) \right]\right)^2 +\textrm{var} \left[Z_{n} \left(\pmb{\mu},\pmb{\theta}_n \right) \right] 
\label{eq:lemma9_7}
\end{align}
where $Z_{n} \left(\pmb{\mu},\pmb{\theta}_n \right)$ is a rv that takes the value $1-p_{1i}(n)$ with probability $\mu_i$ for $i=1,\ldots,r$. Note that
\begin{equation}
\mathbb{E}\left[  Z_{n} \left(\pmb{\mu},\pmb{\theta}_n \right) \right] = \sum_{i=1}^r \mu_i \left(1-p_{1i} \right)=1-\lambda_1(n),
\label{eq:lemma9_1}
\end{equation}
and
\begin{equation}
\lambda_1(n)=\sum_{i=1}^r \mu_i p_{1i}(n)\geq \mu_r p_{1r}
\label{eq:lemma9_2}
\end{equation}
for positive $\pmb{\mu}$. Recalling Proposition~\ref{prop:popo}, and using (\ref{eq:lemma9_1}) and (\ref{eq:lemma9_2}) in  (\ref{eq:lemma9_7}), we get
\begin{align}
&\mathbb{P} \left[ \left( \overline{K_{xj}} \cap \overline{K_{yj}}\right) \given[\bigg] \overline{K_{xy}},t_x=1,t_y=1 \right] \nonumber \\
&\leq \left(1-\lambda_1(n) \right)^2 + \frac{1}{4} \frac{\lambda_1(n)^2}{\mu_r^2} \nonumber \\
&=1-2 \lambda_1(n) + \lambda_1(n)^2 \left( 1+\frac{1}{4 \mu_r^2} \right) 
\label{eq:lemma9_5}
\end{align}
The desired conclusion (\ref{eq:lemma9_6}) follows from (\ref{eq:from_lemma_1}) in view of (\ref{eq:lemma9_3}), (\ref{eq:lemma9_4}), and (\ref{eq:lemma9_5}).

Next, we establish part (b) of the lemma under the assumption that $\lambda_1(n)=o(1)$. Conditioning on $|S_{xy}|=u$ and recalling (\ref{eq:lemma9_7}), we see that
\begin{align}
&\mathbb{P} \left[ \left( \overline{K_{xj}} \cap \overline{K_{yj}}\right) \given[\bigg] \left(|S_{xy}|=u\right),t_x=1,t_y=1 \right] \nonumber \\
&=\sum_{i=1}^r \mu_i \frac{\binom{P_n-2K_{1,n}+u}{K_i}}{\binom{P}{K_i}}
\label{eq:lemma9_8}
\end{align}

Invoking Lemma~\ref{lemma:bounding_fac} and Fact~\ref{fact:2}, we observe that
\begin{align}
&\frac{\binom{P_n-2K_{1,n}+u}{K_{i,n}}}{\binom{P}{K_{i,n}}} \nonumber \\
&\leq \left(1- \frac{2K_{1,n}-u}{P_n}\right)^{K_{i,n}} \nonumber \\
& \leq 1-\frac{K_{i,n} \left( 2K_{1,n}-u \right)}{P_n}+\frac{1}{2} \left( \frac{K_{i,n} \left( 2K_{1,n}-u \right)}{P_n} \right)^2  \nonumber \\
&=1-\frac{K_{i,n} \left( 2K_{1,n}-u \right)}{P_n} + O\left( \left( \frac{K_{1,n}K_{i,n}}{P_n} \right)^2	\right)
\label{eq:lemma9_9}
\end{align}
and
\begin{align}
&\frac{\binom{P_n-2K_{1,n}+u}{K_{i,n}}}{\binom{P}{K_{i,n}}} \nonumber \\
&\geq \left(1- \frac{2K_{1,n}-u}{P_n-K_{1,n}}\right)^{K_{i,n}} \nonumber \\
&\geq 1- \frac{K_{i,n} \left( 2K_{1,n}-u \right)}{P_n-K_{1,n}} \nonumber \\
&=1- \left( \frac{K_{i,n} \left( 2K_{1,n}-u \right)}{P_n-K_{1,n}} - \frac{K_{i,n} \left( 2K_{1,n}-u \right)}{P_n} \right) \nonumber \\
&\quad -\frac{K_{i,n} \left( 2K_{1,n}-u \right)}{P_n} \nonumber \\
&=1-\frac{K_{i,n} \left( 2K_{1,n}-u \right)}{P_n} -O\left( \left( \frac{K_{1,n}K_{i,n}}{P_n} \right)^2 \right)
\label{eq:lemma9_10}
\end{align}

Combining (\ref{eq:lemma9_8}), (\ref{eq:lemma9_9}), and (\ref{eq:lemma9_10}), we notice that
\begin{align} 
&\mathbb{P} \left[ \left( \overline{K_{xj}} \cap \overline{K_{yj}}\right) \given[\bigg] \left(|S_{xy}|=u\right),t_x=1,t_y=1\right] \nonumber \\
&= \sum_{i=1}^r \mu_i \left( 1-\frac{K_{i,n} \left( 2K_{1,n}-u \right)}{P_n} \pm O\left( \left( \frac{K_{1,n}K_{i,n}}{P_n} \right)^2 \right) \right) \nonumber
\end{align}

Recalling Fact~\ref{fact:lambda_o(1)} and Fact~\ref{fact:asym_eq}, we observe that under the enforced assumption $\lambda_1(n)=o(1)$, we have
\begin{align}
&\mathbb{P} \left[ \left( \overline{K_{xj}} \cap \overline{K_{yj}}\right) \given[\bigg] \left(|S_{xy}|=u\right),t_x=1,t_y=1 \right] \nonumber \\
 &= \sum_{i=1}^r \mu_i \left( 1-\frac{2K_{i,n} K_{1,n}}{P_n} + \frac{u K_{i,n}}{P_n}\pm O\left( \left( \frac{K_{1,n}K_{i,n}}{P_n} \right)^2 \right) \right) \nonumber \\
&= \sum_{i=1}^r \mu_i \Bigg( 1-\frac{2K_{i,n} K_{1,n}}{P_n} + \frac{u}{K_{1,n}} \frac{K_{i,n} K_{1,n}}{P_n} \nonumber \\
&\quad \pm O\left( \left( \frac{K_{1,n}K_{i,n}}{P_n} \right)^2 \right) \Bigg) \nonumber \\
&= \sum_{i=1}^r \mu_i \Bigg( 1-2 \left( p_{1i}(n) \pm O\left( p_{1i}(n)^2 \right) \right) \nonumber \\
&\quad + \frac{u}{K_{1,n}} \left( p_{1i}(n) \pm O\left( p_{1i}(n)^2 \right) \right) \pm O\left( p_{1i}(n) \right)^2 \Bigg)  \nonumber \\
&=\sum_{i=1}^r \mu_i \left( 1-2 p_{1i}(n) + \frac{u}{K_{1,n}} p_{1i}(n)  \pm O\left( p_{1i}(n) \right)^2 \right) \nonumber \\
&=1-2\lambda_1(n)+\frac{u}{K_{1,n}} \lambda_1(n) \pm O\left( \sum_{i=1}^r \mu_i  \left( p_{1i}(n) \right)^2 \right)  
\label{eq:lemma9_11}
\end{align}

Next, we note that
\begin{align}
\mathbb{E}\left[Z_n \left(\pmb{\mu},\pmb{\theta}_n\right)^2 \right] &=\sum_{i=1}^r \mu_i \left( 1-p_{1i}(n) \right)^2 \nonumber \\
&=\sum_{i=1}^r \mu_i \left( 1-2p_{1i}(n)+ \left( p_{1i}(n)\right)^2 \right) \nonumber \\
&=1-2\lambda_1(n)+\sum_{i=1}^r \mu_i \left( p_{1i}(n)\right)^2  \nonumber
\end{align}

Now, we recall from (\ref{eq:lemma9_5}) that
\begin{equation} \nonumber
\mathbb{E}\left[Z_n \left(\pmb{\mu},\pmb{\theta}_n\right)^2 \right] \leq 1-2 \lambda_1(n) + \lambda_1(n)^2 \left( 1+\frac{1}{4 \mu_r^2} \right),
\end{equation}
and it follows that
\begin{align}
\sum_{i=1}^r \mu_i \left( p_{1i}(n)\right)^2  &\leq \lambda_1(n)^2 \left( 1+\frac{1}{4 \mu_r^2} \right) \nonumber \\
&=O\left( \lambda_1(n)^2 \right)
\label{eq:lemma9_12}
\end{align}

Combining (\ref{eq:lemma9_3}), (\ref{eq:lemma9_4}), (\ref{eq:lemma9_11}), and (\ref{eq:lemma9_12}), the conclusion
\begin{align}
&\mathbb{P} \left[ \left( K_{xj} \cap K_{yj}\right) \given[\bigg] \left(|S_{xy}|=u\right),t_x=1,t_y=1 \right] \nonumber \\
&= \frac{u}{K_{1,n}} \lambda_1(n) \pm O\left( \lambda_1(n)^2 \right),
\label{eq:lemma9_14}
\end{align}
follows.

Next, we establish (\ref{eq:lemma9_13}). We know that
\begin{align}
&\hspace{-5mm} \mathbb{P} \left[ E_{xj \cup yj} \given[\bigg] \left(|S_{xy}|=u\right),t_x=1,t_y=1 \right]\nonumber \\ 
&=\mathbb{P} \left[ E_{xj} \given[\bigg] \left(|S_{xy}|=u\right),t_x=1,t_y=1 \right] \nonumber \\
& +\mathbb{P} \left[ E_{yj} \given[\bigg] \left(|S_{xy}|=u\right),t_x=1,t_y=1 \right] \nonumber \\
& -\mathbb{P} \left[ E_{xj \cap yj} \given[\bigg] \left(|S_{xy}|=u\right),t_x=1,t_y=1 \right].
\label{eq:lemma9_18}
\end{align}

Now, since $E_{xj}=C_{xj} \cap K_{xj}$ and $E_{yj}=C_{yj} \cap K_{yj}$ , it is clear that $E_{xj}$ and $E_{yj}$ are each independent of the event $|S_{xy}|=u$. It follows that
\begin{align}
\mathbb{P} \left[ E_{xj} \given[\bigg] \left(|S_{xy}|=u\right),t_x=1,t_y=1\right]&=\mathbb{P} \left[ E_{xj}  \given[\big] t_x=1 \right] \nonumber \\
&= \Lambda_1(n),
\label{eq:lemma9_15}
\end{align}
and similarly
\begin{equation}
\mathbb{P} \left[ E_{yj} \given[\bigg] \left(|S_{xy}|=u\right),t_x=1,t_y=1 \right]=\Lambda_1(n).
\label{eq:lemma9_16}
\end{equation}

Finally,
\begin{align}
&\mathbb{P} \left[ E_{xj \cap yj} \given[\bigg] \left(|S_{xy}|=u\right),t_x=1,t_y=1\right] \nonumber \\
&=\mathbb{P} \left[ C_{xj} \cap C_{yj} \right] \cdot \nonumber \\
& \cdot  \mathbb{P} \left[ K_{xj} \cap K_{yj} \given[\bigg] \left(|S_{xy}|=u\right),t_x=1,t_y=1\right] \nonumber \\
&=\alpha_n^2 \mathbb{P} \left[ K_{xj} \cap K_{yj} \given[\bigg] \left(|S_{xy}|=u\right),t_x=1,t_y=1 \right] \nonumber \\
&=\frac{\alpha_n u}{K_{1,n}} \Lambda_1(n) \pm O \left( \Lambda_1(n)^2 \right)
\label{eq:lemma9_17}
\end{align}
by virtue of (\ref{eq:lemma9_14}). Combining (\ref{eq:lemma9_18}), (\ref{eq:lemma9_15}), (\ref{eq:lemma9_16}), and (\ref{eq:lemma9_17}), the conclusion (\ref{eq:lemma9_13}) follows.
\end{proof}

{
\lemma 
\label{lemma_4}
Consider a scaling $K_1,\ldots,K_r,P: \mathbb{N}_0 \rightarrow \mathbb{N}_0^{r+1}$ such that (\ref{scaling_condition_K}) holds, a scaling $\alpha: \mathbb{N}_0 \rightarrow (0,1)$, $\Lambda_1(n)=\frac{\log n+(k-1)\log \log n+\gamma_n}{n}$, with $\lim_{n \to \infty} \gamma_n=-\infty$. Let $m_1$, $m_2$, and $m_3$ be non-negative integer constants. We define event $\mathcal{F}$ as follows.
\begin{equation}
\mathcal{F}:=\left[|N_{xy}|=m_1\right] \cap \left[|N_{x \overline{y}}|=m_2\right] \cap \left[|N_{\overline{x} y}|=m_3\right].
\label{eq:lemma4_1}
\end{equation}

Then, given $u$ in $\{0,1,\ldots,K_{1,n}\}$ and $\Lambda_1(n)=o(\frac{1}{\sqrt n})$ under $\lim_{n \to \infty} \gamma_n=- \infty$, we have
\begin{align}
&\mathbb{P}\left[ \mathcal{F} \given[\bigg] \left(|S_{xy}|=u\right),t_x=1,t_y=1 \right] \nonumber \\
& \sim  \frac{n^{m_1+m_2+m_3}}{m_1! m_2! m_3!} e^{-2n \Lambda_1(n)+\frac{u \alpha_n }{K_{1,n}}n\Lambda_1(n)} \cdot   \nonumber \\
&\cdot \left( \mathbb{P}\left[ E_{xj \cap yj} \given[\bigg] \left( |S_{xy}|=u \right),t_x=1,t_y=1 \right] \right)^{m_1} \nonumber \\
&\cdot \left( \mathbb{P}\left[ E_{xj \cap \overline{yj}} \given[\bigg] \left( |S_{xy}|=u \right),t_x=1,t_y=1 \right] \right)^{m_2} \nonumber \\
&\cdot  \left( \mathbb{P}\left[ E_{\overline{xj} \cap yj} \given[\bigg] \left( |S_{xy}|=u \right),t_x=1,t_y=1 \right] \right)^{m_3} \nonumber
\end{align} 
with $j$ distinct from $x$ and $y$.
}

\begin{proof}
The proof of Lemma~\ref{lemma_4} is very similar with \cite[Lemma~4]{Jun/K-Connectivity}; in fact, it would follow directly from \cite[Eq. (212)-(213)]{Jun/K-Connectivity} if we show that
\begin{align}
&\left( \mathbb{P}\left[ E_{\overline{xj} \cap \overline{yj}} \given[\Big] \left( |S_{xy}|=u\right) ,t_x=1,t_y=1 \right] \right)^{n-m_1-m_2-m_3-2}
\nonumber \\
 \label{eq:lemma4_6}
 &\sim e^{-2n\Lambda_1(n)+\frac{u \alpha_n}{K_{1,n}}n \Lambda_1(n)}.
\end{align}
Recalling Lemma~\ref{lemma:lemma9} and the fact that $\Lambda_1(n) \leq \frac{\log n+(k-1)\log \log n}{n}$ for all $n$ sufficiently large under $\lim_{n \to \infty} \gamma_n=-\infty$, we get
\begin{align}
& \mathbb{P}\left[ E_{\overline{xj} \cap \overline{yj}} \given[\Big] \left( |S_{xy}|=u\right) ,t_x=1,t_y=1 \right] \nonumber \\
 &= 1 - \mathbb{P}\Big[ E_{xj \cup yj}  \given[\Big] \left( |S_{xy}|=u\right) ,t_x=1,t_y=1 \Big]
 \label{eq:lemma4_2}
 \\
&=1- \left(2 \Lambda_1(n) - \frac{\alpha_n u}{K_{1,n}} \Lambda_1(n) \pm O\left( \left( \Lambda_1(n) \right)^2 \right)\right) \nonumber \\
&= 1- O\left(\frac{\log n}{n}\right)=1-o(1).
\label{eq:lemma4_3}
\end{align} 
Also, 
\begin{align}
&\left(n-m_1-m_2-m_3-2\right) \cdot \nonumber \\
&\quad \cdot \left( \mathbb{P}\left[ E_{xj \cup yj} \given[\Big] \left( |S_{xy}|=u\right) ,t_x=1,t_y=1 \right]\right)^2 \nonumber \\
&= \left(n-m_1-m_2-m_3-2\right) \left[ O \left( \frac{\log n}{n} \right) \right]^2 =o(1)
\label{eq:lemma4_4}
\end{align}

Invoking Fact~\ref{fact:3} for (\ref{eq:lemma4_2}), and using (\ref{eq:lemma4_3}) and (\ref{eq:lemma4_4}), we get
\begin{align}
&\left(\mathbb{P}\left[ E_{\overline{xj} \cap \overline{yj}} \given[\Big] \left( |S_{xy}|=u\right) ,t_x=1,t_y=1 \right] \right)^{n-m_1-m_2-m_3-2} \nonumber \\
&\sim e^{\left(n-m_1-m_2-m_3 \right) \mathbb{P}\left[E_{xj \cup yj} \given[\big] \left( |S_{xy}|=u\right) ,t_x=1,t_y=1 \right]} \nonumber \\
&\sim e^{-n \left[ 2 \Lambda_1(n) - \frac{\alpha_n u}{K_{1,n}} \Lambda_1(n) \pm o\left( \frac{1}{n} \right) \right]} e^{\left( m_1+m_2+m_3+2 \right) o(1)} \nonumber \\
&\sim e^{-2n\Lambda_1(n)+\frac{u \alpha_n}{K_{1,n}}n \Lambda_1(n)}.
\label{eq:lemma4_5}
\end{align}
This gives (\ref{eq:lemma4_6}) and Lemma~\ref{lemma_4} is established in view of 
\cite[Lemma~4]{Jun/K-Connectivity}.
\end{proof}

{
\lemma [{\cite[Lemma~10]{Jun/K-Connectivity}}]
\label{lemma_10}
If $P_n \geq 2 K_{1,n}$, we have
\begin{equation} \nonumber
\mathbb{P} \left[ |S_{xy}|=u \given[\Big] t_x=1,t_y=1 \right] \leq \frac{1}{u!} \left( \frac{K_{1,n}^2}{P_n-K_{1,n}} \right)^u
\end{equation}
}

{
\lemma
\label{lemma:kconn_2}
With $m \geq 2$ and $\Lambda_1(n)=o(1)$, we have
\begin{equation} \nonumber
\mathbb{E}\left[ \frac{\binom{P_n-Q(\nu_m)}{|\Sigma|}}{\binom{P_n}{|\Sigma|}} \right] \leq e^{- \left( 1+\frac{\epsilon}{2}\right) \Lambda_1(n)},
\end{equation}
for all $n$ sufficiently large and any $\epsilon \in (0,1)$,  where we define
\[
Q(\nu_m)=K_{1,n} \pmb{1}\left[ |\nu_m|=1 \right]+\left(\left \lfloor{\left(1+\epsilon\right) K_{1,n}}\right \rfloor+1 \right)\pmb{1}\left[ |\nu_m|>1 \right]. 
\]
}
\begin{proof}
Consider fixed $\pmb{K}, P$. We have
\begin{align}
Q(\nu_m) 
&\geq K_1 \left( \pmb{1}\left[ |\nu_m|=1 \right] + (1+\epsilon) \pmb{1}\left[ |\nu_m|>1 \right] \right) \nonumber
\end{align}
Thus, by recalling (\ref{eq:combinatorial_bound}), we get
\begin{align}
\mathbb{E}\left[ \frac{\binom{P-Q(\nu_m)}{|\Sigma|}}{\binom{P}{|\Sigma|}} \right] &\leq \mathbb{E} \left[ \frac{\binom{P- K_1}{|\Sigma|}}{\binom{P}{|\Sigma|}}^{\pmb{1}\left[ |\nu_m|=1 \right] + (1+\epsilon) \pmb{1}\left[ |\nu_m|>1 \right] } \right] \nonumber \\
&= \mathbb{E}\left[ Z^{\pmb{1}\left[ |\nu_m|=1 \right] + (1+\epsilon) \pmb{1}\left[ |\nu_m|>1 \right] } \right]  \nonumber
\end{align}
where $Z=\frac{\binom{P-K_1}{|\Sigma|}}{\binom{P}{|\Sigma|}}$. Taking the expectation over $|\nu_m|$, we get
\begin{align}
&\mathbb{E}\left[ \frac{\binom{P-Q(\nu_m)}{|\Sigma|}}{\binom{P}{|\Sigma|}} \right] \nonumber \\
&\leq \mathbb{E} \Big[ \left( 1-\alpha \right)^m + m \alpha \left( 1-\alpha \right)^{m-1} Z \nonumber \\
&\quad +\left(1-\left( 1-\alpha\right)^m-m\alpha\left(1-\alpha\right)^{m-1} \right) Z^{1+\epsilon} \Big] \nonumber \\
& \leq \mathbb{E} \Big[ \left( 1-\alpha \right)^2 + 2 \alpha \left( 1-\alpha \right) Z \nonumber \\
& \quad +\left(1-\left( 1-\alpha\right)^2-2\alpha\left(1-\alpha\right) \right) Z^{1+\epsilon} \Big] \nonumber \\
&= \left( 1-\alpha \right)^2 + 2 \alpha \left( 1-\alpha \right)\mathbb{E} [Z] + \alpha^2 \mathbb{E} \left[ Z^{1+\epsilon} \right] \nonumber
\end{align}
by virtue of the fact that 
\begin{align*}
& \left( 1-\alpha \right)^m + m \alpha \left( 1-\alpha \right)^{m-1} T \nonumber \\
&\quad +\left(1-\left( 1-\alpha\right)^m-m\alpha\left(1-\alpha\right)^{m-1} \right) T^{1+\epsilon}
\end{align*}
is monotonically decreasing in $m$ (see \cite[Lemma~12]{Jun/K-Connectivity}).

Next, we have
\begin{align}
\mathbb{E}\left[ Z \right]
&=\sum_{j=1}^r \mu_j \frac{\binom{P-K_1}{K_j}}{\binom{P}{K_j}}=1-\lambda_1 \nonumber
\end{align}
Also by recalling Fact~\ref{fact:2}, we get
\begin{align}
\mathbb{E}\left[ Z^{1+\epsilon} \right]&=\mathbb{E}\left[ \left( \frac{\binom{P-K_1}{|\Sigma|}}{\binom{P}{|\Sigma|}}\right)^{1+\epsilon} \right] \nonumber \\
&=\sum_{j=1}^r \mu_j \left( \frac{\binom{P-K_1}{K_j}}{\binom{P}{K_j}}\right)^{1+\epsilon} \nonumber \\
&=\sum_{j=1}^r \mu_j (1-p_{1j}) (1-p_{1j})^\epsilon \nonumber \\
&\leq \sum_{j=1}^r \mu_j (1-p_{1j}) (1-\epsilon p_{1j}) \nonumber \\
&=1-\lambda_1 (1+\epsilon)+\epsilon \sum_{j=1}^r \mu_j p_{1j}^2. \nonumber
\end{align}
Note that
\begin{equation} \nonumber
\sum_{j=1}^r \mu_j \left(1-p_{1j} \right)^2=1-2\lambda_1+\sum_{j=1}^r \mu_j p_{1j}^2
\end{equation}
and we have from (\ref{eq:lemma9_7}) and (\ref{eq:lemma9_5}) that
\begin{equation} \nonumber
\sum_{j=1}^r \mu_j \left(1-p_{1j} \right)^2 \leq 1-2 \lambda_1+ \lambda_1^2 \left(1+\frac{1}{4 \mu_r^2} \right)
\end{equation}
This gives
\begin{equation} \nonumber
\sum_{j=1}^r \mu_j p_{1j}^2 \leq \lambda_1^2 \left(1+\frac{1}{4 \mu_r^2} \right)
\end{equation}
and we get
\begin{align}
&\mathbb{E}\left[ \frac{\binom{P-Q(\nu_m)}{|\Sigma|}}{\binom{P}{|\Sigma|}} \right] \nonumber \\
&\leq \left( 1-\alpha \right)^2 + 2 \alpha \left( 1-\alpha \right) (1-\lambda_1) \nonumber \\
&\quad + \alpha^2 \left( 1-\lambda_1 \left(1+\epsilon \right) + \epsilon \lambda_1^2 \left( 1+\frac{1}{4 \mu_r^2} \right) \right) \nonumber \\
&=1-\Lambda_1 \left( 2-(1-\epsilon)\alpha-\epsilon \left( 1+\frac{1}{4 \mu_r^2} \right) \Lambda_1\right) \nonumber
\end{align}

Now, consider a scaling such that $\Lambda_1(n)=o(1)$. We have
$\Lambda_1(n) \leq \frac{4 \mu_r^2}{2 \left( 4 \mu_r^2+1 \right)}$
for all $n$ sufficiently large. Given also that $\alpha_n \leq 1$, we  get
\begin{align}
\mathbb{E}\left[ \frac{\binom{P_n-Q(\nu_m)}{|\Sigma|}}{\binom{P_n}{|\Sigma|}} \right] &\leq 1-\Lambda_1 \left(2 -(1-\epsilon)- \frac{\epsilon}{2}\right)  \leq e^{- \left(1+ \frac{\epsilon}{2} \right) \Lambda_1(n)} \nonumber
\end{align}
for all $n$ sufficiently large. This completes the proof.
\end{proof}

\section{Proof of Lemma~\ref{lemma:3}}
\label{app:proof_of_lemma_3}
The law of total probability gives
\begin{align}
&\mathbb{P}\left[ {D}_{x,\ell} \cap {D}_{y,\ell} \right]\nonumber \\
&\quad =\mathbb{P}\left[ {D}_{x,\ell} \cap {D}_{y,\ell} \cap \overline{E_{xy}} \right]+\mathbb{P}\left[ {D}_{x,\ell} \cap {D}_{y,\ell} \cap E_{xy} \right].
\label{eq:lemma3_1}
\end{align}
Thus, Lemma~\ref{lemma:3} will be established upon showing the next two results.
{
\prop 
\label{prop:1}
Consider scalings $K_1,\ldots,K_r,P: \mathbb{N}_0 \rightarrow \mathbb{N}_0^{r+1}$ and $\alpha: \mathbb{N}_0 \rightarrow (0,1)$, such that  $\lambda_1(n)=o(1)$ and
(\ref{scaling_condition_KG}) holds with $\lim_{n \to \infty} \gamma_n=-\infty$. The following hold
%
%

(a) If $n \Lambda_1(n) = \Omega(1)$, then for any non-negative integer constant $\ell$ and any two distinct nodes $v_x$ and $v_y$, we have
\begin{equation}
\mathbb{P}\left[ {D}_{x,\ell} \cap {D}_{y,\ell} \cap \overline{E_{xy}}\right] \sim \mu_1^2 \left( \ell ! \right)^{-2} \left( n \Lambda_1(n) \right)^{2\ell} e^{-2n \Lambda_1(n)}
\label{eq:lemma3_2}
\end{equation}

(b) For any two distinct nodes $v_x$ and $v_y$, we have
\begin{equation}
\mathbb{P}\left[ {D}_{x,0} \cap {D}_{y,0} \cap \overline{E_{xy}}\right] \sim  \mu_1^2 e^{-2n \Lambda_1(n)}
\label{eq:lemma3_3}
\end{equation}
}

{
\prop 
\label{prop:2}
Consider scalings $K_1,\ldots,K_r,P: \mathbb{N}_0 \rightarrow \mathbb{N}_0^{r+1}$ and $\alpha: \mathbb{N}_0 \rightarrow (0,1)$, such that  $\lambda_1(n)=o(1)$ and
(\ref{scaling_condition_KG}) holds with $\lim_{n \to \infty} \gamma_n=-\infty$.
If  $n \Lambda_1(n) = \Omega(1)$, then for any non-negative integer $\ell$ and any  distinct nodes $v_x$ and $v_y$, we have
\begin{equation}
\mathbb{P}\left[ {D}_{x,\ell} \cap {D}_{y,\ell} \cap E_{xy} \right] =o \left( \mathbb{P}\left[ {D}_{x,\ell} \cap \mathcal{D}_{y,\ell} \cap \overline{E_{xy}}\right] \right)
\label{eq:lemma3_4}
\end{equation}
}

We establish Propositions~\ref{prop:1} and~\ref{prop:2} in the following two subsections respectively. Next, we show why Lemma~\ref{lemma:3} follows from Propositions~\ref{prop:1} and~\ref{prop:2}. If  $n \Lambda_1(n) = \Omega(1)$, then for any non-negative integer constant $\ell$, we observe that (\ref{eq:lemma3_conc1}) follows from (\ref{eq:lemma3_2}) and (\ref{eq:lemma3_4}) in view of (\ref{eq:lemma3_1}). Now, considering the case when $\ell=0$, we see that (\ref{eq:lemma3_3}) directly implies (\ref{eq:lemma3_conc2}) by virtue of (\ref{eq:lemma3_1}) and the fact that $\mathbb{P}\left[ {D}_{x,0} \cap {D}_{y,0} \cap E_{xy} \right]=0$ since it is impossible for nodes $v_x$ and $v_y$ to be adjacent to each other (i.e., under $E_{xy}$) when both nodes have zero degree.

\subsection{Proof of Proposition~\ref{prop:1}}
\label{sec:prop1}
Consider the vertex set $\mathcal{V}=\{v_1,\ldots, v_n\}$. For each node $v_i \in \mathcal{V}$, we define $N_i$ as the set of neighbors of node $v_i$. 
Also, for any pair of vertices $v_x, v_y$, we let $N_{xy}$ be the set of nodes in $\mathcal{V} \setminus \{v_x , v_y\}$ that are neighbors of both $v_x$ and $v_y$; i.e., $N_{xy} = N_x \cap N_y$. We also let $N_{x\overline{y}}$ denote the set of nodes in $\mathcal{V} \setminus \{v_x , v_y\}$ that are neighbors of $v_x$, but are not neighbors of $v_y$. Similarly, $N_{\overline{x}y}$ is defined as the set of nodes in $\mathcal{V} \setminus \{v_x , v_y\}$ that are not neighbors of $v_x$, but are neighbors of $v_y$. Finally, $N_{\overline{xy}}$ is the set of nodes in $\mathcal{V} \setminus \{v_x , v_y\}$ that are not connected to either $v_x$ or $v_y$. We also define $S_{xy}= \Sigma_x \cap \Sigma_y$.

We start by defining the series of events $A_h$ as follows
\begin{equation} \nonumber
A_h:=\left[|N_{xy}|=h\right] \cap \left[|N_{x \overline{y}}|=\ell-h\right] \cap \left[|N_{\overline{x} y}|=\ell-h\right].
\end{equation}
It is simple to see that
\begin{equation} \nonumber
{D}_{x,\ell} \cap {D}_{y,\ell} \cap \overline{E_{xy}}=\bigcup_{h=0}^\ell \left( A_h \cap \overline{E_{xy}} \cap  \left[t_x=1 \right] \cap \left[t_y=1\right]\right),
\end{equation}
whence  we get
\begin{align}
&\mathbb{P} \left[ {D}_{x,\ell} \cap {D}_{y,\ell} \cap \overline{E_{xy}} \right] \nonumber \\
&= \sum_{h=0}^\ell \mathbb{P} \left[ A_h \cap \overline{E_{xy}} \cap \left[ t_x=1\right] \cap \left[t_y=1\right] \right]
\label{eq:degree_18}
\end{align}
since the events $\{A_h, h=0,\ldots,\ell \}$ are mutually exclusive.

Furthermore, since $\overline{E_{xy}}=\overline{K_{xy}} \cup \overline{C_{xy}}=\overline{K_{xy}} \cup \left(K_{xy} \cap \overline{C_{xy}} \right)$
and
\[
K_{xy} \cap \left[t_x=1\right] \cap \left[t_y=1\right] =\cup_{u=1}^{K_{1,n}} \left( |S_{xy}|=u \right)
\]
we have under $t_x=t_y=1$ that
\begin{align}
\overline{E_{xy}} &=\overline{K_{xy}} \cup \left \{ \left[ \bigcup_{u=1}^{K_{1,n}} \left( |S_{xy}|=u \right) \right] \cap \overline{C_{xy}} \right\} \nonumber \\
&=\overline{K_{xy}} \cup \left( \bigcup_{u=1}^{K_{1,n}} \mathcal{X}_u \right)
\label{eq:degree_17}
\end{align}
where we define
 the event $\mathcal{X}_u$ as
\begin{equation}
\mathcal{X}_u=\left( |S_{xy}|=u \right) \cap \overline{C_{xy}}, \quad u=1,\ldots,K_{1,n}
\label{eq:degree_16}
\end{equation}

Now, we get
\begin{align}
&\mathbb{P}\left[ A_h \cap \overline{E_{xy}} \cap \left[t_x=1\right] \cap \left[t_y=1\right] \right] \nonumber \\
&=\mathbb{P}\left[ A_h  \cap \overline{K_{xy}} \cap \left[t_x=1\right] \cap \left[t_y=1\right] \right] \nonumber \\
 & ~ + \sum_{u=1}^{K_{1,n}} \mathbb{P}\left[ A_h \cap \mathcal{X}_u \cap \left[t_x=1\right] \cap \left[t_y=1\right]  \right], 
\label{eq:degree_19}
\end{align}
by virtue of (\ref{eq:degree_17}) and the fact that the events $\overline{K_{xy}}, \mathcal{X}_1, \mathcal{X}_2, \ldots, \mathcal{X}_{K_{1,n}}$ are mutually disjoint. Combining (\ref{eq:degree_18}) and (\ref{eq:degree_19}) we obtain
\begin{align}
&\mathbb{P}\left[ {D}_{x,\ell} \cap {D}_{y,\ell} \cap \overline{E_{xy}}\right] \nonumber \\
& =\mu_1^2 \sum_{h=0}^\ell \mathbb{P}\left[ A_h \cap \overline{K_{xy}} \given[\Big] t_x=1,t_y=1  \right] \nonumber \\
 & ~~ + \mu_1^2 \sum_{h=0}^\ell \sum_{u=1}^{K_{1,n}} \mathbb{P}\left[ A_h \cap \mathcal{X}_u \given[\Big] t_x=1,t_y=1  \right].
\label{eq:degree_20}
\end{align}

Proposition~\ref{prop:1} is established by virtue of (\ref{eq:degree_20}) and the following two results.

{
\prop
\label{prop:1.1}

Consider scalings $K_1,\ldots,K_r,P: \mathbb{N}_0 \rightarrow \mathbb{N}_0^{r+1}$ and $\alpha: \mathbb{N}_0 \rightarrow (0,1)$, such that  $\lambda_1(n)=o(1)$ and
(\ref{scaling_condition_KG}) holds with $\lim_{n \to \infty} \gamma_n=-\infty$. Then
for any non-negative integer $\ell$, we have
\begin{align}
&\sum_{h=0}^\ell \mathbb{P}\left[ A_h \cap \overline{K_{xy}}\given[\big] t_x=t_y=1  \right] \sim \left( \ell! \right)^{-2} \hspace{-1mm}\left( n \Lambda_1(n)\right)^{2 \ell} \hspace{-1mm}e^{-2n \Lambda_1(n)}
\label{eq:prop1.1_conc}
\end{align}

}

{
\prop
\label{prop:1.2}
Consider scalings $K_1,\ldots,K_r,P: \mathbb{N}_0 \rightarrow \mathbb{N}_0^{r+1}$ and $\alpha: \mathbb{N}_0 \rightarrow (0,1)$, such that  $\lambda_1(n)=o(1)$ and
(\ref{scaling_condition_KG}) holds with $\lim_{n \to \infty} \gamma_n=-\infty$. If $n \Lambda_1(n) = \Omega(1)$,  then 
\begin{align}
&\sum_{h=0}^\ell  \sum_{u=1}^{K_{1,n}}  \mathbb{P}\left[ A_h \cap \mathcal{X}_u \given[\Big] t_x=1,t_y=1 \right] \nonumber \\
&=o \left( \sum_{h=0}^\ell \mathbb{P}\left[ A_h \cap \overline{K_{xy}}\given[\Big] t_x=1,t_y=1\right] \right)
\label{eq:prop1.2_conc1}
\end{align}
for any $\ell=0,1,\ldots$. Furthermore, we have (\ref{eq:prop1.2_conc1}) for $\ell=0$ without requiring
the condition $n \Lambda_1(n) = \Omega(1)$.
%
}

Before we prove Propositions~\ref{prop:1.1} and~\ref{prop:1.2}, we explain why Proposition~\ref{prop:1} follows from these two results. Combining (\ref{eq:prop1.1_conc}) and (\ref{eq:prop1.2_conc1}) we establish (\ref{eq:lemma3_2}) in view of (\ref{eq:degree_20}). Furthermore, by using (\ref{eq:prop1.1_conc}) and
 (\ref{eq:prop1.2_conc1}) with $\ell=0$, we readily obtain (\ref{eq:lemma3_3}) in view of (\ref{eq:degree_20}). This establishes Proposition~\ref{prop:1}. 

\subsubsection
{Proof for Proposition~\ref{prop:1.1}} 
We write
\begin{align} \nonumber
&\sum_{h=0}^\ell \mathbb{P}\left[ A_h \cap \overline{K_{xy}}\given[\Big] t_x=1,t_y=1 \right] \nonumber \\ \nonumber
&= \sum_{h=0}^\ell \mathbb{P}\left[ A_h \given[\Big] \overline{K_{xy}},t_x=1,t_y=1\right]   \mathbb{P} \left[ \overline{K_{xy}} \given[\Big] t_x=1,t_y=1 \right],
\end{align}
where
\begin{align}
\mathbb{P} \left[ \overline{K_{xy}} \given[\Big] t_x=1,t_y=1 \right]=1-p_{11}(n) \sim 1 \label{eq:p_11_o_1}
\end{align}
under the assumption $\lambda_1(n)=o(1)$ and Fact~\ref{fact:lambda_o(1)}. 
Also, using Lemma~\ref{lemma_4} with $u=0$, $m_1=h$, and $m_2=m_3=\ell-h$, we see that
\begin{align}
&\mathbb{P}\left[ A_h \given[\Big] \overline{K_{xy}},t_x=1,t_y=1\right] \nonumber \\
&\sim \frac{n^{2\ell -h }}{h! \left( \left(\ell -h\right)! \right)^2} e^{-2n \Lambda_1(n)} \cdot \nonumber \\
&  ~~~ \cdot \left( \mathbb{P} \left[ E_{xj \cap yj} \given[\Big] \overline{K_{xy}},t_x=1,t_y=1\right] \right)^h
\nonumber \\
& ~~~ \cdot \left(\mathbb{P} \left[  E_{xj \cap \overline{yj}} \given[\Big] \overline{K_{xy}},t_x=1,t_y=1  \right] \right)^{\ell-h} \nonumber \\
&  ~~~ \cdot \left( \mathbb{P} \left[  E_{\overline{xj} \cap yj} \given[\Big] \overline{K_{xy}},t_x=1,t_y=1  \right] \right)^{\ell-h}.
\label{eq:prop1.1_1}
\end{align}

Next, we evaluate the three probability terms appearing in (\ref{eq:prop1.1_1}). We know that
\begin{align}
&\mathbb{P} \left[ E_{xj \cap yj} \given[\Big] \overline{K_{xy}},t_x=1,t_y=1\right] \nonumber \\
&=\mathbb{P} \left[ C_{xj} \cap C_{yj} \right] \cdot \mathbb{P} \left[ K_{xj} \cap K_{yj} \given[\Big] \overline{K_{xy}},t_x=1,t_y=1 \right] \nonumber \\
&=\alpha_n^2 \mathbb{P} \left[ K_{xj} \cap K_{yj} \given[\Big] \overline{K_{xy}},t_x=1,t_y=1 \right] \nonumber \\
&\leq \left(1+\frac{1}{4 \mu_r^2} \right) \Lambda_1(n)^2
\label{eq:prop1.1_2}
\end{align}
by virtue of Lemma~\ref{lemma:lemma9}. We also see that
\begin{align}
&\mathbb{P} \left[ E_{xj \cap \overline{yj}} \given[\Big] \overline{K_{xy}},t_x=1,t_y=1\right] \nonumber \\
&=\mathbb{P} \left[ E_{xj} \given[\Big] \overline{K_{xy}},t_x=1,t_y=1\right] \nonumber \\
&\quad -\mathbb{P} \left[ E_{xj \cap yj} \given[\Big] \overline{K_{xy}},t_x=1,t_y=1\right] \nonumber \\
&=\mathbb{P} \left[ E_{xj} \given[\Big] t_x=1 \right]-\mathbb{P} \left[ E_{xj \cap yj} \given[\Big] \overline{K_{xy}},t_x=1,t_y=1\right] \nonumber \\
&=\Lambda_1(n)-O\left( \Lambda_1(n)^2 \right) \nonumber \\
&\sim \Lambda_1(n)
\label{eq:prop1.1_3}
\end{align}
as we invoke (\ref{eq:prop1.1_2}) and use the fact that $\Lambda_1(n)=o(1)$ under $\lim_{n \to \infty} \gamma_n = -\infty$. It is also easy to see that 
\begin{equation}
\mathbb{P} \left[ E_{\overline{xj} \cap yj} \given[\Big] \overline{K_{xy}},t_x=1,t_y=1\right] \sim \Lambda_1(n)
\label{eq:prop1.1_4}
\end{equation}
via similar arguments. 

For $h=1,2,\ldots,\ell$, we observe from (\ref{eq:prop1.1_1}), (\ref{eq:prop1.1_2}), (\ref{eq:prop1.1_3}), and (\ref{eq:prop1.1_4})
that 
\begin{align}
&\frac{\mathbb{P}\left[ A_h \given[\Big] \overline{K_{xy}},t_x=1,t_y=1\right]}{\mathbb{P}\left[ A_0 \given[\Big] \overline{K_{xy}},t_x=1,t_y=1\right]} \nonumber \\
&\sim \frac{n^{-h} \left( \ell ! \right)^2}{h! \left( \left(\ell-h \right)! \right)^2} \Biggr( \frac{\mathbb{P} \left[ E_{xj \cap yj} \given[\Big] \overline{K_{xy}},t_x=1,t_y=1\right]}{\mathbb{P} \left[  E_{xj \cap \overline{yj}} \given[\Big] \overline{K_{xy}},t_x=1,t_y=1  \right] } \cdot \nonumber \\
& \cdot \frac{1}{\mathbb{P} \left[  E_{\overline{xj} \cap yj} \given[\Big] \overline{K_{xy}} ,t_x=1,t_y=1 \right]}\Biggr)^h \nonumber \\
& \leq \frac{n^{-h} \left( \ell ! \right)^2}{h! \left( \left(\ell-h \right) !\right)^2} \left( \frac{\left( 1+\frac{1}{4 \mu_r^2}\right) \Lambda_1(n)^2}{\Lambda_1(n)^2 (1-o(1))} \right)^h \nonumber \\
&=o(1)
\label{eq:prop1.1_5}
\end{align}
Similarly, setting $h=0$, we obtain
\begin{equation}
\mathbb{P}\left[ A_0 \given[\Big] \overline{K_{xy}},t_x=1,t_y=1\right] \sim \left( \ell ! \right)^{-2} \left(n \Lambda_1(n) \right)^{2 \ell} e^{-2n\Lambda_1(n)}
\label{eq:prop1.1_6}
\end{equation}

The conclusion (\ref{eq:prop1.1_conc}) follows by combining (\ref{eq:p_11_o_1}), (\ref{eq:prop1.1_5}), (\ref{eq:prop1.1_6}), and noting that $\ell$ is constant.

\subsubsection{Proof of Proposition~\ref{prop:1.2}}
Our approach is to find an upper bound to the left hand side of (\ref{eq:prop1.2_conc1}) and show that this upper bound is $o \left( \sum_{h=0}^\ell \mathbb{P}\left[ A_h \cap \overline{K_{xy}}\given[\Big] t_x=1,t_y=1\right] \right)$. It will be clear that the condition $n \Lambda_1(n) = \Omega(1)$ needed to establish (\ref{eq:prop1.2_conc1}) is not needed for the case when $\ell=0$.

We know that
\begin{align}
&\mathbb{P}\left[ A_h \cap \mathcal{X}_u \given[\Big] t_x=1,t_y=1\right]\nonumber \\
&=\mathbb{P}\left[ A_h \cap |S_{xy}|=u \cap \overline{C_{xy}} \given[\Big] t_x=1,t_y=1\right] \nonumber \\
&\leq \mathbb{P}\left[ A_h \cap |S_{xy}|=u \given[\Big] t_x=1,t_y=1\right] \nonumber
\end{align}
%

Thus,
\begin{align}
&\sum_{h=0}^\ell  \sum_{u=1}^{K_{1,n}}  \mathbb{P}\left[ A_h \cap \mathcal{X}_u \given[\Big] t_x=1,t_y=1\right] \nonumber \\
 &\leq \sum_{h=0}^\ell  \sum_{u=1}^{K_{1,n}}  \mathbb{P}\left[ A_h \cap |S_{xy}|=u \given[\Big] t_x=1,t_y=1\right] \nonumber \\
&= \sum_{u=1}^{K_{1,n}} \mathbb{P}\left[ |S_{xy}|=u \given[\Big] t_x=1,t_y=1\right]  \cdot \nonumber \\
&\quad \cdot \sum_{h=0}^\ell  \mathbb{P}\left[ A_h \given[\Big] \left( |S_{xy}|=u\right) ,t_x=1,t_y=1\right] \nonumber
\end{align}

Now, since $E_{xj}=C_{xj} \cap K_{xj}$ and $E_{yj}=C_{yj} \cap K_{yj}$ , it is clear that $E_{xj}$ and $E_{yj}$ are each independent of the event $|S_{xy}|=u$. It follows that
\begin{align}
&\mathbb{P}\left[ E_{xj \cap yj} \given[\bigg] \left( |S_{xy}|=u \right),t_x=1,t_y=1 \right] \nonumber \\
&\leq \mathbb{P}\left[ E_{xj} \given[\bigg] \left( |S_{xy}|=u \right),t_x=1,t_y=1 \right] \nonumber \\
&=\Lambda_1(n).
\label{eq:prop_1.2_1}
\end{align}
Similarly, we have
\begin{align}
\mathbb{P}\left[ E_{xj \cap \overline{yj}} \given[\bigg] \left( |S_{xy}|=u \right),t_x=1,t_y=1 \right] \leq \Lambda_1(n)
\label{eq:prop_1.2_2}
\end{align}
and
\begin{align}
&\mathbb{P}\left[ E_{\overline{xj} \cap yj} \given[\bigg] \left( |S_{xy}|=u \right),t_x=1,t_y=1 \right] \leq \Lambda_1(n)
\label{eq:prop_1.2_3}
\end{align}

Now, using Lemma~\ref{lemma_4} with $m_1=h$, and $m_2=m_3=\ell-h$, (\ref{eq:prop_1.2_1}), (\ref{eq:prop_1.2_2}), and (\ref{eq:prop_1.2_3}), it follows that
\begin{align} \nonumber
&\mathbb{P}\left[ A_h \given[\Big] \left( |S_{xy}|=u\right) ,t_x=1,t_y=1 \right] \nonumber \\
&\leq 2n^{2\ell-h} e^{-2n \Lambda_1(n)+\frac{u \alpha_n}{K_{1,n}} n \Lambda_1(n)} \left( \Lambda_1(n) \right)^{2\ell -h}
\end{align}
for all $n$ sufficiently large. Thus, we get
\begin{align}
&\sum_{h=0}^\ell  \sum_{u=1}^{K_{1,n}}  \mathbb{P}\left[ A_h \cap \mathcal{X}_u \given[\Big] t_x=1,t_y=1\right] \nonumber \\ 
&\leq \sum_{u=1}^{K_{1,n}} \Bigg( \mathbb{P}\left[ |S_{xy}|=u \given[\Big] t_x=1,t_y=1\right] \cdot \nonumber \\
&\quad \cdot 2e^{-2n \Lambda_1(n)+\frac{u \alpha_n}{K_{1,n}} n \Lambda_1(n)} \sum_{h=0}^\ell \left( n \Lambda_1(n) \right)^{2 \ell -h}  \Bigg)
\label{eq:prop_1.2_4}
\end{align}

Now, if $n \Lambda_1(n)= \Omega(1)$ it follows that
\begin{equation}
\sum_{h=0}^\ell \left( n \Lambda_1(n) \right)^{2 \ell -h} = O \left( \left( n \Lambda_1(n) \right)^{2 \ell} \right).
\label{eq:prop_1.2_5}
\end{equation}
Note that (\ref{eq:prop_1.2_5}) follows trivially for $\ell=0$ with no condition on $n \Lambda_1(n)$. Combining (\ref{eq:prop_1.2_4}), (\ref{eq:prop_1.2_5}) and Lemma~\ref{lemma_10}, we get
\begin{align}
&\sum_{h=0}^\ell  \sum_{u=1}^{K_{1,n}}  \mathbb{P}\left[ A_h \cap \mathcal{X}_u \given[\Big] t_x=1,t_y=1\right] \label{eq:prop_a_3_before_final} \\
& \leq O \left( \left( n \Lambda_1(n) \right)^{2 \ell}  e^{-2n \Lambda_1(n)}\hspace{-.3mm} \right) \sum_{u=1}^{K_{1,n}} \hspace{-.6mm} \left(  \frac{K_{1,n}^2}{P_n-K_{1,n}} e^{\frac{\alpha_n}{K_{1,n}} n \Lambda_1(n)}{\hspace{-.7mm}} \right)^{\hspace{-.5mm}u} \nonumber 
\end{align}
In view of Proposition~\ref{prop:1.1} (and the fact that $\ell$ is constant),
 we will immediately establish the desired result (\ref{eq:prop1.2_conc1}) from  (\ref{eq:prop_a_3_before_final}) if we show that
\begin{equation}
 \frac{K_{1,n}^2}{P_n-K_{1,n}} e^{\frac{\alpha_n}{K_{1,n}} n \Lambda_1(n)} =o(1).
 \label{eq:prop_1.2_7}
\end{equation}

Next, we establish (\ref{eq:prop_1.2_7}). 
From  (\ref{scaling_condition_K}),
we get for all $n$ sufficiently large that 
\begin{equation} \nonumber
\frac{K_{1,n}^2}{P_n-K_{1,n}} \leq 2\frac{K_{1,n}^2}{P_n} \leq 4 p_{11}(n)
\end{equation}
where the last bound used the fact that $\frac{K_{1,n}^2}{P_n}\sim p_{11}(n)$
when $p_{11}(n)=o(1)$ (e.g., see {\cite[Lemma~4.2]{Yagan/Inhomogeneous}}); this in turn follows from the assumption that $\lambda_1(n)=o(1)$ in view of Fact~\ref{fact:lambda_o(1)}. It is also clear from the definition 
$\lambda_1(n) =\sum_{i=1}^r \mu_i p_{1i}(n)$ that $p_{11}(n) \leq \frac{1}{\mu_1}\lambda_1(n)$. Thus, for all $n$  large, we get 
\begin{equation} 
\frac{K_{1,n}^2}{P_n-K_{1,n}} \leq   \frac{4}{\mu_1} \lambda_1(n).
\label{eq:prop_1.2_8}
\end{equation}
Now, with $\Lambda_1(n) \leq \frac{\log n+(k-1)\log \log n}{n}$ for all $n$ sufficiently large under $\lim_{n \to \infty} \gamma_n=-\infty$, we see that 
\begin{equation}
n \Lambda_1(n) = n \alpha_n \lambda_1(n) \leq \frac{3}{2}\log n 
\label{eq:prop_1.2_9}
\end{equation}
for all $n$ sufficiently large. Combining (\ref{eq:prop_1.2_8}) and (\ref{eq:prop_1.2_9}) 
and the fact that $K_{1,n} \geq 2$, we obtain
\begin{equation}
\frac{K_{1,n}^2}{P_n-K_{1,n}} e^{\frac{\alpha_n}{K_{1,n}} n \Lambda_1(n)} = O(1) \lambda_1(n) e^{\frac{3}{4} \alpha_n \log n}.
\label{eq:prop_1.2_10}
\end{equation}

Next, we define $F(n)= \lambda_1(n) e^{\frac{3}{4} \alpha_n \log n}$. Fix $n$ sufficiently large such that (\ref{eq:prop_1.2_8}) and (\ref{eq:prop_1.2_9}). We consider the cases when $\alpha_n \leq \frac{1}{\log n}$ and $\alpha_n > \frac{1}{\log n}$.  In the former case, $F(n)\leq \lambda_1(n) e^{3/4}$ follows directly. In the latter case we use (\ref{eq:prop_1.2_9}) to get
\begin{equation} \nonumber
F(n) \leq \frac{3}{2} \frac{\log n}{n \alpha_n} e^{\frac{3}{4} \alpha_n \log n} \leq \frac{3}{2} \frac{\left( \log n \right)^2}{n} n^{\frac{3}{4}}
\end{equation}
by virtue of the fact that $\alpha_n \log n \leq \log n$. Combining the two bounds, we have
\begin{equation} \nonumber
F(n) \leq \max \left\{ \lambda_1(n) e^{0.75}, 1.5 n^{-0.25} \left( \log n \right)^2 \right\}
\end{equation}
for all $n$ sufficiently large. In view of $\lambda_1(n)=o(1)$ this immediately gives
$
\lim_{n \to \infty} F(n)=0$, 
and the conclusion (\ref{eq:prop_1.2_7}) follows in view of (\ref{eq:prop_1.2_10}). The desired result (\ref{eq:prop1.2_conc1}) is now established from (\ref{eq:prop_a_3_before_final}) and (\ref{eq:prop_1.2_7}) for constant $\ell$. Note that for $\ell=0$,
we have (\ref{eq:prop1.2_conc1}) without requiring
 $n \Lambda_1(n) = \Omega(1)$,
since that extra condition is used only once in obtaining (\ref{eq:prop_1.2_5}) which holds trivially for $\ell=0$. This establishes Proposition~\ref{prop:1.2}.

\subsection{Proof of Proposition~\ref{prop:2}}
\label{sec:prop2}

Recalling Proposition~\ref{prop:1.2} and (\ref{eq:degree_20}), Proposition~\ref{prop:2} will follow if we show that
\begin{align}
&\mathbb{P}\left[ {D}_{x,\ell} \cap {D}_{y,\ell} \cap E_{xy} \right] \nonumber \\
&=o \left( \sum_{h=0}^\ell \mathbb{P}\left[ A_h \cap \overline{K_{xy}} \given[\Big] t_x=1,t_y=1\right]  \right),
\label{eq:degree_21}
\end{align}
for each $\ell=1,\ldots$. To establish (\ref{eq:degree_21}), we define the series of events $B_h$ as follows
\begin{equation} \nonumber
B_h:=\left[|N_{xy}|=h\right] \cap \left[|N_{x \overline{y}}|=\ell-h-1\right] \cap \left[|N_{\overline{x} y}|=\ell-h-1\right],
\end{equation}
for each $h=0,1,\ldots,\ell-1$. Now, it is easy to see that
\begin{equation}
{D}_{x,\ell} \cap {D}_{y,\ell} \cap E_{xy}=\bigcup_{h=0}^{\ell-1} \left( B_h \cap E_{xy} \cap \left[t_x=1\right] \cap \left[t_y=1\right]\right).
\label{eq:degree_22}
\end{equation}

Note that $h$ varies from $0$ to $\ell-1$ in (\ref{eq:degree_22}) because given the event $E_{xy}$, nodes $x$ and $y$ are adjacent; thus, they could have at most $\ell-1$ nodes in common when their degrees are $\ell$. Since the events $B_h$ are mutually exclusive for $h=0,\ldots,\ell-1$, we get
\begin{align} 
&\mathbb{P}\left[ {D}_{x,\ell} \cap {D}_{y,\ell} \cap E_{xy} \right] \nonumber \\
&= \sum_{h=0}^{\ell-1} \mathbb{P}\left[ B_h \cap E_{xy} \cap \left[t_x=1\right] \cap \left[t_y=1\right]\right] \nonumber
\end{align}

Thus, the proof of Proposition~\ref{prop:2}
will be completed upon showing
\begin{align}
&\sum_{h=0}^{\ell-1} \mathbb{P}\left[ B_h \cap E_{xy} \cap \left[t_x=1\right] \cap \left[t_y=1\right]\right] \nonumber \\
&=o \left( \sum_{h=0}^\ell \mathbb{P}\left[ A_h \cap \overline{K_{xy}} \given[\Big] t_x=1,t_y=1\right] \right)
\label{eq:osy_new_to_do_for_prop_2}
\end{align}
under the enforced assumptions of Proposition~~\ref{prop:2}, namely, with $\lim_{n \to \infty} \gamma_n=-\infty$, and $n \Lambda_1(n)=\Omega(1)$.
%
%
Proceeding as before, and noting that $\mathbb{P}[E_{xy}]= \alpha \mathbb{P}[K_{xy}]$ we write
\begin{align}
&\sum_{h=0}^{\ell-1}   \mathbb{P}\left[ B_h \cap E_{xy}\cap \left[t_x=1\right] \cap \left[t_y=1\right] \right]  \label{eq:prop_2.1_1}
 \\
&= \mu_1^2 \alpha \sum_{h=0}^{\ell-1}  \sum_{u=1}^{K_{1,n}} \mathbb{P}\left[ B_h \cap  \left( |S_{xy}|=u \right) \given[\Big] t_x=1,t_y=1 \right] \nonumber \\
& \leq \mu_1^2 \sum_{u=1}^{K_{1,n}}  \mathbb{P} \left[ \left( |S_{xy}|=u \right) \right] \sum_{h=0}^{\ell-1} \mathbb{P} \left[ B_h \given[\Big]  |S_{xy}|=u, t_x=t_y=1 \right]
\nonumber
\end{align}

Next, by recalling Lemma~\ref{lemma_4} with $m_1=h$, $m_2=m_3=\ell-h-1$, we get
\begin{align}
&\mathbb{P}\left[ B_h \given[\Big] \left( |S_{xy}|=u \right),t_x=1,t_y=1 \right] \nonumber \\
&\sim \frac{n^{2\ell -h-2 }}{h! \left( \left(\ell -h-1\right)! \right)^2} e^{-2n \Lambda_1(n) + \frac{u \alpha_n \Lambda_1(n)}{K_{1,n}} n} \nonumber \\
& \times \left\{ \mathbb{P} \left[ E_{xj \cap yj} \given[\Big] \left( |S_{xy}|=u \right),t_x=1,t_y=1\right] \right\}^h
\nonumber \\
& \times \left\{ \mathbb{P} \left[  E_{xj \cap \overline{yj}} \given[\Big] \left( |S_{xy}|=u \right),t_x=1,t_y=1 \right] \right\}^{\ell-h-1} \nonumber \\
& \times \left\{ \mathbb{P} \left[  E_{\overline{xj} \cap yj} \given[\Big] \left( |S_{xy}|=u \right)  ,t_x=1,t_y=1\right] \right\}^{\ell-h-1}. \nonumber
\end{align}
Recalling (\ref{eq:prop_1.2_1}), (\ref{eq:prop_1.2_2}), and (\ref{eq:prop_1.2_3}), we get
\begin{align}
&\mathbb{P}\left[ B_h \given[\Big] \left( |S_{xy}|=u \right),t_x=1,t_y=1 \right] \nonumber \\
&\leq 2e^{-2n \Lambda_1(n) + \frac{u \alpha_n}{K_{1,n}} n \Lambda_1(n)} \left( n \Lambda_1(n) \right)^{2 \ell-h-2}
\label{eq:prop_2.1_2}
\end{align}
for all $n$ sufficiently large. Using (\ref{eq:prop_2.1_2}) in (\ref{eq:prop_2.1_1}), we get for all $n$ sufficiently large that
\begin{align}
&\sum_{h=0}^{\ell-1}  \mathbb{P}\left[ B_h \cap E_{xy} \cap \left[t_x=1\right] \cap \left[t_y=1\right] \right] \nonumber \\
&\leq \mu_1^2 \sum_{u=1}^{K_{1,n}}
\Bigg(\mathbb{P}\left[ |S_{xy}|=u \given[\Big] t_x=1,t_y=1\right] 
\cdot \nonumber \\
&\quad \cdot 2e^{-2n \Lambda_1(n)+\frac{u \alpha_n}{K_{1,n}} n \Lambda_1(n)} \sum_{h=0}^\ell \left( n \Lambda_1(n) \right)^{2 \ell -h-2}  \Bigg)
\nonumber \\
&= \mu_1^2 \left( n \Lambda_1(n) \right)^{-2} \times \text{ right hand side of } (\ref{eq:prop_1.2_4}) \nonumber \\
&=O \left(  \text{ right hand side of } (\ref{eq:prop_1.2_4}) \right)
\label{eq:prop_2.1_3}
\end{align}
since $n \Lambda_1(n)=\Omega(1)$. We have shown in the proof of Proposition~\ref{prop:1.2} that 
\[
 \textrm{right hand side of } (\ref{eq:prop_1.2_4})\hspace{-1mm}=\hspace{-1mm}o \left( \sum_{h=0}^\ell \mathbb{P}\left[ A_h \cap \overline{K_{xy}} \given[\big] t_x=t_y=1\right] \right)
 \]
Together with (\ref{eq:prop_2.1_3}) this establishes (\ref{eq:osy_new_to_do_for_prop_2}) and the proof of Proposition~\ref{prop:2} is complete.

\section{Confining $\gamma_n$}
\label{subsection:confining}
{
In this section, we show that establishing the one-law of Theorem~\ref{theorem:kconnectivity} under the additional constraint 
\begin{equation}
\gamma_n=o(\log n)
\label{eq:degree_gamma}
\end{equation}
establishes the one-law for the case when that additional constraint is not present. Namely, we will show that for any scaling that satisfies conditions  (\ref{eq:conn_Pn}), (\ref{eq:conn_KrPn}), (\ref{eq:conn_Kr_K1}), and  (\ref{scaling_condition_KG}) with $\lim_{n \to \infty} \gamma_n=+\infty$, there exists a scaling that satisfies the same conditions with $\lim_{n \to \infty} \gamma_n=+\infty$ { \emph {and}} $\gamma_n=o(\log n)$, such that the probability of $k$-connectivity under the latter scaling (with $\gamma_n=o(\log n)$) is less than or equal to that under the former scaling.

Firstly, consider a probability distribution $\pmb{\mu}=\{\mu_1,\ldots,\mu_r\}$ with $\mu_i >0$ for $i=1,\ldots,r$, a scaling $K^*_1,K^*_2,\ldots,K^*_r,P^*: \mathbb{N}_0 \rightarrow \mathbb{N}_0^{r+1}$, and a scaling $\alpha^*:\mathbb{N}_0 \rightarrow (0,1)$ such that
\begin{equation}
\Lambda^*_1(n)=\alpha^*_n \lambda^*_1(n) = \frac{\log n + (k-1)\log \log n+\gamma^*_n}{n}, \label{scaling_condition_KG_star}
\end{equation}
 for each $n=1,2, \ldots$. Assume that 
\begin{align}
P^*_n &= \Omega(n), \quad  
\frac{K^*_{r,n}}{P^*_n} =o(1), \quad \textrm{and} \quad
\frac{K^*_{r,n}}{K^*_{1,n}} =o(\log n)
\end{align} 
and that we have  $\lim_{n \to \infty} \gamma^*_n=+\infty$; i.e., the $^\ast$-scaling satisfies all conditions enforced by part (b) of Theorem~\ref{theorem:kconnectivity}.

Now, with the same distribution $\pmb{\mu}$, consider a scaling $\hat{K}_1,\hat{K}_2,\ldots,\hat{K}_r,\hat{P}:\mathbb{N}_0 \rightarrow \mathbb{N}_0^{r+1}$ and a scaling $\hat{\alpha}: \mathbb{N}_0 \rightarrow (0,1)$ such that $\hat{P}_n=P^*_n$ and $\hat{\pmb{K}}_n=\pmb{K}^*_n$. Obviously, we have $\hat{\lambda}_1(n)=\lambda^*_1(n)$ by recalling (\ref{eq:osy_edge_prob_type_ij}) and (\ref{eq:osy_mean_edge_prob_in_RKG}) and also that 
\begin{align}
\hat{P}_n & =  \Omega(n), \quad
\frac{\hat{K}_{r,n}}{\hat{P}_n} =o(1),  \quad \textrm{and} \quad
\frac{\hat{K}_{r,n}}{\hat{K}_{1,n}} =o(\log n). \nonumber
\end{align}
Next, let $\hat{\gamma}_n :=\min \left(\gamma^*_n,\log \log n \right)$ and define $\hat{\alpha}_n$  through
\begin{equation}
\hat{\alpha}_n \hat{\lambda}_1(n)=\frac{\log n+(k-1) \log \log n+\hat{\gamma}_n}{n}.
\label{scaling_condition_KG_hat}
\end{equation}
Clearly, we have  $\hat{\gamma}_n=o(\log n)$ and $\lim_{n \to \infty} \hat{\gamma}_n=+\infty$. This establishes that for any scaling satisfying the conditions of part (b) of 
of Theorem~\ref{theorem:kconnectivity}, there exists another scaling (with the same 
$\pmb{\mu}, \pmb{K}_n$, and $P_n$) that satisfies all of the same conditions {\em and}
(\ref{eq:degree_gamma}). In addition, this latter scaling has a smaller probability of a channel being {\em on} than the original scaling; i.e., we have
\begin{equation} 
\hat{\alpha}_n \leq \alpha^*_n, \qquad n=2, 3, \ldots
\label{eq:coupling_1}
\end{equation}
 by virtue of the fact that $\hat{\gamma}_n \leq \gamma^*_n$ for all $n$.
 
In view of the above, we will establish that
part (b) of Theorem~\ref{theorem:kconnectivity} under $\gamma_n=o(\log n)$
 implies Theorem~\ref{theorem:kconnectivity}
%
if we show that 
\begin{equation} 
\mathbb{P} \left[ \begin{split} &\mathbb{H}(n;\pmb{\mu},\pmb{K}^*_n,P^*_n,\alpha^*_n) \\ &\text{is }k- \text{connected} \end{split} \right] \geq \mathbb{P} \left[ \begin{split} &\mathbb{H}(n;\pmb{\mu},\hat{\pmb{K}}_n,\hat{P}_n,\hat{\alpha}_n) \\ &\text{is }k- \text{connected}  \end{split} \right]
\label{eq:coupling:final}
\end{equation} 
This is clear since (\ref{eq:coupling:final}) would ensure that if $\mathbb{H}(n;\pmb{\mu},\hat{\pmb{K}}_n,\hat{P}_n,\hat{\alpha}_n)$ is $k$-connected asymptotically almost surely 
(as would be deduced from Theorem~\ref{theorem:kconnectivity} under $\gamma_n=o(\log n)$), then so would $\mathbb{H}(n;\pmb{\mu},\pmb{K}^*_n,P^*_n,\alpha^*_n)$.



In view of (\ref{eq:coupling_1}), we get (\ref{eq:coupling:final}) by means of an easy coupling argument showing that $\mathbb{H}(n;\pmb{\mu},\hat{\pmb{K}}_n,\hat{P}_n,\hat{\alpha}_n)$ is a spanning subgraph of $\mathbb{H}(n;\pmb{\mu},\pmb{K}^*_n,P^*_n,\alpha_n)$. This follows from the fact that under (\ref{eq:coupling_1}) the corresponding ER graphs satisfy
\[
\mathbb{G}(n; \hat{\alpha}_n) \subseteq \mathbb{G}(n; \alpha^*_n)
\]
meaning that for any monotone increasing graph property $\mathcal{P}$ (e.g., $k$-connectivity), the probability of 
that $\mathbb{G}(n; \alpha^*_n)$ has $\mathcal{P}$ is larger than that of $\mathbb{G}(n; \hat{\alpha}_n)$;
 see \cite[Section~V.B]{Jun/K-Connectivity} for details. 


\section{Proof of Lemma~\ref{lemma:kconn_1}}
\label{app:proof_of_lemma_A4}
From (\ref{eq:kconn_new_scaling}) and the fact that $\beta_{\ell,n}=o(\log n)$, we clearly have
\begin{equation}
\frac{1}{2} \frac{\log n}{n} \leq \Lambda_1(n) \leq 2 \frac{\log n}{n}
\label{eq:interm_step_for_lemma_A4}
\end{equation}
for all $n$ sufficiently large. We also have
\begin{equation*}
 \Lambda_1(n) =\alpha_n \sum_{j=1}^r \mu_j p_{1j} \geq \mu_r \alpha_n p_{1r}(n)
\end{equation*}

Now, since $p_{1j}$ is monotone increasing in $j=1,\ldots,r$ by virtue of (\ref{eq:ordering_of_lambda}), we also see that

\begin{align*}
\Lambda_1(n)&=\alpha_n \sum_{j=1}^r \mu_j p_{1j}(n)\leq \alpha_n p_{1r}(n) \sum_{j=1}^r \mu_j=\alpha_n p_{1r}(n)
\end{align*}

Thus, we obtain that
\begin{equation*}
\Lambda_1 \leq \alpha_n p_{1r}(n) \leq \frac{1}{\mu_r} \Lambda_1
\end{equation*}
and the conclusion (\ref{eq:kconn_range2}) immediately follows by virtue of (\ref{eq:interm_step_for_lemma_A4})
for all $n$ sufficiently large.

Next, we establish (\ref{eq:conn_prr_asym}). Here this  will be established  by showing that
\begin{align}
p_{rr}(n) &\leq \max \left( 2,  4 \frac{\log n}{w_n}\right) p_{1r}(n), \quad n=2,3, \ldots
\label{eq:to_show_p_rr_appendix}
\end{align}
for some sequence $w_n$ such that $\lim_{n \to \infty} w_n = \infty$.
Fix $n=2,3,\ldots.$  We  have either
$p_{1r}(n) > \frac{1}{2}$,
or
$p_{1r}(n) \leq \frac{1}{2}$.
In the former case, it automatically holds that
\begin{equation}
p_{rr}(n) \leq 2 p_{1r}(n)
\label{eq:rig:1st_part}
\end{equation}
by virtue of the fact that $p_{rr}(n) \leq 1$.

Assume now that $p_{1r}(n) \leq \frac{1}{2}$.
We know from \cite[Lemmas~7.1-7.2]{yagan2012zero} that
\begin{equation}
1-e^{-\frac{K_{j,n}K_{r,n}}{P_n}} \leq p_{jr}(n) \leq \frac{K_{j,n}K_{r,n}}{P_n-K_{j,n}}, ~~ j=1,\ldots, r
\label{eq:rig_3}
\end{equation}
and it follows that
\begin{align}
\frac{K_{1,n}K_{r,n}}{P_n} \leq \log \left( \frac{1}{1-p_{1r}(n)} \right) \leq \log 2 < 1.
\label{eq:rig_4}
\end{align}
Using the fact that $1-e^{-x} \geq \frac{x}{2}$ with $x$ in $(0,1)$,
we then get
\begin{equation}
p_{1r}(n) \geq \frac{K_{1,n}K_{r,n}}{2P_n}.  
\label{eq:p_1r_lower_bound}
\end{equation}
In addition, using the upper bound in (\ref{eq:rig_3}) with $j=r$  gives
\[
p_{rr}(n) \leq  \frac{K_{r,n}^2}{P_n-K_{r,n}} \leq 2 \frac{K_{r,n}^2}{P_n}
\]
as we invoke (\ref{scaling_condition_K}). Combining the last two bounds we obtain
\begin{equation}
\frac{p_{rr}(n)}{p_{1r}(n)} \leq  4 \frac{K_{r,n}}{K_{1,n}}
\label{eq:bound_on_prr_by_p1r}
\end{equation}

Next, combining (\ref{eq:conn_Kr_K1}) and (\ref{eq:bound_on_prr_by_p1r}), we get
\begin{equation}
p_{rr}(n) \leq 4 \frac{\log n}{w_n} p_{1r}(n)
\label{eq:rig:2nd_part}
\end{equation}
for some sequence $w_n$ such that $\lim_{n \to \infty} w_n = \infty$. Combining (\ref{eq:rig:1st_part}) and (\ref{eq:rig:2nd_part}), we readily obtain (\ref{eq:to_show_p_rr_appendix}).

It is easy to see that (\ref{eq:conn_p11_asym}) can be established using the same steps with the proof of (\ref{eq:to_show_p_rr_appendix}).

\section{Proof of Lemma~\ref{lemma_kconn_KeyLemma}}
\label{app:proof_of_KeyLemma}
Lemma \ref{lemma_kconn_KeyLemma} will be established by bounding each term in (\ref{eq:kconn_1}).
First, we note from \cite[Proposition~9.1]{Yagan/Inhomogeneous} that
\begin{equation} \nonumber
\mathbb{P}\left[ \mathcal{C}_m \right] \leq m^{m-2} \left( \alpha_n p_{rr}(n) \right)^{m-1}
\end{equation}
Next, we derive upper bounds on the terms $\mathbb{E}\left[ 1-\frac{\binom{P-|\nu_m|K_r}{|\Sigma|}}{\binom{P}{|\Sigma|}} \right] $ and $\mathbb{E}\left[ \frac{\binom{P-L(\nu_m)}{|\Sigma|}}{\binom{P}{|\Sigma|}} \right]$, respectively. It is clear that  Lemma~\ref{lemma_kconn_KeyLemma} will follow if we show that
\begin{align}
\mathbb{E}\left[ 1-\frac{\binom{P_n-|\nu_m|K_{r,n}}{|\Sigma|}}{\binom{P_n}{|\Sigma|}} \right] 
&\leq 1- e^{-3\alpha_n p_{rr}(n) m}
\label{eq:lem_8_1_to_show_1}
\end{align}
for all $m \leq \lfloor \frac{P-K_{r,n}}{2K_{r,n}}\rfloor$
and that
\begin{align} 
&\mathbb{E}\left[ \frac{\binom{P_n-L(\nu_m)}{|\Sigma|}}{\binom{P_n}{|\Sigma|}} \right] 
\label{eq:lem_8_1_to_show_2} \\
&\leq \min \Bigg(1-\Lambda_1(n),e^{- \left( 1+\frac{\epsilon}{2}\right) \Lambda_1(n)}, \nonumber \\
&\quad \min \Big( 1-\mu_r+\mu_r e^{-\alpha_n p_{1r}(n) \zeta m}, e^{-\alpha_n p_{11}(n) \zeta m} \Big) \nonumber \\
&\quad + e^{- \psi K_{1,n}} \pmb{1}\left[ m>m_n \right] \Bigg). \nonumber
\end{align}
We establish (\ref{eq:lem_8_1_to_show_1}) and (\ref{eq:lem_8_1_to_show_2}) in turn in the next two sections.
\subsection{Establishing (\ref{eq:lem_8_1_to_show_1}) }
First, with $m \leq \frac{P-K_r}{2K_r}$, we have
$|\nu_m| \leq m \leq \frac{P-K_r}{2K_r}
$
and using Fact~\ref{fact:fact5} we get
\begin{align}
\mathbb{E}\bigg[\hspace{-.3mm}1\hspace{-.5mm}-\hspace{-.5mm}\frac{\binom{P-|\nu_m|K_r}{|\Sigma|}}{\binom{P}{|\Sigma|}} \bigg] 
\hspace{-.5mm}\leq \mathbb{E}\bigg[\hspace{-.5mm}1\hspace{-.5mm}-\hspace{-.5mm}\left(\hspace{-.7mm}\frac{\binom{P-K_r}{|\Sigma|}}{\binom{P}{|\Sigma|}}\hspace{-.7mm} \right)^{\hspace{-1mm}2|\nu_m|} \bigg]\hspace{-.5mm}=\hspace{-.5mm}1\hspace{-.5mm}-\hspace{-.5mm}\mathbb{E}\left[W^{2|\nu_m|} \right] \label{eq:final_step_lemma_8_a}
\end{align}
where we set $W=\frac{\binom{P-K_r}{|\Sigma|}}{\binom{P}{|\Sigma|}}$. We also have
\begin{align}
\mathbb{E}\left[ W^{2|\nu_m|} \right]&=\mathbb{E}\left[ \sum_{j=0}^m \binom{m}{j} \alpha^j \left(1-\alpha \right)^{m-j} W^{2j} \right]\nonumber \\
&=\mathbb{E}\left[ \left( 1-\alpha \left( 1-W^2\right) \right)^m \right]\nonumber \\
& \geq \mathbb{E}\left[ \left( 1-2\alpha \left( 1-W\right) \right)^m \right] \label{eq:osy_also_know}
\end{align}
using Fact~\ref{fact:2} in the last step. We also know that
\begin{align}
W &= \frac{\binom{P-K_r}{|\Sigma|}}{\binom{P}{|\Sigma|}}  \geq \frac{\binom{P-K_r}{K_r}}{\binom{P}{K_r}}=1-p_{rr} \label{eq:osy_also_know2}
\end{align}
Thus,
\begin{equation} \nonumber
\alpha_n (1-W_n) \leq \alpha_n p_{rr}(n) \leq \frac{1}{4}
\end{equation}
for all $n$ sufficiently large by virtue of (\ref{eq:conn_prr_asym}) and that $\beta_{\ell,n}=o\left(\log n \right)$. Using the fact that
$1-2x \geq e^{-3x}$ for all $0 \leq x \leq \frac{1}{4}$, 
we then get from (\ref{eq:osy_also_know})  and (\ref{eq:osy_also_know2}) that
\begin{align}
\mathbb{E}\left[ W_n^{2|\nu_m|} \right]& \geq \mathbb{E}\left[e^{-3 \alpha_n  (1-W_n) m} \right] \geq e^{-3 \alpha_n p_{rr}(n) m } \nonumber
\end{align}
for all $n$ sufficiently large. 
The  desired conclusion (\ref{eq:lem_8_1_to_show_1})
now follows immediately by means of (\ref{eq:final_step_lemma_8_a}).

\subsection{Establishing (\ref{eq:lem_8_1_to_show_2})}
Let $\pmb{Y}$ be defined as follows
\begin{equation} \nonumber
Y_i=
\begin{cases} 
       \left \lfloor{i \zeta K_{1,n}}\right \rfloor    \hfill & i=2,\ldots,m_n \\
       \left \lfloor{\psi P_n}\right \rfloor \hfill & i=m_n+1,\ldots,n \\
\end{cases}
\end{equation}
where $\zeta \in (0,\frac{1}{2})$ selected small enough such that (\ref{eq:conn_zeta}) holds, and $\psi \in (0,\frac{1}{2})$ selected small enough such that (\ref{eq:conn_psi}) holds. Recalling (\ref{eq:conn_X}), we see that
\begin{equation} \nonumber
J_i=
\begin{cases} 
       \max \left(\left \lfloor{\left(1+\epsilon\right) K_{1,n}}\right \rfloor,Y_i \right)    \hfill & i=2,\ldots,m_n \\
       Y_i & i=m_n+1,\ldots,n \\
\end{cases}
\end{equation}
Next, we let
\begin{align}
 & M(\nu_m) 
\nonumber \\ 
 & = K_{1,n} \pmb{1}\left[ |\nu_m|=1 \right] + \max \left( K_{1,n},Y_{|\nu_m|}+1 \right)
 \pmb{1}\left[ |\nu_m|>1 \right],
 \nonumber
\end{align}
and
\begin{equation}
Q(\nu_m)=K_{1,n} \pmb{1}\left[ |\nu_m|=1 \right] + \left( \left \lfloor{\left(1+\epsilon\right) K_{1,n}}\right \rfloor +1\right) \pmb{1}\left[ |\nu_m|>1 \right].
\nonumber
\end{equation}
We also recall that
\begin{equation} \nonumber
L(\nu_m)=\max \left( K_{1,n} \pmb{1}\left[ |\nu_m|>0 \right], \left( J_{|\nu_m|}+1 \right) \pmb{1}\left[ |\nu_m|>1 \right]\right)
\end{equation}
Let's consider the following three cases
\begin{enumerate}
\item $|\nu_m|=0$: In this case we have $L(\nu_m)=M(\nu_m)=Q(\nu_m)=0$.
\item $|\nu_m|=1$: In this case we have $L(\nu_m)=M(\nu_m)=Q(\nu_m)=K_{1,n}$.
\item $|\nu_m| \geq 2$: In this case we have
\begin{itemize}
\item[--] $L(\nu_m)=\max \left(K_{1,n},J_{|\nu_m|}+1 \right)$.
\item[--] $M(\nu_m)=\max \left(K_{1,n},Y_{|\nu_m|}+1 \right)$.
\item[--] $Q(\nu_m)=\left \lfloor{\left(1+\epsilon\right) K_{1,n}}\right \rfloor +1$.
\end{itemize}
\end{enumerate}
More specifically, considering the case when $|\nu_m|=2,3,\ldots,m_n$, we have
\begin{equation} \nonumber
J_{|\nu_m|}=\max \left( (1+\epsilon)K_{1,n} , Y_{|\nu_m|} \right)
\end{equation}
and it follows that
\begin{align}
L(\nu_m)&=\max \left( K_{1,n},\lfloor(1+\epsilon)K_{1,n}\rfloor+1,Y_{|\nu_m|}+1 \right) \nonumber \\
&=\max \left( \lfloor(1+\epsilon)K_{1,n}\rfloor+1,M(\nu_m) \right) \nonumber \\
&=\max \left( Q(\nu_m), M(\nu_m) \right) \nonumber
\end{align}

Also, when $|\nu_m|=m_n+1,\ldots,n$, we clearly have $J_{|\nu_m|}=Y_{|\nu_m|}$, and thus
\begin{equation} \nonumber
L(\nu_m)=M(\nu_m)=\max \left( K_{1,n} , \left \lfloor{\psi P_n}\right \rfloor +1 \right).
\end{equation}
Since $K_{1,n} \leq K_{r,n}=o(P_n)$ in view of (\ref{eq:conn_KrPn}), we have
\begin{equation} \nonumber
\left \lfloor{\psi P_n}\right \rfloor \geq \left \lfloor{\left(1+\epsilon\right) K_{1,n}}\right \rfloor
\end{equation}
for all $n$ sufficiently large. Thus, we can rewrite $L(\nu_m)$ as 
\begin{align}
L(\nu_m)&=\max \left( K_{1,n} , \left \lfloor{\psi P_n}\right \rfloor +1 ,\left \lfloor{\left(1+\epsilon\right) K_{1,n}}\right \rfloor+1\right) \nonumber \\
&=\max \left( Q(\nu_m), M(\nu_m) \right).
\nonumber
\end{align}
Combining, we conclude that it always holds  
that
$
L(\nu_m)=\max \left( Q(\nu_m), M(\nu_m) \right)$,  
whence
\begin{equation}  \label{eq:final_lemma_8_1}
\mathbb{E}\left[ \frac{\binom{P-L(\nu_m)}{|\Sigma|}}{\binom{P}{|\Sigma|}} \right] \leq \min \left( \mathbb{E}\left[ \frac{\binom{P-M(\nu_m)}{|\Sigma|}}{\binom{P}{|\Sigma|}} \right],\mathbb{E}\left[ \frac{\binom{P-Q(\nu_m)}{|\Sigma|}}{\binom{P}{|\Sigma|}} \right] \right)
\end{equation}
Note that it was shown in \cite[Lemma~7.2]{Rashad/Inhomo} that 
\begin{align} 
&\mathbb{E}\left[ \frac{\binom{P-M(\nu_m)}{|\Sigma|}}{\binom{P}{|\Sigma|}} \right] \nonumber \\
&\leq \min \Big( 1-\Lambda_1(n) , \min \left( 1-\mu_r+\mu_r e^{-\alpha_n p_{1r}(n) \zeta m},e^{-\alpha_n p_{11}(n) \zeta m}\right) \nonumber \\
&\quad + e^{- \psi K_{1,n}} 1\left[ m>m_n \right] \Big) \nonumber
\end{align}
for all $n$ sufficiently large. 
On the same range, we also get from Lemma~\ref{lemma:kconn_2} that
\begin{equation} \nonumber
\mathbb{E}\left[ \frac{\binom{P_n-Q(\nu_m)}{|\Sigma|}}{\binom{P_n}{|\Sigma|}} \right] \leq e^{- \left( 1+\frac{\epsilon}{2}\right) \Lambda_1(n)}
\end{equation}
upon noting that $\Lambda_1(n)=o(1)$ under (\ref{eq:kconn_new_scaling}) with $\beta_{\ell,n}=o(\log n)$. 
Reporting the last two bounds into (\ref{eq:final_lemma_8_1}), we establish (\ref{eq:lem_8_1_to_show_2}).

\end{document}